\documentclass[12pt,preprint]{aastex}
\usepackage{emulateapj5}
\usepackage{epsfig}

\newcommand{\etal}{\mbox{et al.}}
\newcommand{\ergcms}{erg cm$^{-2}$ s$^{-1}$}
\newcommand{\ergsec}{erg s$^{-1}$}
\newcommand{\phcms}{ph cm$^{-2}$ s$^{-1}$}

\newcommand{\msun}{$M_{\odot}$}

\newcommand{\chandra}{{\it Chandra}}
\newcommand{\rosat}{{\it ROSAT}}

\newcommand{\asca}{{\it ASCA}}

\newcommand{\bepposax}{{\it BeppoSAX}}
\newcommand{\xmm}{{\it XMM-Newton}}
\newcommand{\integral}{{\it INTEGRAL}}

\newcommand{\sgrastar}{\mbox{Sgr A$^*$}}
\newcommand{\program}[1]{{\tt {#1}}}
\newcommand{\html}[1]{{\tt http://#1}}

\newcommand{\gcllb}{\mbox{GRS~1741.9$-$2853}}

\makeatletter
\newenvironment{inlinefigure}{%
\def\@captype{figure}%
\noindent\begin{minipage}{0.999\linewidth}\begin{center}}
{\end{center}\end{minipage}\smallskip}

\newcommand\newtablebreak{\cr\ptable@@split}
\makeatother

\shortauthors{Muno \etal}
\shorttitle{Properties of Galactic Center X-ray Sources}

\begin{document}
\title{The Spectra and Variability of X-ray Sources in a Deep \chandra\ Observation of the Galactic Center}
\author{M. P. Muno,\altaffilmark{1,2} 
J. S. Arabadjis,\altaffilmark{3}
F. K. Baganoff,\altaffilmark{3} M. W. Bautz,\altaffilmark{3}
W. N. Brandt,\altaffilmark{4} P. S. Broos,\altaffilmark{4}
E. D. Feigelson,\altaffilmark{4} G. P. Garmire,\altaffilmark{4}
 M. R. Morris,\altaffilmark{1} and
G. R. Ricker\altaffilmark{3}}

\altaffiltext{1}{Department of Physics and Astronomy, University of California,
Los Angeles, CA 90095; mmuno@astro.ucla.edu}
\altaffiltext{2}{Hubble Fellow}
\altaffiltext{3}{Center for Space Research,
Massachusetts Institute of Technology, Cambridge, MA 02139}
\altaffiltext{4}{Department of Astronomy and Astrophysics, 
The Pennsylvania State University, University Park, PA 16802}

\begin{abstract}
We examine the X-ray spectra and variability of the sample of X-ray
sources with $L_{\rm X} \approx  10^{31} - 10^{33}$~\ergsec\ identified 
within the inner 9\arcmin\ of the Galaxy by \citet{mun03}. Very few of 
the sources exhibit intra-day or inter-month variations. We find that 
the spectra of the point sources near the 
Galactic center are very hard 
between 2--8~keV, even after accounting for absorption.
When modeled as power laws the median photon index is $\Gamma = 0.7$, 
while when modeled as thermal plasma we can only obtain lower limits
to the temperature of $kT > 8$~keV.
The combined spectra of the point sources is similarly hard, with a photon
index of $\Gamma = 0.8$. Strong line emission is observed from low-ionization,
He-like, and H-like Fe, both in the average spectra and in the brightest
individual sources. The line ratios of the highly-ionized Fe in the 
average spectra
are consistent with emission from a plasma in thermal equilibrium. This
line emission is observed whether average spectra are examined as a function
of the count rate from the source, or as a function of the hardness ratios of 
individual sources. This suggests that the hardness of the spectra may in 
fact be to due local absorption that partially-covers the X-ray emitting 
regions in the Galactic center systems. We suggest that most of these 
sources are intermediate polars, which 
(1) often exhibit hard spectra with
prominent Fe lines, (2) rarely exhibit either flares on short time scales
or changes in their mean X-ray flux on long time scales, and 
(3) are the most numerous 
hard X-ray sources with comparable luminosities in the 
Galaxy. 
\end{abstract}


\section{Introduction\label{sec:intro}}

Recent deep \chandra\ observations of the inner 9\arcmin\ around the
super-massive black hole at the Galactic center have revealed 
over 2000 individual point-like X-ray sources \citep{mun03}. The sources 
have luminosities between $10^{31}$ and $10^{33}$~\ergsec 
\citep[for a distance of 8 kpc to the Galactic center; see][]{mcn00}, and thus 
they probably represent some combination of young stellar objects,
Wolf-Rayet and early O stars, interacting binaries (RS CVns), cataclysmic
variables (CVs), young pulsars, and black holes and neutron 
stars accreting from binary companions (low- and high-mass X-ray binaries;
LMXBs and HMXBs). 
However, the spectra of the Galactic center sources are very hard in the
energy range of 2--8 keV. Spectra that are similarly hard 
have only been observed previously from magnetically accreting 
CVs (mCVs) and HMXB pulsars. 
Moreover, seven of the hard sources exhibit X-ray modulations with 
periods between 300~s and 4.5~h, which also suggests that they are 
magnetized white dwarfs or neutron stars \citep{mun03c}. These basic 
observations are a good first step toward determining the natures of 
the point sources. However, if their natures can be determined 
conclusively, the large number of sources in the
field would make it possible to study two important pieces of astrophysics:
(1) the history of star formation at the Galactic center, and (2) 
the physics of X-ray production in accreting stellar remnants.

How stars form at the Galactic center is still a mystery, because the
strong tidal forces and milliGauss magnetic fields there should prevent 
all but the most massive molecular clouds from collapsing. Nonetheless,
it appears that star formation has occurred recently, because 
three massive stellar clusters younger than 
$10^{7}$ years old lie within $\approx 30$~pc of the Galactic 
center: IRS 16, the Arches, and the Quintuplet \citep{kra95,pau01,fig99}. 
However, it is still a matter of debate as to whether the star formation is
continuous or episodic, and whether it occurs only in localized regions or is 
relatively uniform throughout the Galactic center.
\citet{fig03} addressed this question by modeling the evolution of the
population of luminous infrared stars, and concluded that the star
formation is probably continuous. 
The X-ray sources at the Galactic center could provide an additional, 
independent constraint on the star formation history there, because
they should be dominated by accreting 
stellar remnants.

The size of the sample of X-ray sources --- an order of magnitude 
larger than the numbers of known LMXBs, HMXBs, and magnetic CVs --- 
also makes it a valuable database for studying the physics of X-ray production.
Several outstanding questions could be addressed with the current data.
If the sample contains large numbers of magnetic CVs, it 
could be used to determine the duty cycle of 
bright accretion states \citep[e.g.,][]{gs88} and the fraction of such 
systems that exhibit 
hard spectral components \citep[e.g.][]{ram03}. If there is a significant
number of neutron star HMXBs, it may be possible to determine whether 
material accreted at rates far below the Eddington limit can penetrate the 
neutron star's 
magnetosphere and reach its surface \citep{neg00,cam02,orl03}. 
Finally, the large sample of Galactic center X-ray sources 
would be useful for identifying systems with unusual properties. Previous
hard X-ray surveys of the Galactic plane have identified several 
slowly-rotating accreting neutron stars and white dwarfs
\citep{kin98,tor99,oos99,sak00,sug00}, magnetic CVs with
extremely strong emission lines from He-like Fe
\citep{mis96,ish98,ter99}, 
and accreting stellar remnants with high intrinsic absorption
\citep{pat03,mg03,wal03}. 
These systems could represent
resting points for stellar remnants that have not been observed previously, 
and are therefore important for 
calculating the formation rate of such remnants in the Galaxy.

In this paper, we take a further step toward the above goals by 
using the properties of the X-ray emission from the point sources near the 
Galactic center \citep{mun03} to constrain better their natures. 
In Sections 2.1--2.3, we examine the spectra of the point sources both 
individually and averaged together, in order to determine the temperatures 
of the emitting regions. In Section 2.4, we search for short-term 
variability, which is often seen from coronal X-ray sources, and long-term 
variability, which is common in some accreting X-ray sources. In Section~3, 
we compare the properties of the observed sources with those of known
classes of X-ray source. Finally, in Section~4, we briefly explore 
the future prospects for definitively identifying the natures of these 
sources.

\section{Observations and Data Analysis\label{sec:obs}}

Twelve separate pointings toward the Galactic center have been carried out 
using the Advanced CCD Imaging Spectrometer imaging array (ACIS-I) 
aboard the {\it Chandra X-ray Observatory} \citep{wei02}
in order to monitor \sgrastar\ (Table~\ref{tab:obs}).
The ACIS-I is a set of four 1024-by-1024 pixel CCDs, covering
a field of view of 17\arcmin\ by 17\arcmin. When placed on-axis at the focal
plane of the grazing-incidence X-ray mirrors, the imaging resolution 
is determined primarily by the pixel size of the CCDs, 0\farcs492. 
The CCDs also 
measure the energies of incident photons within a 
calibrated energy band of 0.5--8~keV, with a resolution of 50--300 eV 
(depending on photon energy and distance from the read-out node). 
The CCD frames are read out every 3.2~s,
which provides the nominal time resolution of the data.

The methods we used to create a combined image of the field,
to identify point sources, and to compute the photometry for each source
are described in \citet{mun03} and \citet{tow03}. In brief, for 
each observation we corrected the pulse heights of the events for
position-dependent charge-transfer inefficiency \citep{tow02b}, excluded 
events that did not pass the standard ASCA grade filters and \chandra\ X-ray
center (CXC) good-time 
filters, and removed intervals during which the background rate flares to
$\ge 3\sigma$ above the mean level. The final total live time was 626 ks. 
In order to produce a single composite image, we then applied a correction to 
the 
absolute astrometry of each pointing using three Tycho sources detected 
strongly in each \chandra\ observation \citep[compare][]{bag03},
and re-projected the sky coordinates of each event to the tangent plane at the 
radio position of \sgrastar.
The image (excluding the first half of ObsID 1561, during which the 
$10^{-10}$~\ergcms\ transient \gcllb\ was observed; see Muno et al. 2003c) 
was searched for point 
sources using \program{wavdetect} \citep{fre02} in three energy 
bands: 0.5--8~keV, 0.5--1.5~keV, and 4--8~keV. 
We used a significance threshold of $10^{-7}$,
which corresponds to the chance probability of detecting a spurious source 
within a beam defined by the point spread function (PSF). 
We detected a total of 2357 X-ray point sources. Of these, 
281 were detected in the soft band (124 exclusively in the soft band), and 
so are located in the foreground of the Galactic center. The remaining
sources, of which 1792 were detected in the full band, 
and 1832 in the hard band (441 exclusively in the hard band) are most likely 
located near or beyond the Galactic center.

We computed photometry for each source in the 0.5--8.0~keV band using 
the \program{acis\_extract} routine from the Tools for 
X-ray Analysis (TARA).\footnote{\html{www.astro.psu.edu/xray/docs/TARA/}}
We extracted event lists for each source for each observation, using a 
polygonal region generally chosen to match the contour of 90\% encircled energy
from the PSF, although smaller regions were used if 
two sources were nearby in the field. We used a region defined by the PSF  
for 1.5~keV photons
for foreground sources, and a larger extraction area corresponding 
to the PSF for 4.5 keV photons for Galactic center sources. 
A background event list 
was extracted for each source from a circular region 
centered on the point source, excluding from the event list ({\it i}) 
counts in circles circumscribing 
the 95\% contour of the PSF around any point sources and ({\it ii}) the 
bright, filamentary structures noted by \citet{par04}. The background region 
was unique for each observation. It was chosen to include a fraction 
of $\approx 1200$ total counts, where the number of 
counts from each observation was scaled to the fraction of the total 
exposure time. The photometry for the complete sample of sources is 
listed in the electronic version of Table~3 from \citet{mun03}.

We then extracted spectra and background estimates for 
each of the sources from the same regions from which we computed the 
photometry. We summed the source and background spectra from all 
12 observations. Then, we grouped the source spectra between 
0.5--8.0~keV so that each spectral bin contained at least 20 total counts.
Next, we computed the effective area function at the position of each 
source for each observation. This was corrected to account for the fraction 
of the PSF enclosed by the extraction region and for the time-varying 
hydrocarbon build-up on the 
detectors.\footnote{\html{www.astro.psu.edu/users/chartas/xcontdir/xcont.html}}
We estimated the detector response for each source in each observation
using position-dependent response files that accounted for the 
corrections we made to undo partially the charge-transfer inefficiency 
\citep{tow02a}.
Finally, to create composite functions
\begin{figure*}[ht]
\centerline{\epsfig{file=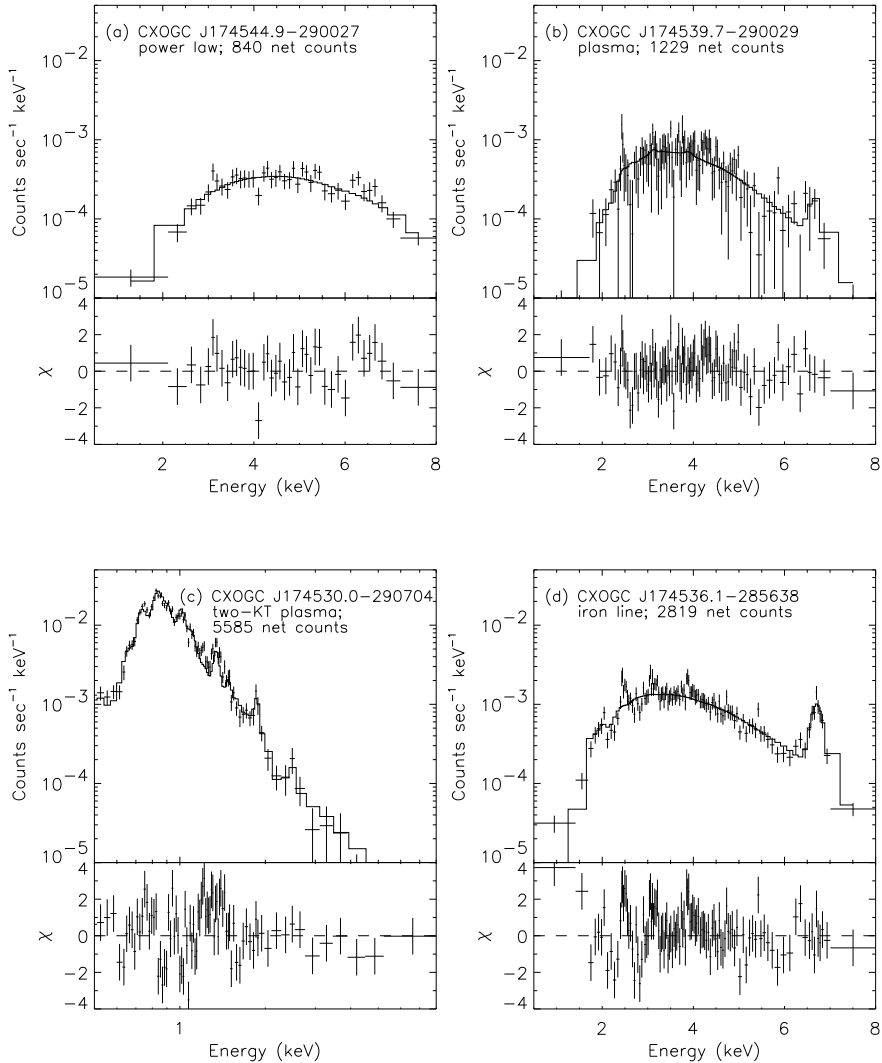,width=0.7\linewidth}}
\caption{Example spectra of four bright sources, illustrating the 
signal-to-noise available from the \chandra\ observations. The 
{\it top panels} have the spectrum in 
units of detector counts s$^{-1}$ keV$^{-1}$ as a function of energy
in keV, so that the varying effective area of the detector is convolved
with the spectrum. The solid histograms represent the best-fit spectra. The
{\it bottom panels} show the residuals to the fit, in units of the 
difference between the model and the data divided by the 
uncertainties on the data.
{\it Panel a} displays a Galactic center source with a $\Gamma = 0.2$ 
power-law spectrum. 
{\it Panel b} illustrates a Galactic center source with a spectrum consistent
with a $kT = 2~keV$ plasma. We note that this source is far off axis, so its 
PSF is $\approx 50$ times larger than for the other sources in 
the figure. As a result, the relative fraction of background counts is 
larger, and the uncertainty on the net counts also larger.
{\it Panel c} exhibits a foreground source that can be modeled with 
a two-temperature thermal plasma.
{\it Panel d} displays a Galactic center source with an Fe line at 6.7~keV
with a 2~keV equivalent width, as well as lines from ionized Si, S, and 
Ar.}
\label{fig:spec}
\end{figure*}
 for the full
data set, we averaged both the response and effective area functions, 
weighted by the number of counts detected from each source in each observation.
Four example spectra are displayed in Figure~\ref{fig:spec}.\footnote{The 
spectra, response functions, effective area functions, background estimates, 
and event lists for each source are available from 
\html{www.astro.psu.edu/users/niel/galcen-xray-data/galcen-xray-data.html}.}

We have confirmed that the spectra of the point sources were 
not contaminated by the diffuse X-ray emission in the field by repeating 
the analysis above for a subset 
of sources using an extraction region that enclosed only 50\% of the PSF
at 4.5~keV. The spectra were indistinguishable for the larger and smaller 
extraction regions, 
which 
confirms that we have successfully removed the background emission from the
point-source spectra.

\subsection{Spectra of Individual Sources}

We modeled the X-ray spectra of those sources with at least 80 net counts,
which provided four or more independent spectral bins. 
To provide a rough characterization of the spectrum, we used either
a power-law or thermal plasma continuum absorbed at low energies by 
gas and dust. To model the thermal plasma, we used 
\program{mekal} in \program{XSPEC} \citep{mew86}. 
We assumed that the elemental abundances were 0.5 
solar, which is consistent with the values derived from the average spectra
of the point sources (see Section~\ref{sec:av:vapec}), as well as the 
Fe abundances often observed from CVs 
\citep[e.g.,][]{do97,fi97,ish97}.\footnote{We 
note that the abundance parameter merely measures the relative strengths of 
the lines and the continuum. If the continuum is non-thermal or the lines are 
produced by photo-ionization, the abundance parameter will not measure the 
physical abundances of metals in the plasma.}
We accounted for 
gas absorption using the model {\tt phabs}, and the dust scattering using 
a modified version of the model {\tt dust}
\begin{figure*}[ht]
\centerline{\epsfig{file=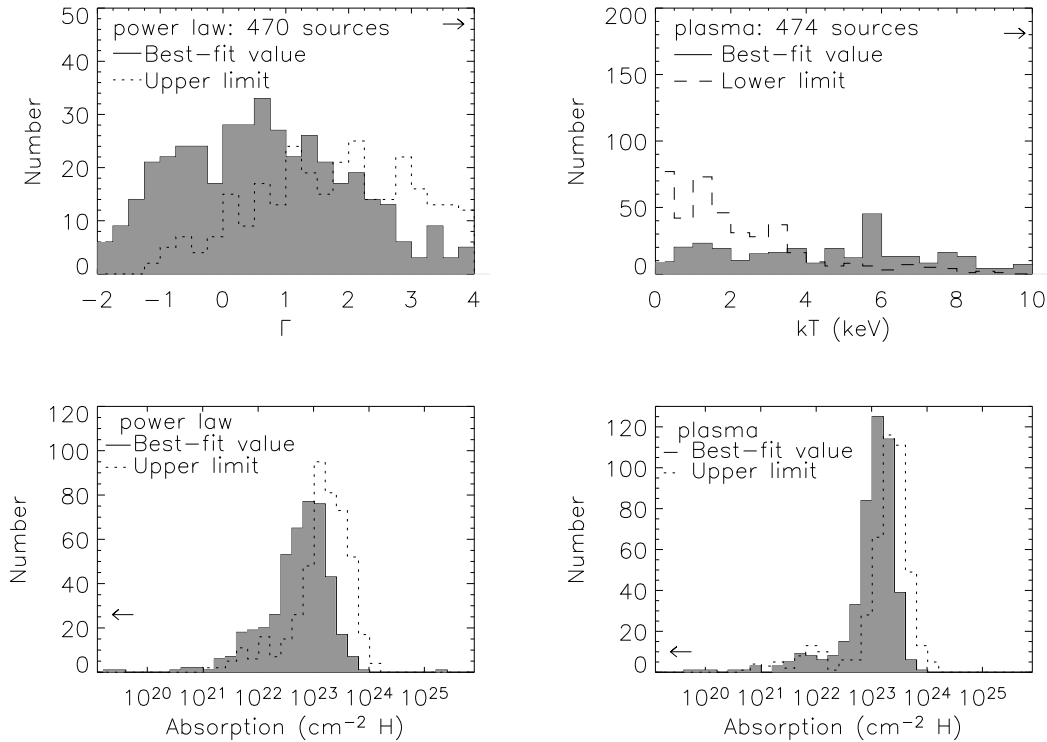,width=0.8\linewidth}}
\caption{
The distribution of the photon indices $\Gamma$, plasma temperatures $kT$,
and absorption columns $N_{\rm H}$ derived from spectral fits to sources
with at least 80 
net counts. We have only included those sources that are consistent with 
the spectral models at the 95\% confidence level. The filled histogram 
represents the best-fit values. The upper limits from the 90\%
uncertainty (single parameter of interest) on $\Gamma$ and $N_{\rm H}$
are indicated with the {\it dotted} histograms, and the lower limit on $kT$
with the {\it dashed} histogram. The arrows indicate the numbers of sources
with best-fit values outside the ranges of the plots. 
Most of the sources have hard spectra:
the median photon index is $\Gamma = 0.7$, while we were only able to obtain
lower limits on $kT$ from the plasma modes. The median absorption column
is $6\times10^{22}$ cm$^{-2}$.}
\label{fig:dist}
\end{figure*}
in which we removed the assumption that the dust was optically thin.
The column depth of dust was set to 
$\tau = 0.485 \cdot N_{\rm H}/(10^{22} {\rm cm}^{-2})$, and the halo size 
to 100 times the PSF size \citep{bag03}. In Table~\ref{tab:indiv} we list 
the parameters of the best-fit spectral models: the column densities
$N_{\rm H}$, either the power-law slope $\Gamma$ or the temperature $kT$,
the observed and de-absorbed 2--8~keV fluxes, and the reduced $\chi^2$. The 
uncertainties are 90\% confidence
intervals ($\Delta \chi^2 = 2.71$). We also indicate sources from
which the spectra should be viewed with caution, of which there
are three categories: confused sources for which 
the radii of their PSF overlap those of a nearby source by more than 25\%, 
sources that were near chip edges, and sources with variability 
(Section~\ref{sec:var}). About 25\% of the sources are flagged in this manner.

We consider a spectrum to be adequately reproduced by a model if the chance
probability of obtaining the derived value of $\chi^2$ is greater than 5\%.
Of the 566 sources that we modeled in this manner, the spectra of 
470 could be modeled with an absorbed power law, and 469 could be modeled with
an absorbed, collisionally-ionized plasma. Both spectral models 
were consistent with the data for 440 sources, because of limited statistics 
and the small bandpass over which photons are detected
(for most sources, Galactic absorption prevents photons $< 2$~keV from 
reaching the detector, while the effective area of the ACIS-I is small above 
8~keV).
Only 30 sources could be modeled with a power law but not 
a thermal plasma, while 29 sources could be modeled by a thermal plasma 
but not a power law. Finally, 
67 spectra deviated from both continuum models at the 95\% level. About 25
of these resemble statistical fluctuations, 30 are bright sources that appear
to require multiple continuum components, and 12 exhibited strong iron 
emission between 6 and 7 keV.

For individual sources, the uncertainties on the spectral parameters are 
rather large.
However, it is possible to draw some general conclusions about
the population of X-ray sources from Table~\ref{tab:indiv}. 
In Figure~\ref{fig:dist}, we plot the distributions of the best-fit
absorption columns and photon indices or temperatures for all of those 
sources that
\begin{inlinefigure}
\centerline{\epsfig{file=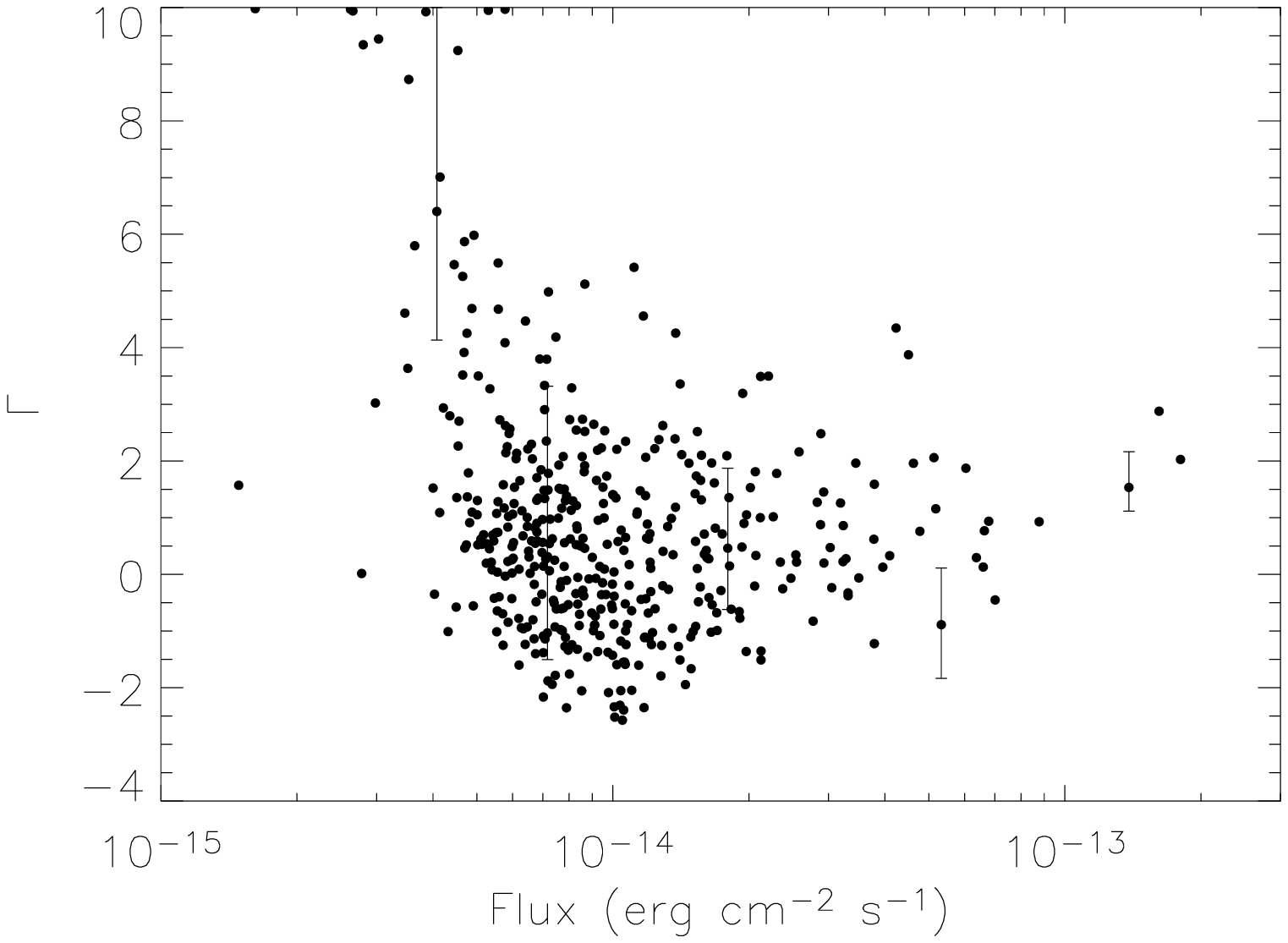,width=0.95\linewidth}}
\caption{
Plot of the best-fit power-law spectral index ($\Gamma$) as a function of the 
observed 2--8 keV flux. Representative uncertainties on $\Gamma$ are 
indicated for a few points. There is no correlation between $\Gamma$ and 
the flux.}
\label{fig:gvf}
\end{inlinefigure}

\noindent
 were adequately fit with the simple absorbed continuum model. 
As was mentioned in \citet{mun03}, 
265 out of 470 of the point sources have power-law spectra 
with $\Gamma < 1$, even after accounting for the absorption column.
 However,
only 77 sources have 90\% confidence limits for which $\Gamma < 1$, and
some of the apparent hardness of their spectra is due to line emission from 
Fe between 6--7 keV (see Section~\ref{sec:fe}). Not 
surprisingly, the hard sources also have high best-fit 
temperatures under the thermal plasma
 model. However, because of the poor 
statistics and small bandpass, 
the temperatures are unconstrained in one-third of the sources, and there 
are only 21 sources with 90\% lower limits on the temperatures that are
$>10$~keV. The slopes of the spectra are not correlated with the intensities
of the sources (Figure~\ref{fig:gvf}).

The median absorption column toward the 
the sources is $6 \times 10^{22}$~cm$^{-2}$ for a power-law continuum,
and $11 \times 10^{22}$~cm$^{-2}$ for a thermal plasma continuum.
The median value for the power-law continuum is identical to the column 
that is inferred from the $K$-band extinction toward 
\sgrastar\ \citep[see][for a summary]{td03},
while that for a thermal plasma continuum is twice as high.
The
 difference between the median values results from the facts that 
(1) for a faint, hard source the values of $N_{\rm H}$ and the spectral slope 
cannot be determined independently, and (2) the thermal plasma continuum,
can appear no harder than a $\Gamma \approx 1.5$ power law. Therefore,
the thermal plasma models produce a higher median value of $N_{\rm H}$.
Under either model, 30\% of the sources are inferred to have a
absorption column of $> 10^{23}$~cm$^{-2}$. The one
source that appears to be absorbed by more than $10^{24}$~cm$^{-2}$, 
CXOGC J174539.3--290027, is relatively faint, and the spectrum is 
poorly constrained.

Finally, we note that these spectra have been modeled assuming that all
of the X-rays from a given source are absorbed by a single column of
material. If we assume instead that a fraction of the X-ray emitting region 
is absorbed by a higher column (so-called ``partial covering'' models) 
an acceptable fit can be obtained for an arbitrary range of continuum 
shapes, because the bandpass over which we measure the spectrum is limited.

%
\begin{inlinefigure}
\centerline{\epsfig{file=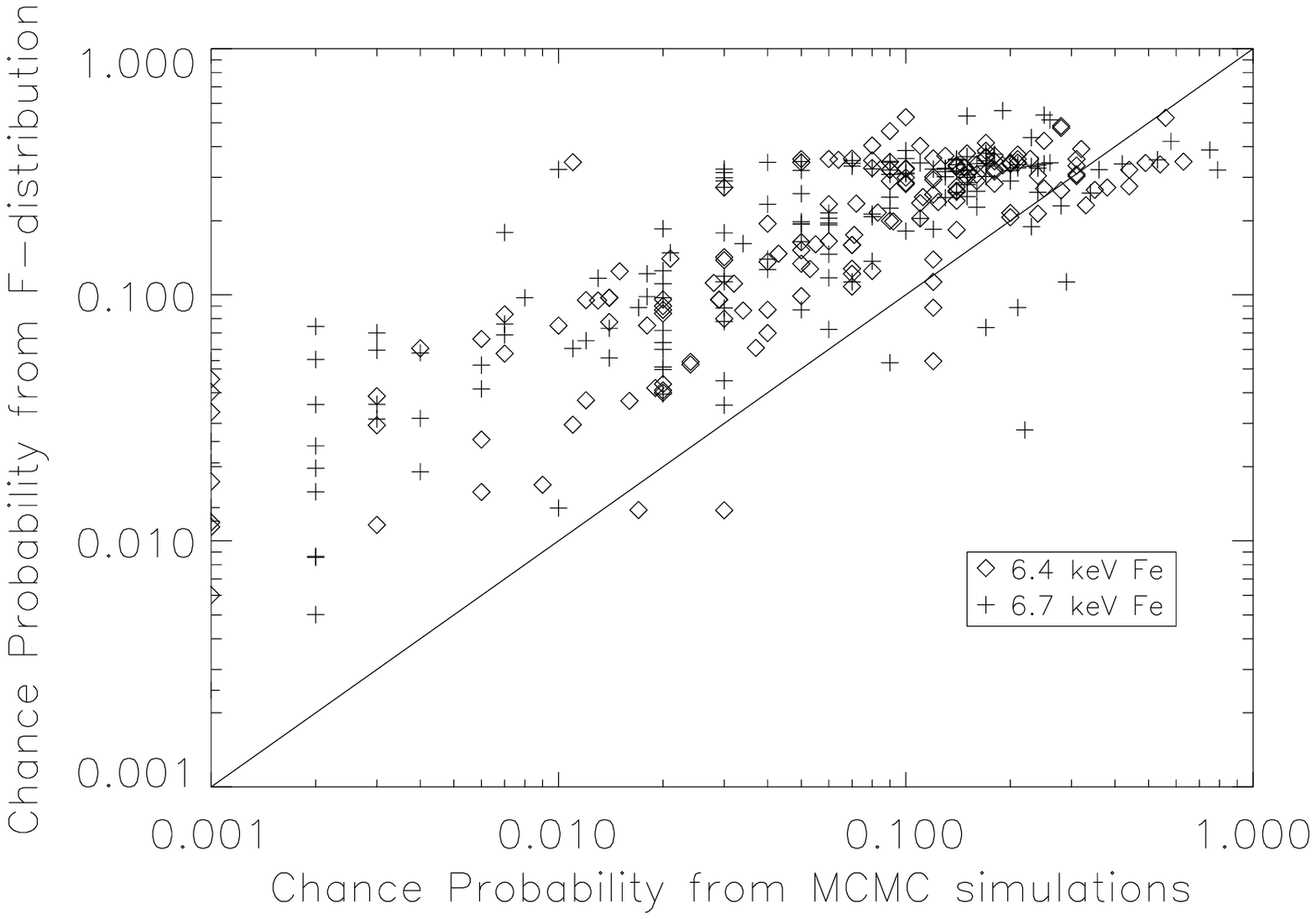,width=0.95\linewidth}}
\caption{Comparisons of the chance probability that an iron line would be 
detected from sources, derived using Markov-Chain Monte Carlo simulations and
using the $F$-distribution. The $F$-distribution generally over-estimates the
chance probability of the line, which would cause the significance of a line 
to be under-estimated.
}
\label{fig:mcmc}
\end{inlinefigure}

\noindent

\subsection{Iron Emission\label{sec:fe}}

Visual inspection of the 67 sources for which the simple continuum 
failed to reproduce their spectra indicates that $\approx 20$\% 
exhibit residuals between 6--7~keV that may represent line emission from 
Fe. Therefore, we have performed 
a uniform search for 
Fe line emission from the brightest sources with hard spectra. We selected
only those sources with more than 160 net counts, because we found that 
fainter sources were unlikely to provide more than one spectral bin 
between 6--7 keV. Likewise, we selected only
 those sources that were best-fit
by absorbed power laws with $\Gamma < 5$, because sources with steeper 
spectra were typically background-dominated above 6~keV. There were 
183 sources that met both of these criteria. We modeled each source with
an absorbed power law plus a Gaussian line that was fixed at either 6.4 keV
to search for low-ionization (``neutral'') Fe emission, or at 
6.7 keV to search for He-like 
Fe. The widths of the lines were fixed at 100 eV, to account for the fact that 
both the low-ionization and He-like lines are actually a blend of multiple 
transitions \citep[e.g.,][]{nag94}.

To evaluate the significance of the added line, we computed a statistic
$f$ from the reduction in $\chi^2$ provided by
 the more complex model:
\begin{equation}
f = {{(\chi^2_s - \chi^2_c)} \over {(\nu_s - \nu_c)}} 
{{\nu_s} \over {\chi^2_s}},
\end{equation}
where $\chi^2_c$ and $\chi^2_s$ are the values with and without the line,
and $\nu_c$ and $\nu_s$ are the numbers of degrees of freedom for the 
fits ($\nu_s - \nu_c = 1$, here). Unfortunately, because the null result of
our more complex model, a line of zero flux, lies on a boundary of the 
parameter space we are considering (i.e., 
an emission line with negative
flux is not physical, so the hard lower limit to the line flux is 0),
$f$ is not distributed according to the $F-$distribution \citep{pro02}.
Therefore, we have simulated the expected distribution of $f$ for each 
of our sources using
the Markov-Chain Monte Carlo technique described in \citet{ara03}. This 
technique allows us to evaluate the chance probability of observing a
value of $f$ given (1) the statistical distributions of 
$N_{\rm H}$, $\Gamma$, and 
\begin{figure*}[pth]
\centerline{\epsfig{file=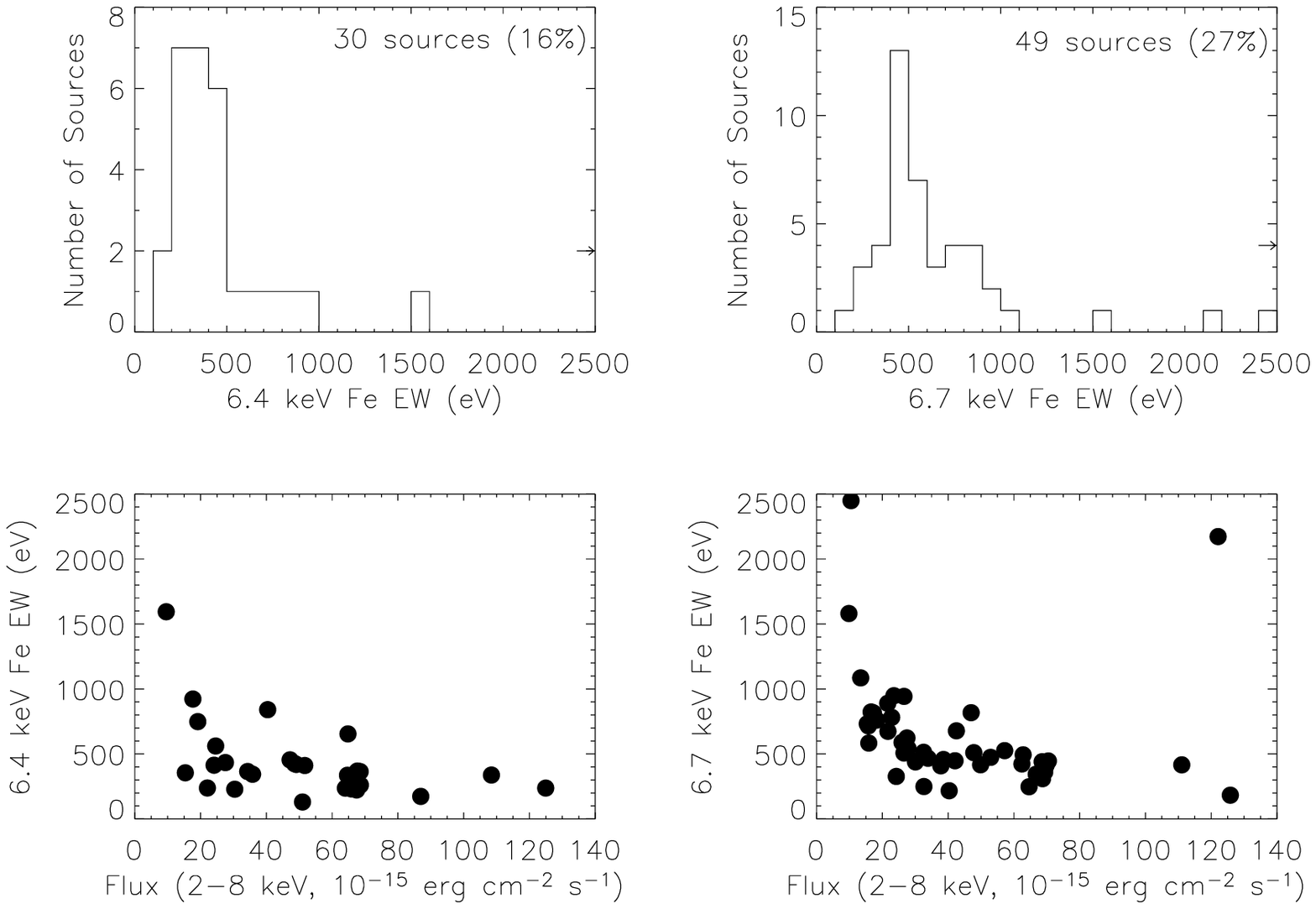,width=0.8\linewidth}}
\caption{The strength of line emission at {\it left panels:} 6.4 keV from
low-ionization Fe, and {\it right panels:} 6.7 keV from He-like Fe. 
The {\it top panels} display histograms of the equivalent widths 
of the iron lines. The rightward-pointing arrow denotes the
the four sources that may have iron emission with equivalent width
$> 2500$~keV. Most of these sources have very steep ($\Gamma > 4.5$) 
spectra, and so the excess emission between 6--7 keV could represent either 
a hard continuum component, or poor background subtraction. However, the line
from one of the sources with 6.4 keV iron emission with equivalent width 
$> 2500$ eV appears to be real (CXOGC J174617.2--285449).
The {\it bottom panels} illustrate how the
measured equivalent widths are correlated with the photon flux from a source.
Iron emission can only be detected in the fainter sources if it has a high
equivalent width.
}
\label{fig:iron}
\end{figure*}
\begin{inlinefigure}
\centerline{\epsfig{file=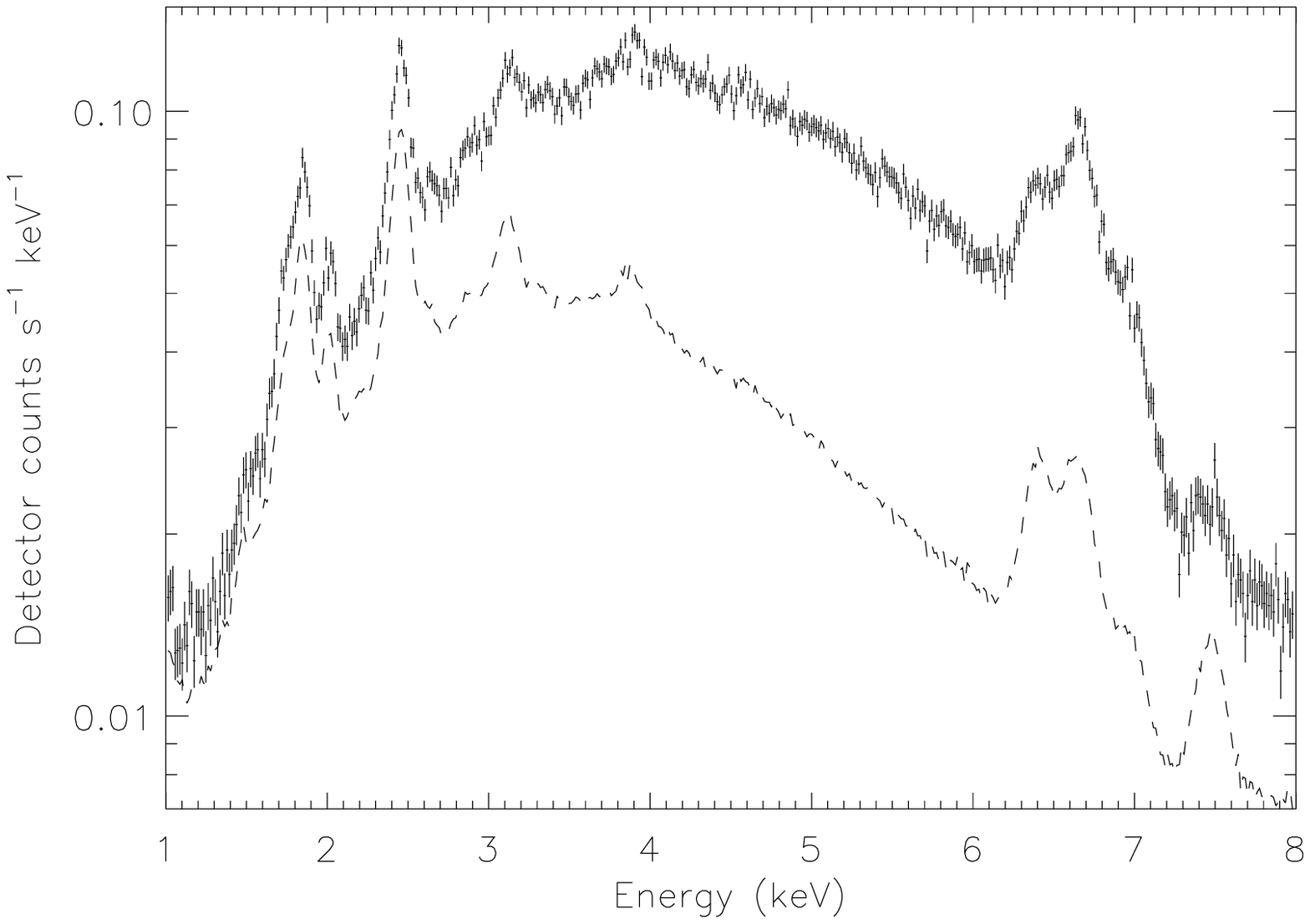,width=0.95\linewidth}}
\caption{
Combined spectrum of all Galactic center point sources with fewer 
than 500 net counts, before background subtraction. The background
spectrum is taken to be that of the diffuse emission, and is indicated with
the dashed curve.}
\label{fig:psraw}
\end{inlinefigure}

\noindent
the normalization from the fit, (2) the method used
to group the spectral bins, and (3) the spectrum assumed for the background
subtraction. Between 100--1000 simulations\footnote{We first computed 100 
simulations. For many of the sources, the $f$ values from the simulations 
exceeded the observed $f$ value more than once, indicating the significance 
of the line was $ 99$\%. For the rest, we ran 1000 simulations, to 
establish whether the line was at least 99\% significant (i.e., $< 10$ $f$ 
values exceeding the observed one).}
 were computed for each source. 
We present 
in Figure~\ref{fig:mcmc} a comparison of the chance probabilities 
derived from the theoretical $F$-distribution and from the Monte Carlo 
simulations. In general, the theoretical $F$-distribution significantly 
under-estimates the significance of a line (the chance probability is deemed
to be too large), although in 10\% of the cases the theoretical distribution 
over-estimates the significance. This demonstrates the necessity of performing
simulations in order to estimate the significance of an added line feature.

We have listed those sources with significant line emission 
in Table~\ref{tab:iron}. We consider a line to be detected if it has less 
than a 1\% chance probability of arising from random variations in the 
continuum, and if the fit produced $\chi^2_c/\nu_c < 2$. 
In total, 35\% (64 out of 181) of the sources that we examined have 
significant Fe line emission. We find that 25\% of sources exhibit a 
significant line near 6.7~keV, while 16\% of sources exhibit emission at 
6.4~keV. There are 15 sources with significant lines at both 6.4 and 
6.7~keV, but these tend to be faint sources in which it is
not possible to constrain the line energy. We plot histograms of the 
equivalent widths and the equivalent widths as a function of the flux in
Figure~\ref{fig:iron}. For both species, the
equivalent widths range from 200 eV to 5 keV. The median equivalent 
widths were 370 eV for the 6.4 keV Fe line, and 
530 eV for the 6.7 keV Fe line. 
Six sources have lines with equivalent widths greater than 1~keV that appear
upon visual inspection to be real.
There are two systems with apparent equivalent widths $> 10$ keV, but 
these have steep continuum spectra with $\Gamma > 4.5$, and the excess 
emission between 6--7 keV represents either a hard continuum 
component or poor background subtraction. 
When line emission is not detected, the median 1-$\sigma$ upper limits to 
the line equivalent widths are 220~eV for
He-like Fe, and 240~eV for low-ionization Fe. 

\begin{inlinefigure}
\centerline{\epsfig{file=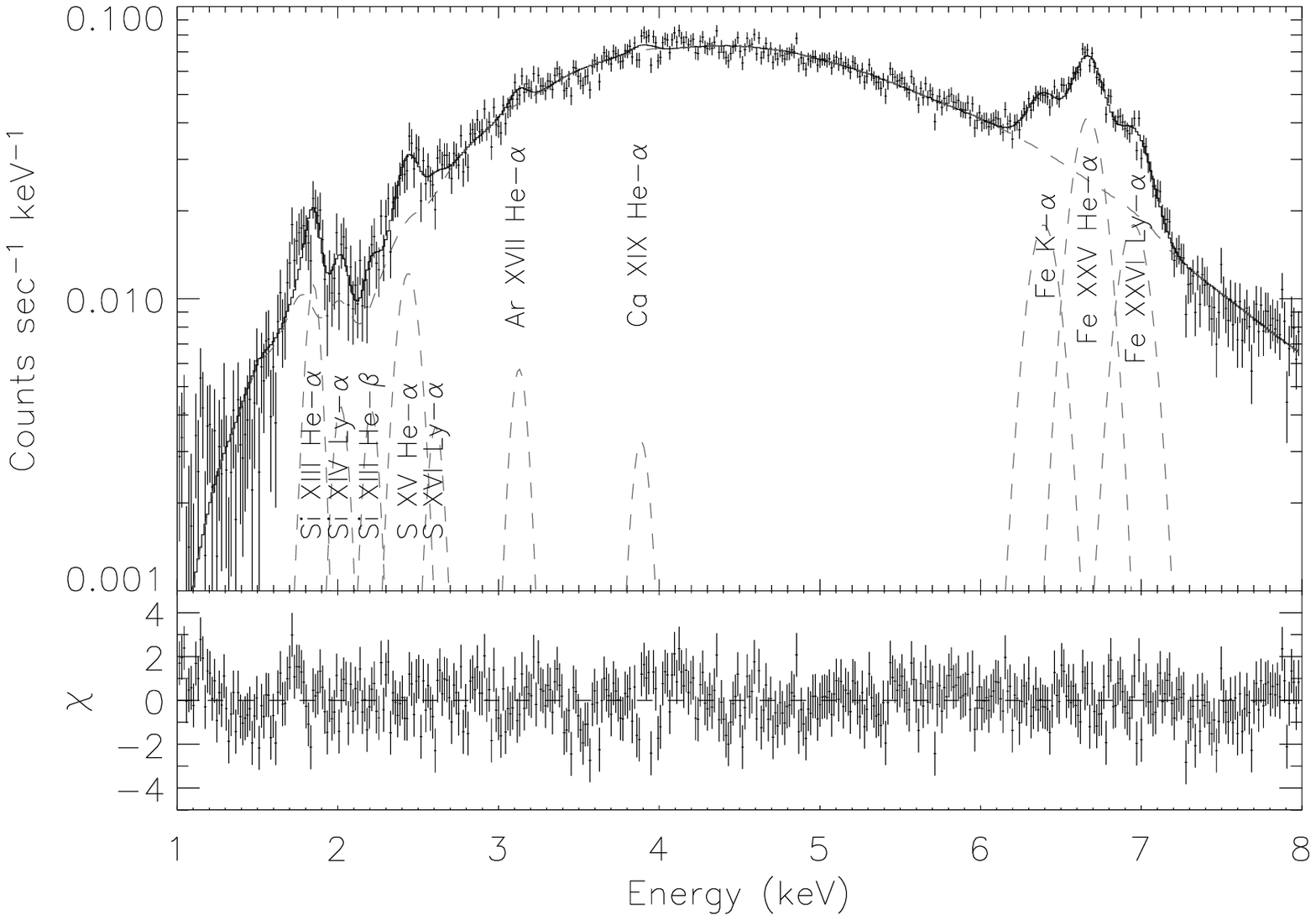,width=0.95\linewidth}}
\caption{
Combined, background-subtracted spectrum of the Galactic center point sources 
with fewer than 500 net counts. The best-fit model of a power-law plus
10 Gaussian lines is over-plotted. The expected centroid energies of 
lines from Si, S, Ar, Ca, and Fe are indicated.}
\label{fig:psmod}
\end{inlinefigure}

\noindent

\subsection{Combined Spectra of Point Sources\label{sec:avspec}}

In order to understand the average spectra of the Galactic center 
point sources, we summed the spectra of sub-groups of the individual
point sources. We selected only 
those sources that were not
detected below 1.5~keV with \program{wavdetect}, as these are most likely
to lie near the Galactic center. We also excluded sources brighter than
500 net counts, because these sources provided individual spectra of good 
quality. We computed average effective area and response functions
by averaging those from the individual sources, weighted by the number of 
counts from each source. 
We estimated the average background by extracting
a spectrum from the rectangular region that traced the orientation of the 
ACIS-I detector during the 500~ks series of observations from 2002 May to 
June.\footnote{We did not simply sum the background spectra obtained for 
individual sources, because doing so would have double-counted events from 
background regions that overlapped.} We excluded
 from the background spectrum 
events that fell within circles circumscribing the 95\% contour of the PSF 
around any point sources.

In Figure~\ref{fig:psraw}, we display the summed spectra of the Galactic 
center point sources with fewer than 500 net counts. The spectrum 
below 2~keV
is dominated by the diffuse emission from the Galactic center.
In Figure~\ref{fig:psmod} we display the background-subtracted spectrum.
The instrumental Ni line at 7.5~keV is absent from 
the spectrum, which indicates that the background subtraction was successful. 
Lines of Si, S, and Ar are also weak or absent in the spectrum of the point 
sources, in contrast to that of the diffuse emission 
(dashed line in Figure~\ref{fig:psraw}). 
Prominent lines from He-like and 
H-line Fe are evident at 6.7 and 6.9~keV, while weak, fluorescent K-$\alpha$ 
emission from low-ionization Fe is evident at 6.4 keV.

We modeled this spectrum using two approaches. First,  we modeled
the emission phenomenologically, using a power-law continuum and Gaussian
line emission from Si, %
\begin{inlinefigure}
\centerline{\epsfig{file=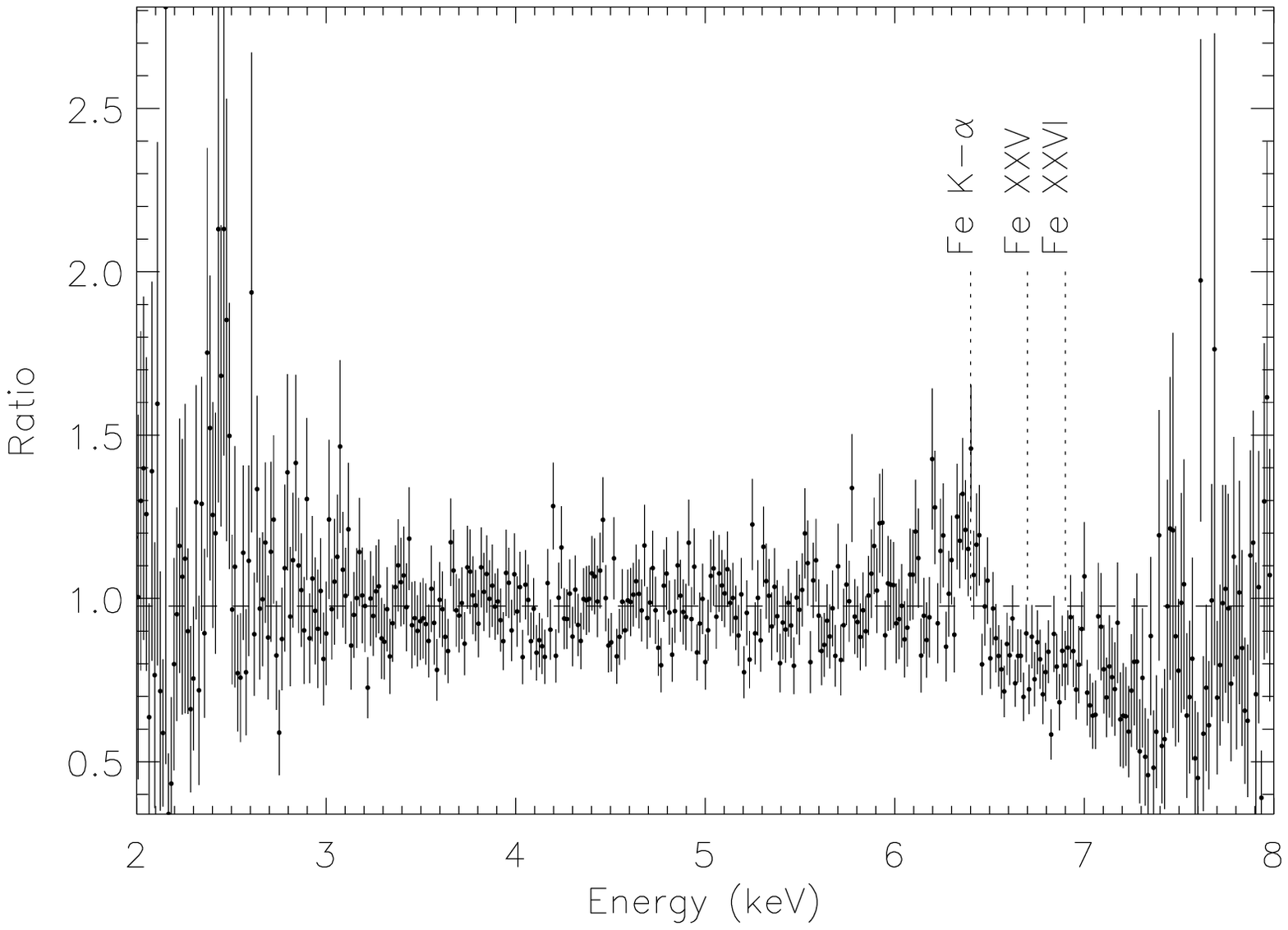,width=\linewidth}}
\caption{
Ratio of the spectra of point sources with  between 80 and 500 net counts 
to those with $< 80$ net counts. The ratio is flat between 2.5--6 keV, 
which indicates that the continuum shapes are nearly identical. The 
deviations at 2.4 keV are due to stronger S emission in the brighter 
sources. The deviations between 6--7 keV are due to the range of strengths
in the neutral, He-like, and H-like lines of Fe.}
\label{fig:rat}
\end{inlinefigure}

\noindent
S, Ar, Ca, and Fe.
The lines we included were chosen by examining whether lines from the 
strongest transitions from Table~1 in Mewe, Gronenschild, \& van den Oord 
(1985)\nocite{mgo85} significantly improved the residuals when comparing
the model to the data. For the final model, we placed lines at the energies 
expected for the He-like 
$n=2-1$ transitions of Si, S, Ar, Ca, and Fe; the He-like $n=3-1$ 
transitions of Si and S; the H-like $n=2-1$ transitions of Si, S, Ar, and 
Fe; and low-ionization Fe K-$\alpha$ at 6.4~keV. 
This model allowed us to measure and
compare the equivalent widths of the lines. We also 
used a model 
consisting of two thermal plasma components, each of which was absorbed by
a separate column of gas. This model is identical to the model used for the 
diffuse emission by \citet{mun04}. We also note that this model is 
qualitatively similar to the 
multi-temperature, multi-absorber models typically used
to model the 
accretion shocks in magnetized CVs \citep[e.g.][]{ram03}.

Several assumptions were required for the models to reproduce the data. First,
we only applied the model between $1.0-8.0$~keV. Below this energy
range the photon
 counts are dominated by foreground diffuse emission, while above 
this range the ACIS-I has a small effective area. 
Second, when modeling individual
lines with Gaussians, the widths of the lines from
He-like Si, S, and Fe and low-ionization Fe 
were allowed to be as large as $\approx 70$~eV to account 
for the fact that the lines are blends of several transitions that 
cannot be resolved with ACIS. 
Third, we allowed for a $\lesssim 1$\% shift in the energy scale in each
spectrum because of uncertainties in gain calibration of our CTI-corrected
data. When fitting Gaussians to the He-like transitions and the 6.4~keV
line of Fe, the line centroids were varied one-by-one until they achieved 
best-fit values, and then frozen.
When using plasma models, the red-shift parameter was used to change the 
energy scale in a similar manner. Finally, a 
3\% systematic uncertainty was added in quadrature to the statistical 
uncertainty in order to account for uncertainties in the ACIS effective
area.

To account for absorption
in each model we assumed 
(1) 
\begin{inlinefigure}
\centerline{\epsfig{file=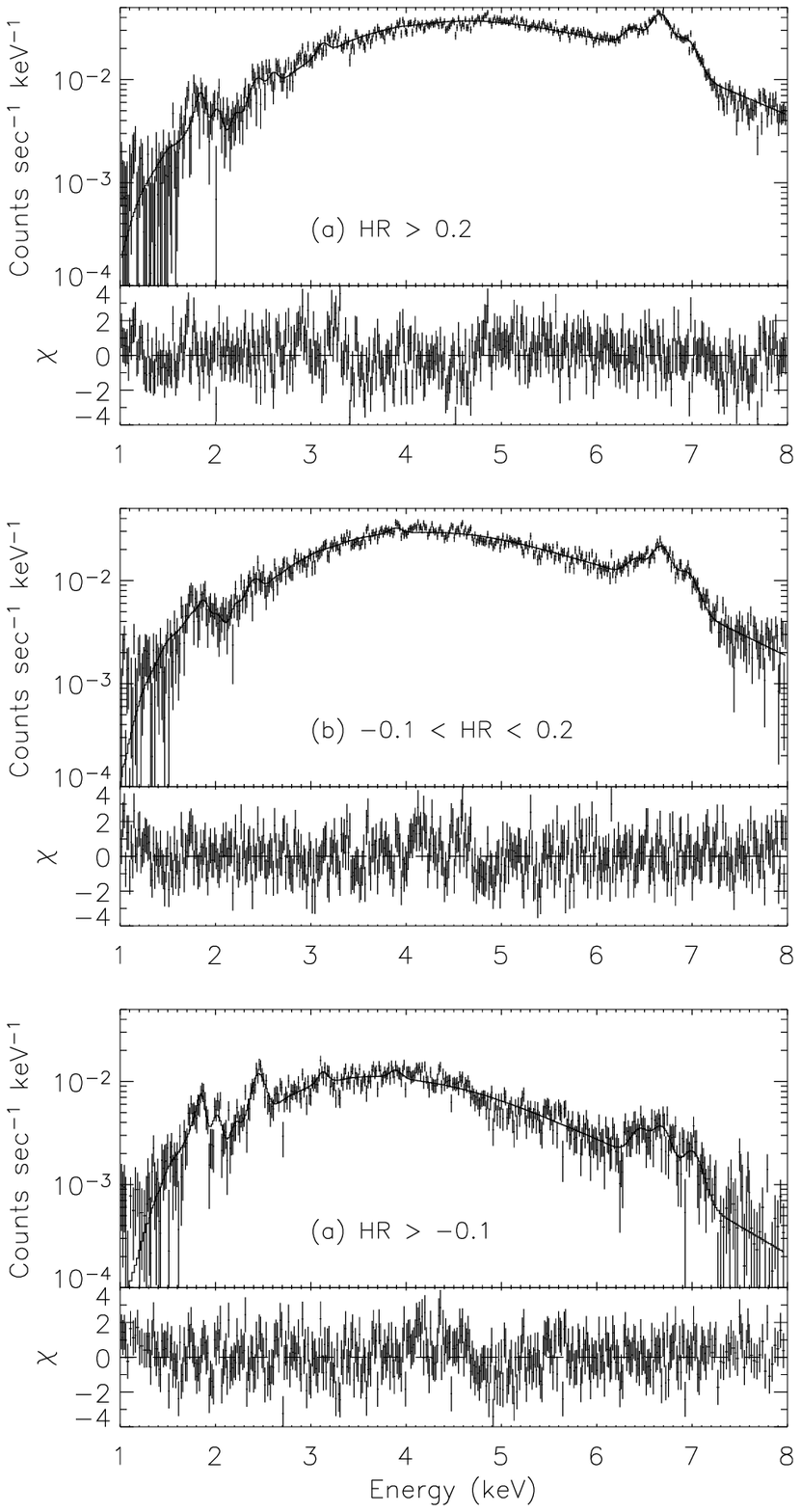,width=0.95\linewidth}}
\caption{
Combined spectrum of Galactic center point sources with fewer 
than 500 net counts and sorted according to their hard color. Although the
observed shape of the continuum varies strongly with hardness ratio, 
lines from He-like and H-like Fe are present in all three groups of
sources. The ratios of the Fe lines are consistent with a $kT \approx 8$~keV 
plasma in
all cases. This suggests that variations in the local absorption column
toward the sources cause the variations in the apparent hardness of
the Galactic center sources.}
\label{fig:spechr}
\end{inlinefigure}

\noindent
that the entire 
region was affected by one column of material that represents the average
Galactic absorption (modeled with 
\program{phabs} in \program{XSPEC}) and (2) that a fraction of each region was 
affected by a second column that represents absorbing material that only
partially-covers the X-ray emitting region (modeled with \program{pcfabs}). 
This partial-covering absorption model produces a low-energy cut-off that is 
less steep than that which would be produced by 
a single absorber. The model
can roughly account for the fact that both the point sources
and absorbing material are distributed along the line of sight.
The mathematical form of the model was
\begin{equation}
e^{-\sigma(E)N_{\rm H}}([1-f_{\rm pc}] + f_{\rm pc}e^{-\sigma(E)N_{\rm pc,H}}),
\label{eq:abs}
\end{equation}
where $\sigma(E)$ is the energy-dependent absorption cross-section, 
$N_{\rm H}$ is the absorption column, $N_{\rm pc,H}$ is the partial-covering
column, and $f_{\rm pc}$ is the partial-covering fraction. 
Dust scattering was not included, because when modeling the spectrum 
its optical depth was degenerate with the partial-covering fraction 
$f_{\rm pc}$.

\subsubsection{Phenomenological Model}

Since the natures of these point sources are uncertain, the most 
straightforward way of modeling their 
spectrum is with an absorbed power-law 
continuum and Gaussian line emission. The average spectrum of
sources with fewer than 500 net counts is displayed along with the model
spectrum in Figure~\ref{fig:psmod}. The model parameters are listed in the 
first column of Table~\ref{tab:psint}. 
After some initial tests, we found that $f_{\rm pc}$ 
was poorly constrained, but the best-fit values were near 0.95. We therefore
fixed $f_{\rm pc}$ to this value. The remaining parameters were allowed to vary.
The total absorption column, 
$N_{\rm H} + N_{\rm pc,H} \approx 9\times10^{22}$~cm$^{-2}$, is slightly higher
than the expected Galactic value. Simulations in \program{XSPEC} indicate that 
this is probably because we did not include dust scattering, which produces 
about 30\% of the total absorption. 
The inferred continuum is flat, with photon 
index $\Gamma = 0.8$, which is similar to the median value from the individual
sources, $\Gamma = 0.7$. Finally, the equivalent width of the He-like
Fe line is 400~eV, which is similar to that observed from individual bright
sources. The strength of the neutral Fe line is considerably lower than
those detected from the individual sources, although this is not surprising
since fewer of the individual sources exhibit 6.4~keV lines than do
6.7~keV lines (see Section~\ref{sec:fe}).

We then examined how the average spectra of the point sources varied with
flux. We compared the summed spectra of sources with fewer than 80 net
counts to those with 80--500 net counts, because these two groups of sources
produce nearly the same numbers of net counts 
($\approx 7.5 \times 10^{5}$). The best-fit parameters of our phenomenological
model of these spectra are listed in the second two columns of 
Table~\ref{tab:psint}. Both the absorption column and photon index were 
slightly larger in the bright sources. However, the continuum shape
looks very similar by eye. To highlight this, in 
Figure~\ref{fig:rat} we plot the ratio of the averaged spectra of sources 
with 80--500 net counts to that of sources with $< 80$ net counts. 
The only differences in the continuum spectra are above 7~keV, which may 
indicate that the faint sources produce slightly more high-energy flux.
Differences in the equivalent widths of the line emission
are also evident in Figure~\ref{fig:rat}.
The equivalent width of the Fe XXV He-$\alpha$ emission was 30\% higher
in the fainter sources (3.7$\sigma$), and the equivalent width of neutral 
Fe K-$\alpha$ was a factor of 2 lower in the faint sources (4.9$\sigma$). 
There is also a factor of 3 less S He-$\alpha$ emission in the faint sources 
(4.3$\sigma$). We also split the data into 
smaller flux intervals, and found similar results. Thus, the continuum 
spectra of the point sources change very little with intensity, but
the S and Fe lines range over a factor of $\ga 2$ in equivalent width.

We also examined the combined spectra as a function of the hard
color, which provides a measure of the steepness of the spectrum. 
The 
hard color is defined as $HR = (h-s)/(h+s)$, where $h$ is the number
of counts in a hard band from 4.7--8.0~keV, and $s$ is the number of counts
in a softer band from 3.3--4.7~keV.
We divided the data using three ranges in hard color, guided by the 
expected power-law index from simulations with \program{PIMMS} 
\citep{mun03}: $HR > 0.2$ for sources with $\Gamma \approx 0$ spectra,
$-0.1 < HR < 0.2$ for sources with $\Gamma \approx 1.5$ spectra, and
$HR < -0.1$ for sources with $\Gamma > 2$ spectra. 
The spectra are plotted in Figure~\ref{fig:spechr}, and the best-fit
parameters for the phenomenological model are listed in the last three
columns of Table~\ref{tab:psint}.
We find that 
the sources with larger $HR$ have continuum spectra that are 
intrinsically flatter, at least given our model for the interstellar
and intrinsic absorption. We also find significant changes in the line 
strengths, although there is no monotonic trend with $HR$.

\subsubsection{Two-$kT$ Plasma Model\label{sec:av:vapec}}

We next modeled the spectra of the point sources as originating from two 
thermal plasmas, each of which was absorbed by a separate component as 
parameterized in Equation~\ref{eq:abs}, 
as well as emission from low-ionization Fe at 6.4~keV. The parameters of the
best fit models to the average spectra for the groups of sources used in the 
previous section are all listed in Table~\ref{tab:twokt}. The values
of $\chi^2_\nu$ are generally larger under the plasma models than under
the phenomenological models, because the plasma models predict too little 
flux above 7~keV. However, the plasma models do provide a good qualitative
description of the data. The weak lines from Si, S, and Ar require cool 
plasmas with $kT_1 \approx 0.5$~keV; this component appears to be visible only
in sources with $HR < -0.1$. The abundances of Si and S 
appear to be significantly below solar values in the soft sources, where the
lines are most clearly detected. 
The He-like and H-like Fe lines require hotter 
plasmas with $kT_2 \approx 8$~keV. The abundances of Fe are also
generally about 50\% solar, although they are consistent with the solar ones
in the faintest and softest sources.

The temperatures of these plasma components are very similar to those inferred
from the diffuse emission, because emission lines from the same set of ions 
are present in both spectra \citep{mun04}. 
However, in contrast to the diffuse emission,
the cooler plasma in the point sources is heavily absorbed and 
contributes little to the observed flux. 

More importantly, the presence of prominent He-like and H-like Fe emission
from $kT = 7-9$ keV plasma in all groups of sources suggests that the 
X-ray emission is produced by similar physical mechanisms. 

\subsection{Search for Variability\label{sec:var}}

We searched for variability in the 0.5--8.0 keV bandpass from the entire 
sample of X-ray sources from \citet{mun03}, in order to identify flux 
changes that occurred within 
single observations, and long-term variations in the mean flux between 
observations.

\subsubsection{Short-term Variability}

\begin{figure*}[pth]
\centerline{\epsfig{file=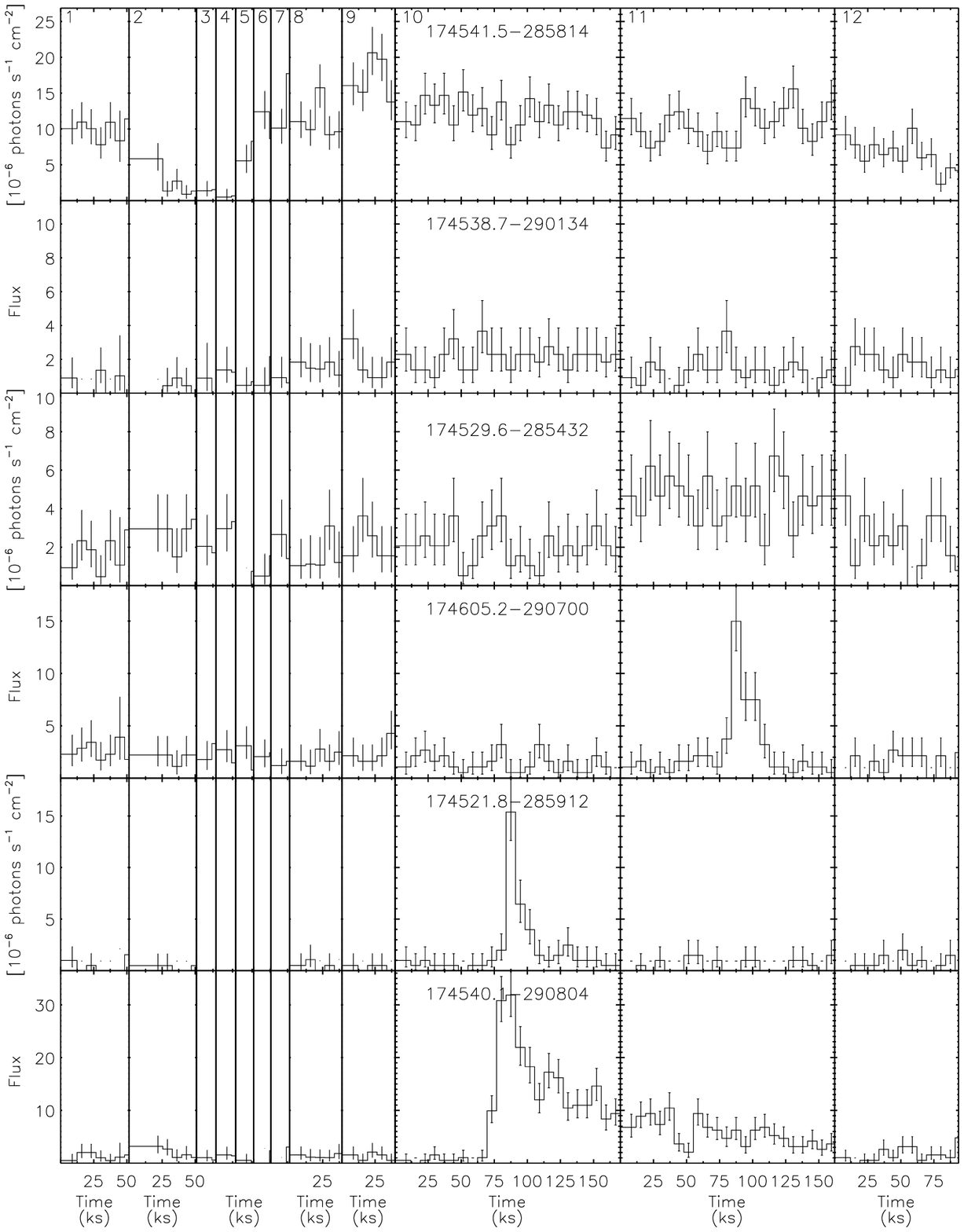,width=0.8\linewidth}}
\caption{
Example light curves of variable sources in the field. To compute the 
resulting photon fluxes, the total 
counts received
from each source as a function of time have been divided by the mean value
of the detector effective area from each observation.}
\label{fig:lcurves}
\end{figure*}

To search for short-term
variability, we applied a Kolmogorov-Smirnov (KS) test to the un-binned 
arrival times of the events during each observation. 
Before performing the search, we removed events flagged as
potential cosmic ray after-glows. 
We also excluded data received near the edges of the detector chips, 
and data from the first part of ObsID 1561 for sources that were within 
5.5\arcmin\ of the bright transient \gcllb\ \citep{mun03b}.
If the cumulative distribution 
of the arrival times differed
\begin{figure*}[th]
\centerline{\epsfig{file=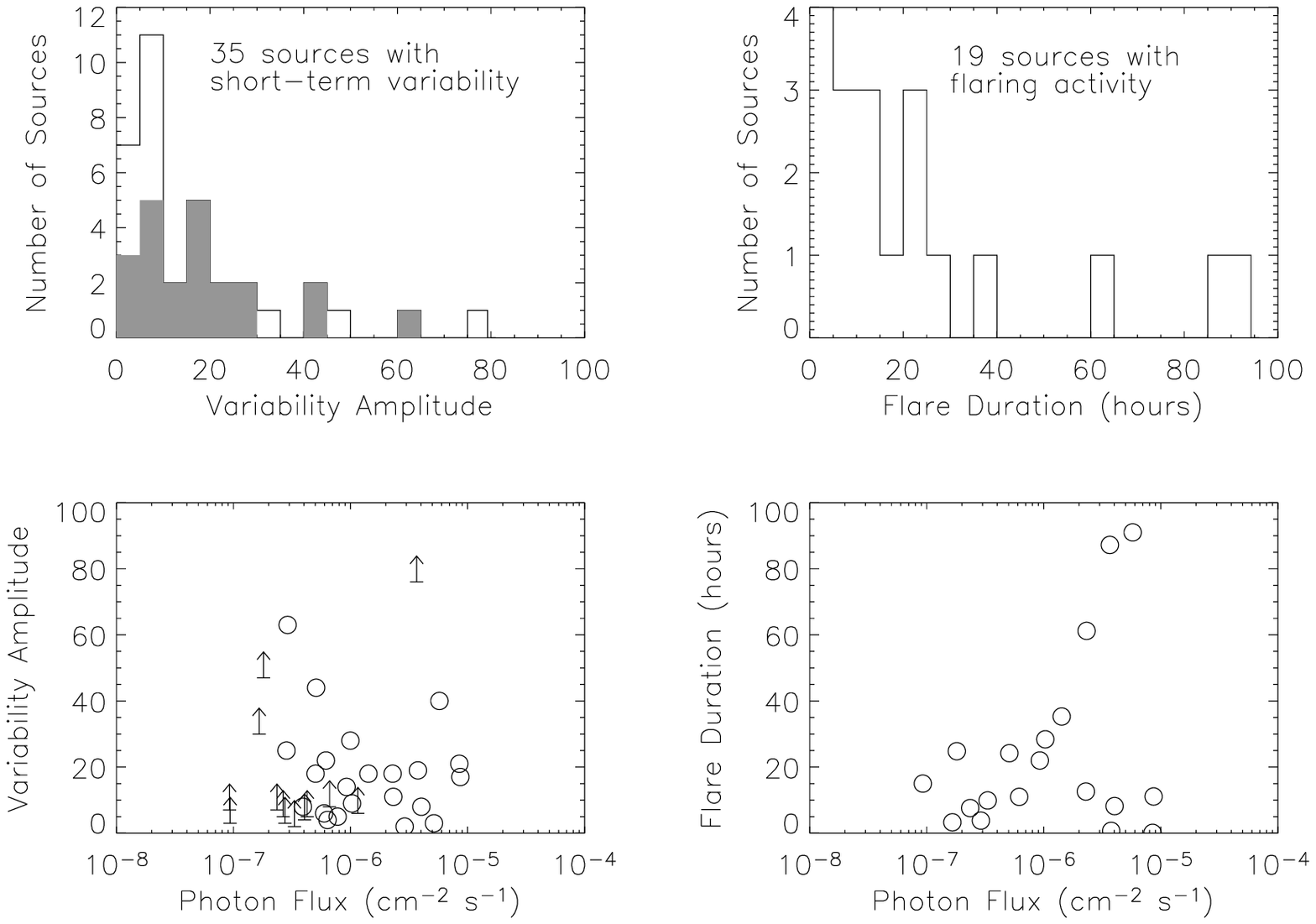,width=0.8\linewidth}}
\caption{Summary of the amplitudes and durations of the variability in
sources detected toward the Galactic center. The {\it top left} panel 
displays a histogram of the ratio of the maximum to the minimum flux
from each variable source. Sources that were detectable at minimum are
indicated with solid histogram, while lower limits on the variability 
amplitude are illustrated with the open histogram. The {\it bottom left}
displays the variability amplitude (or lower limits thereto) as a function
of the mean photon flux from the source. Not surprisingly, small-amplitude
variability can only be detected if a source is bright. The {\it top right}
panel displays a histogram of the durations of flare-like events. These
range from $< 1$ ks to nearly 100 ks. The {\it bottom right} panel
illustrates that the observed flare duration is not strongly correlated
with the mean photon flux from a source.
}
\label{fig:shortterm}
\end{figure*}
 from a uniform distribution (which would 
imply a constant flux) with greater than 99.9\% confidence in any observation,
we considered the source to vary on short time-scales.  
We find that 18 foreground
sources and 21 Galactic center sources are 
variable on short time scales according to the KS test. We list these
sources in Table~\ref{tab:shortvar}. Examples of 
short-term variability are shown in the bottom three panels of 
Figure~\ref{fig:lcurves}.

To characterize the duration and amplitude of the variability, we have 
applied the``Bayesian Blocks'' algorithm of
\citet{sca98} \citep[see also][]{eck04}. 
The algorithm is based on a parametric maximum-likelihood
model of a Poisson process that divides the data into sequential segments, 
each of which has a constant count rate. The segments were identified by 
dividing the events into sub-intervals, and computing the odds ratio that 
the count rate has varied. If variability was found, then each interval was 
split further into sub-intervals, in order to track the structure of the 
variability. We found that by using an odds ratio corresponding to a 67\% 
chance that the variation is real, we could identify changes in the flux from 
all but four of the variable sources. 
If the Bayesian Blocks code identified only two intervals with differing count
rates, then the variability was classified as a ``step'' function.
If it identified more
than two intervals with differing count rates, we defined the variability
as a flare; a search of the data revealed no instances of dips, which are 
unlikely to be detected in sources with low count rates. We computed the 
background-subtracted maximum and minimum count
rates for each variable observation, and divided them by the mean value of 
the effective area function for that source to convert them into photon 
fluxes. The maximum and minimum fluxes are listed in Table~\ref{tab:shortvar}.
We also list the durations of the flares in kiloseconds, and either the ratio 
between the maximum and minimum flux, or the lower limit thereto if the
baseline flux is an upper limit. The photon fluxes in the table can 
be converted to energy fluxes by the factors  
1 \phcms $= 3\times10^{-9}$ \ergcms\ (0.5--8.0 keV) for foreground sources, 
and 
1 \phcms $= 8\times10^{-9}$ \ergcms\ (2--8 keV) for Galactic center sources.
The peak luminosities of the variable Galactic center sources typically
range from $6\times10^{31}$ to $1.7 \times 10^{33}$~\ergsec. However, 
one flare from CXOGC J174552.2--290744 consists of 5 photons received 
in 100 s and has a peak luminosity of $10^{34}$~\ergsec. None of the 
events in the apparent flare were flagged as cosmic rays by the CXC pipeline,
and tests of
\begin{inlinefigure}
\centerline{\epsfig{file=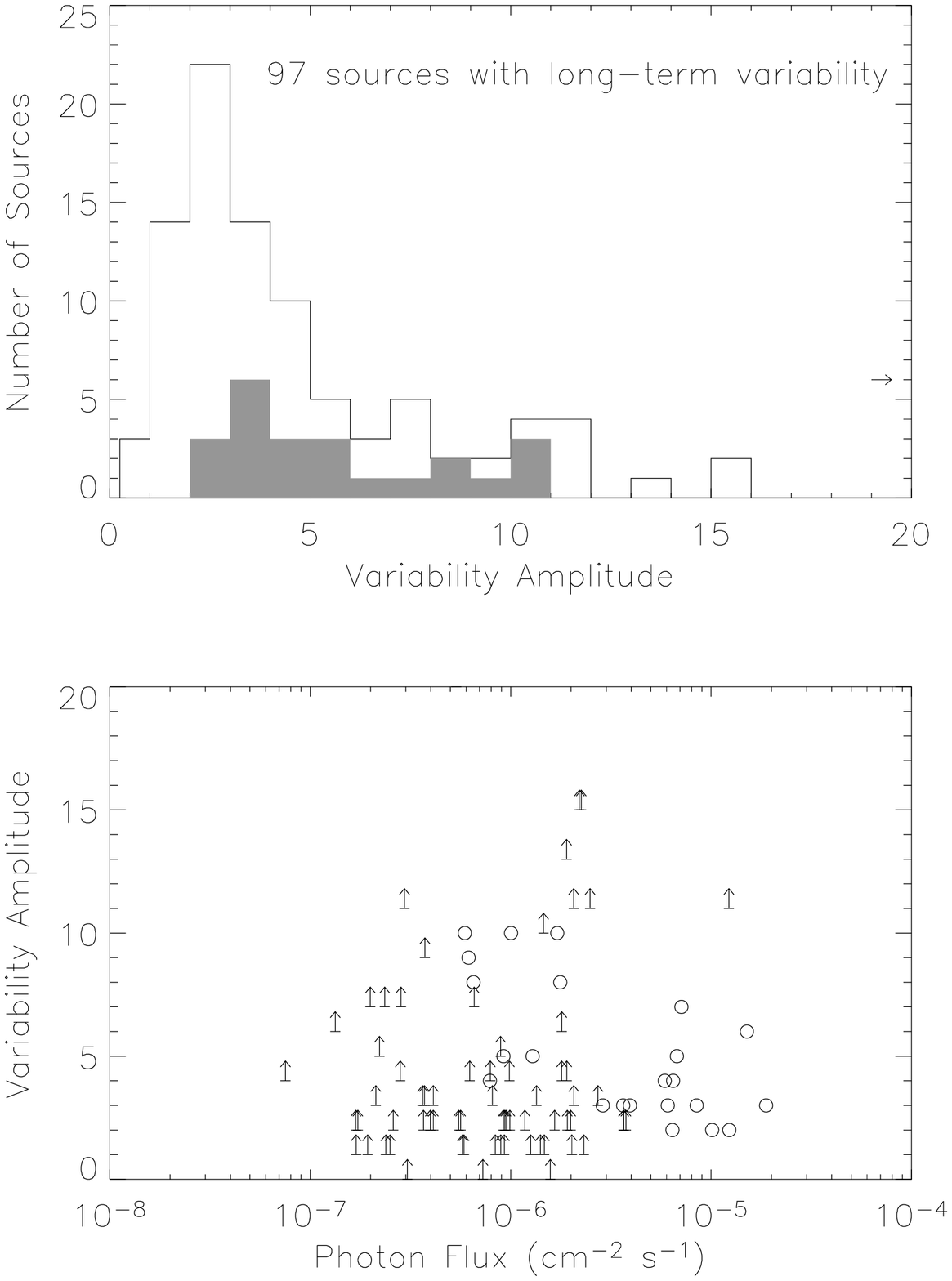,width=0.95\linewidth}}
\caption{Summary of the amplitudes and durations of the variability observed
between observations in
sources detected toward the Galactic center. The {\it top} panel 
displays a histogram of the ratio of the maximum to the minimum flux
from each source with long-term variability. Sources that were detectable 
at minimum are
indicated with solid histogram, while lower limits on the variability 
amplitude are illustrated with the open histogram. The {\it bottom}
displays the variability amplitude (or lower limits thereto) as a function
of the mean photon flux from the source. Once again, small-amplitude
variability can only be detected if a source is bright. 
}
\label{fig:longterm}
\end{inlinefigure}

\noindent
 events from elsewhere on the detector that were flagged as 
cosmic rays indicate that they deposit their energy on time scales of 
$\la 20$ s, so we consider the flare real.
No source exhibits flares similar to those seen about once a day
from \sgrastar, with durations of 
$\approx 1$ h and $L_{\rm X} \ga 10^{34}$~\ergsec.

Figure~\ref{fig:shortterm} displays histograms of the variability amplitude
and the durations of the flares. Nearly half of the variability has a peak
flux over 10 times the quiescent 
level. In the bottom-left panel, we plot
the amplitude of the variability as a function of the mean flux from each
source. The amplitudes of the variability are not a strong function of the 
mean flux from the source. However, we are unable to detect the faintest 
sources when they are in their low-flux states, so for these
sources we only can report lower limits to the variability amplitude.
The flare durations are 
spread fairly evenly, with a median duration of about 20 ks. As can
be seen from the bottom-right panel of Figure~\ref{fig:shortterm}, the flare
durations show no correlation with the mean flux from a source. 

In order to quantify our sensitivity to short-term variations, we need to 
examine the probability that a change in count rate could be detected. If
we assume a baseline count rate $r_l$ persists for a time $t_l$, and 
that a flare occurs with count rate $r_h$ lasting $t_h$, then the total
number of counts in each interval follows the Poisson distribution. 
Therefore, the joint probability
that the measured baseline count rate $N_l/t_l$ is less than the measured
flare count rate $N_h/t_h$ is
\begin{eqnarray}
\nonumber P(N_h > N_l t_h/t_l) = \\ 
\sum_{N_l = 0}^\infty \left( {{(r_l t_l)^{N_l} e^{- r_l t_l}} \over {N_l!}}
\sum_{N_h > N_l t_h/t_l}^\infty {{(r_h t_h)^{N_h} e^{- r_h t_h}} \over {N_h!}}
\right).
\label{eq:poiss}
\end{eqnarray}
This probability represents the chance that a flare of amplitude 
$r_h/r_l$ would be detected.

The median net counts from the sources in the catalog of \citet{mun03} 
was 49, with a background of 52 counts. These values translate to count
rates of $n = 7.8\times10^{-5}$ count s$^{-1}$ net and $b = 8.3 \times 10^{-5}$
count s$^{-1}$ background. If we use $r_l = n + b$ in 
Equation~\ref{eq:poiss}, a 36~ks flare during the long 150~ks 
observation could be detected with an amplitude a factor of $\approx 10$. 
Such a flare could be detected from half of the sources in our sample. 
On the other hand, a flare that reaches twice the quiescent flux level 
for 36~ks could only be detected if the 
quiescent count rate was $r_l = 2 \times 10^{-3}$ count s$^{-1}$. Only 
17 sources are this bright, so such a small-amplitude flare would generally
be unobservable. Not surprisingly, all of the short-time scale variability with
amplitudes $< 3$ are long duration, step-like changes in flux.

\subsubsection{Long-term Variability}

To search for long-term variability, we computed the value of $\chi^2$ 
for the photon fluxes in each observation under the assumption that the 
mean rate was constant. We computed the approximate total (source plus
background) photon flux by dividing 
the total number of counts detected by the live time and the mean value of the 
effective area function. To compute a net flux, we then subtracted a 
background count rate, which was estimated in
the same manner as for the 
spectrum.
We considered a source as variable if the photon fluxes both
before and after background subtraction were inconsistent 
with a constant mean value with more than 99\% confidence.\footnote{The 
requirement on the total flux was designed to ensure that the 
variability was not due systematic changes in the background estimate. Such 
changes occurred where there were gradients in the diffuse 
emission, because the regions in which the background were estimated were 
not identical for each observation (Section~\ref{sec:obs}).}
We excluded sources with short-term variability when searching for long-term
variability. We also excluded data from
the first part of ObsID 1561 for 
sources that were within 5.5\arcmin\ of the bright transient \gcllb. Long-term 
variability is illustrated in the top three panels of 
Figure~\ref{fig:lcurves}. 

We find that 20 foreground sources and 77 Galactic center sources vary
on long time scales. We list in Table~\ref{tab:longvar} the minimum and 
maximum background-subtracted photon fluxes for the variable sources.
We present a histogram of the ratio of the maximum to minimum fluxes in
the top panel of Figure~\ref{fig:longterm}. 
Most of the ratios are upper limits, because
the sources are not detected at their minimum flux levels. The bottom
panel illustrates the variability amplitudes as a function of 
the mean intensity of each source. There is no apparent correlation between
the amplitude and intensity of the source.

We can quantify our sensitivity to long-term variations in the same manner as
for short-term variability, using Equation~\ref{eq:poiss}. The most extreme
form of long-term variability is that of a source that is bright for
some portion of the observations lasting a total time $t_h$, and decreases 
below the background level for the remaining observations lasting a time
$t_l$. We therefore assume that the total counts from 
a source is consistent with the background $b = 8.3 \times 10^{-5}$ 
count s$^{-1}$ during the time $t_l$ when it is faint. 
If the source is ``off'' during one of the
12~ks observations, we could detect this decrease in count rate 
with 99\% 
confidence if during the remaining 614~ks of observations
the source was brighter than $3 \times 10^{-4}$ 
net count $s^{-1}$. Approximately 9\% of the sources are this bright.
We could detect variability from a source that is ``off'' during all but the 
500~ks monitoring campaign (2002 May--June) if the count rate at maximum was 
$8\times 10^{-5}$ count s$^{-1}$, which is valid for half of the sources 
searched.

\section{Discussion\label{sec:disc}}

The large number of sources detected toward the Galactic center is 
most likely a product of the large density of stars there. 
The 17\arcmin\ by 17\arcmin\ field spans a physical 
distance of 20 pc in projection from \sgrastar, and therefore probes
the inner regions of the Nuclear Bulge that was studied 
extensively by Launhardt, Zylka, \& Mezger (2002)\nocite{lzm02}. 
Using their models, we estimate that 
$1.3\times10^8$ \msun\ of stars lie within a cylinder of radius 20 pc 
and depth 440 pc that is centered on the Galactic center. Thus, our 
observation encompasses up to 0.1\% of the total Galactic stellar mass, 
which is
$\sim 10^{11}$ \msun. However, we have also found that the surface density 
of the X-ray sources falls off as $\theta^{-1}$ away from \sgrastar, so it 
is possible that most of the X-ray sources lie in an isothermal sphere of 
radius 20 pc \citep{mun03}. Such a sphere would contain 
$3\times 10^7$ \msun\ of stars, or 0.03\% of the
\begin{figure*}[ht]
\centerline{\epsfig{file=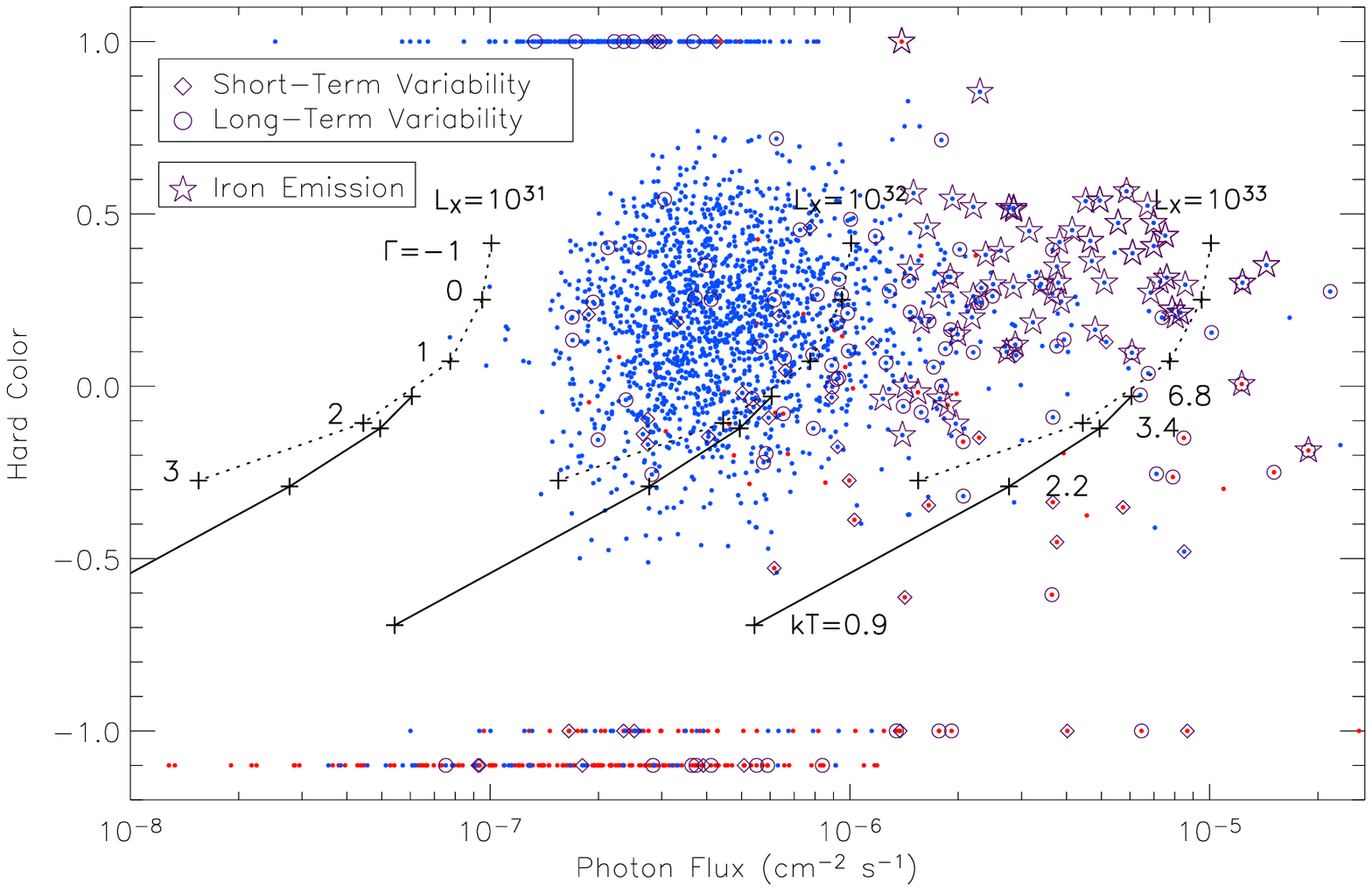,width=0.8\linewidth}}
\caption{Summary of the spectral and variability properties
of the X-ray sources detected toward the Galactic center. For all of the
sources in the sample, we plot the hard color against the photon flux.
The hard color is defined as $HR = (h-s)/(h+s)$, where $h$ is the number
of counts in a hard band from 4.7--8.0~keV, and $s$ is the number of counts
in a softer band from 3.3--4.7~keV.
We were able to derive hard colors for 1848 out of 2357 sources.
Sources that are only detected in the 4.7--8.0~keV band are given values of 1. 
Sources that are only detected in the 3.3--4.7~keV band are given values of 
$-1$. Sources detected in neither band are given hard colors of $-1.1$.
Sources located at or beyond the Galactic center are indicated in blue, while
foreground sources are indicated in red. 
We use solid lines to illustrate the hard colors and photon fluxes 
expected for thermal plasma with a range of temperatures and luminosities 
(bolometric for $d=8$~kpc), and the dotted lines for power laws with a range 
of photon 
indices and luminosities (0.5--8 keV). 
We also indicate which sources exhibit iron lines (stars)
and short-term variability (diamonds)  or long-term variability (circles).
Note that sources with short-term variability were not searched for long-term
variability.}
\label{fig:indiv}
\end{figure*}
 Galactic stellar mass. 
For comparison, the shallower survey carried 
out by \citet{wgl02} covered the entire Nuclear Bulge (albeit with a 
factor of 5 less sensitivity), and thus sampled $\sim 1$\% of the mass of 
stars in the Galaxy. The stellar density at the location of the 
X-ray sources is between 240--900 \msun pc$^{-3}$ (for a 20 by 440 pc 
cylinder and a 20 pc sphere, respectively), compared to 
0.1 \msun\ pc$^{-3}$ in the local stellar neighborhood \citep[][p. 16]{bt94}. 
We will keep these numbers in mind as we consider the likely natures of the
Galactic center point sources. 

The sample identified as part of the \chandra\ observations of the 
\sgrastar\ field is unique, because the long exposure
time allows us to detect faint sources 
($F_{\rm X} = (3-100)\times 10^{-15}$~\ergcms, 2--8~keV), whereas the 
strong diffuse X-ray emission and the high absorption toward the 
Galactic center prevent us from observing X-ray sources unless they are
prominent in the 4--8 keV band. 
Prior to \chandra, the most sensitive hard X-ray survey of the Galactic 
plane was taken with \asca. That survey identified only 163 sources, with 
a detection limit of $\approx 3\times 10^{-13}$~\ergcms\ 
\citep[2--10~keV;][]{sug01}. The Galactic
center sources are on average much harder than those detected in the 
\asca\ survey. The brighter \asca\ 
sources had a median photon index of $\Gamma = 2.5$, with only 15\% of
the sources having $\Gamma < 1$, while the fainter \chandra\ sources
have a median $\Gamma = 0.7$ (Figure~\ref{fig:dist}). The difference in 
hardness of the two samples is probably a selection effect caused by the high 
absorption and the
strong diffuse emission toward the Galactic center. 
The X-ray sources that we have 
studied in this paper probably sample only the hardest examples of the 
population identified with \asca.


Likewise, \chandra\ observations of globular clusters have identified 
a couple hundred X-ray sources with $L_{\rm X} = 10^{29} - 10^{33}$~\ergsec
\citep{gri01, pool02, bec03, hei03, hei03b}.
These luminosities overlap those inferred for the sources 
near the Galactic center in our sample. However, only a few of the
$\approx 60$ globular cluster sources with spectral information 
are best modeled by a $\Gamma < 1$ power-law. Most have steeper spectra
that are consistent with $\Gamma > 1.5$ power-laws or 
$kT \approx 1-20$~keV thermal plasmas.

In addition to the hardness of their spectra, the X-ray sources detected
toward the Galactic center share several other interesting properties. 
Most notable is line emission from low-ionization, He-like, and
H-like Fe (Figures~\ref{fig:iron} and \ref{fig:psmod}). On average, 
the low-ionization Fe lines have equivalent widths of 100--230~eV, while
the He-like Fe lines have equivalent widths of 350--450~eV. The 
strengths of these lines range over a factor of two 
when considering sources with a 
range of intensity and spectral hardness (Table~\ref{tab:psint}).
However, in all cases the average ratios of the 
He-like and H-like lines are consistent with those expected from a 
thermal plasma of $kT \approx 8$~keV (Table~\ref{tab:twokt}). 
The presence of these Fe lines in a large fraction of 
the sources suggests that they could be dominated by a single population of 
sources.

However, the 
emission from such a plasma should produce a much steeper continuum spectrum, 
with $\Gamma \approx 1.5$ instead of $\Gamma \approx 0.7$. Unfortunately,
it is not possible to determine unambiguously the physical process producing
the X-ray emission from the continuum and iron lines alone. For instance, if 
the X-ray emitting regions are partially absorbed by material local to the 
X-ray sources, the observed spectra can be much harder than the intrinsic ones.
Alternatively, the line emission could be produced in photo-ionized plasmas,
although the large equivalent widths of the lines indicates that the continuum
emission exciting them must not be observed directly \citep[e.g.][]{muk03}.
In either of these cases, the intrinsic X-ray luminosity would be significantly
higher than is inferred from the flux received with \chandra. 

While the average spectral properties provide an overview of the 
characteristics of the X-ray sources near the Galactic center, it is still 
important to examine the properties of individual sources to determine 
how various classes of sources contribute to the population there. 
Therefore, in Figure~\ref{fig:indiv} we display the 
spectra and intensities of individual sources
by plotting the hard color of each source against its photon flux.
These quantities are measures 
of the physical quantities of interest, the intrinsic spectral shape and the 
luminosity of a system.\footnote{As we
discuss in Section~\ref{sec:avspec}, if there is local 
X-ray absorption that affects only a fraction of the emitting region,
the inferred spectrum can seem artificially flat, and the inferred 
luminosity would be too low. 
Therefore, the hardness ratios and photon fluxes could potentially be 
misleading. This is nearly impossible to avoid. Indeed, the spectral models
we applied to the individual sources 
suffer from the same shortcoming, because we also assume that the
emitting region is absorbed uniformly. Using hardness ratios and photon
fluxes is the best option, since unlike the parameters of the spectral models,
the former can be derived for almost all of the sources in 
our sample.}  We also indicate in Figure~\ref{fig:indiv} the expected 
hardness ratios and photon fluxes for sources over a range of luminosities and
with either (1) thermal plasmas over a range of temperatures $kT$ 
({\it solid lines}), or power laws over a range of photon indices 
$\Gamma$ ({\it dotted lines}). In all cases, we have assumed the sources 
are absorbed by the median column density from the spectral models, 
$6 \times 10^{22}$ cm$^{-2}$ of gas and dust. Approximately 75\% of the
2000 Galactic center X-ray sources are detected with 90\% confidence in both 
the 3.3--4.7 and 4.7--8.0 keV bands, and therefore have hard colors in
Figure~\ref{fig:indiv}.

We also indicate in Figure~\ref{fig:indiv} which sources exhibit 
line emission from Fe. Nearly all of the sources brighter than 
$4\times10^{-6}$ \phcms\ and harder than $HR = 0$ exhibit line emission
from He-like Fe. This is not surprising, given the prominence of line
emission in the average spectra. Fe emission is detected less often in 
fainter sources, but this is probably due to lower signal-to-noise. 

Finally, we indicate which sources exhibit variability. 
The sources that are identified as 
variable tend to be brighter, because the signal-to-noise is better. 
Soft sources are most likely to exhibit short-term variability.

In the following sections, we use these properties to guide our discussion
of the natures of the sources. The luminosities of the Galactic center 
sources are consistent with those of young stellar 
objects (YSOs), interacting binaries (RS CVns), Wolf-Rayet (WR) and early O 
stars, cataclysmic variables, quiescent black hole and neutron star 
X-ray binaries, and possibly the ejecta of recent supernova that are 
interacting with molecular clouds. We consider each in turn.

\subsection{Sources with Active Stellar Coronae}

Many stars produce X-rays in their magnetic coronae. In particular, K and
M dwarfs are so numerous that they contribute significantly to heating
the ISM \citep[e.g.,][]{schl02}. However, individually their X-ray emission is 
faint, with $L_{\rm X} < 10^{29}$~\ergsec, and cool, with $kT < 1$~keV 
\citep[e.g.,][]{kri01}. 
Although, most of the foreground sources are probably low-mass main 
sequence stars (P. Zhao, in preparation), few of the Galactic center sources
should be. 
YSOs and RS CVns are significantly brighter, with $L_{\rm X} \approx 
10^{29}$ to $10^{32}$~\ergsec \citep[e.g.,][]{fei02,dem93a}, and 
are therefore more likely to be seen at the Galactic center.

\subsubsection{Young Stellar Objects}

The number of YSOs at the Galactic center will depend upon
whether low-mass stars have formed there recently. For instance, if star 
formation proceeds at the Galactic center in a similar manner as it 
has in the Orion nebula, then the two-dozen 
massive, emission-line stars in the central parsec of the Galaxy could 
conceivably be accompanied by tens of thousands of low-mass YSOs
\citep[e.g.,][]{fei02}. However,
the strong tidal forces, milliGauss magnetic fields, and turbulent 
molecular clouds near the Galactic center may prevent low-mass stars from
forming there \citep{mor93}.

YSOs have luminosities between $10^{29} - 10^{31.7}$~\ergsec,
and spectra that can be described by thermal plasma emission
with $kT = 1-10$~keV \citep[e.g.,][]{pz02,koh02,fei02}. 
Therefore, any YSOs that are located near the 
Galactic center should be found in the bottom left of
Figure~\ref{fig:indiv}, with $HR < 0$ and fluxes $<2 \times 10^{-7}$
\phcms. However, the detection threshold for sources with $HR < 0$ is 
approximately $10^{31.5}$~\ergsec, and only $\sim 0.4$\% of YSOs are brighter
than this limit \citep{fei02}. YSOs also commonly exhibit flares lasting 
several hours, but fewer 
than 0.1\% exhibit flares brighter than $10^{32}$~\ergsec\ \citep{gro04,fei04}.
In contrast, the faintest genuine flare from a Galactic center source has 
a peak luminosity of $5\times10^{32}$~\ergsec, whereas only three sources with 
short-term variability have peak luminosities below $10^{32}$~\ergsec. 
Therefore, we believe that even the flaring sources are unlikely to be YSOs, 
and that any population of YSOs remain largely undetected at the Galactic
center.

\subsubsection{Interacting Binaries}

RS CVn systems are among the most numerous hard X-ray sources with 
$L_{\rm X} > 10^{29}$~\ergsec, with a local space density 
of $\approx 5 \times 10^{-5}$ pc$^{-3}$ \citep{fms95}. Using the
models of \citet{lzm02} to scale the local number density to the 
stellar density at the Galactic center (Section~\ref{sec:disc}), we estimate
that the total number of RS CVns within 20 pc of the Galactic center 
is $\approx 1.5\times10^{4}$, while the number within a cylinder of 20 pc
radius extending the length of the nuclear bulge (440 pc) is $7\times10^{4}$.

However, RS CVns would be difficult to detect near the Galactic center. 
They typically have soft spectra, with $kT \approx 0.1-2$~keV, and luminosities
of $L_{\rm X} = 10^{29} - 10^{32}$~\ergsec \citep[e.g.,][]{dem93b,sdw96}. 
Therefore,
RS CVns would have $HR < -0.3$ and photon flux $< 3\times10^{-7}$ \phcms\
in Figure~\ref{fig:indiv}. This portion of the figure is sparsely populated.
Moreover, we are only sensitive to sources with $kT \la 2$ keV if they
are more luminous than $5\times10^{32}$~\ergsec, whereas only $\sim 2\%$ of 
RS CVns are this luminous \citep{dem93a}.
Finally, although RS CVns do exhibit flares lasting several hours with 
amplitudes of up to a factor of ten, they are seldom more luminous 
than $\approx 10^{32.5}$~\ergsec\ \citep[e.g.,][]{tsu89, fpt01, fpt03}.
Therefore, these flares would not be observable in our Galactic center data.
The difficulty of detecting RS CVns and the lack of good candidate objects
indicates we have probably identified only a tiny fraction of the 
RS CVns at the Galactic center.
 
\subsection{Winds from Massive Stars}

There is currently significant debate about the origin of the X-ray
emission from WR and early O stars \citep[e.g.,][]{wc01}, 
but it is generally thought
that the X-rays are produced through shocks in their winds 
\citep[see, e.g.,][]{cg91}. They have 
luminosities of up to $\approx 10^{33.5}$~\ergsec\ in isolation, and 
$\approx 10^{35}$~\ergsec\ when two
such stars are in a colliding-wind binary. 
Their spectra can usually be modeled as thermal
plasma with $kT = 0.1 - 6$~keV \citep[e.g.,][]{pol87,poz02}. These 
systems would lie in the portion of the color-intensity diagram with 
$HR < 0$ in Figure~\ref{fig:indiv}. 

The number of these systems present near the Galactic center is unknown, 
because it is determined by the uncertain star formation history 
\citep[see, e.g.,][]{mor93}. \citet{fig95} and \citet{cot99} 
have identified several massive, emission-line stars associated with HII 
regions near the Galactic center, but none of these has counterparts in 
our X-ray catalog.
The wide-area search that \citet{fig95} conducted failed to turn up 
additional candidates. Still, the unique conditions at the Galactic center 
make it important to understand the number of massive stars there, so 
we suggest that the relatively soft X-ray sources 
would serve as good targets for future searches for 
massive stars. 

\subsection{Millisecond Pulsars}

Isolated millisecond pulsars typically produce X-ray emission from particles
accelerated as they spin down, with $L_{\rm X} = 10^{28} - 10^{31}$~\ergsec\
\citep{pos02}. At these luminosities, millisecond pulsars would be undetectable
at the Galactic center. However,
\citet{cheng04} have predicted that the wind from a millisecond pulsar could
produce $L_{\rm X} = 10^{31} - 10^{33}$~\ergsec\ by interacting with dense 
regions of the ISM ($n \ga 100$ cm$^{-3}$). They suggest that $\sim 100$ 
millisecond pulsars could be present in our field, although the number of 
detectable systems would depend upon the volume of dense gas at the Galactic 
center. \citet{mun04} have
demonstrated that a large fraction of the inner 20 pc of the Galaxy is filled
with hot ($T \sim 10^{8}$ K), low density ($n \approx 0.1$ cm$^{-3}$), 
X-ray emitting plasma, so only a small fraction of isolated millisecond pulsars
may be detectable. Moreover, their spectra 
should be power laws with $\Gamma = 1.5-2.5$, which corresponds to 
$-0.2 < HR < 0.0$ in Figure~\ref{fig:indiv}. This places millisecond pulsars 
on the same portion of the hardness-intensity diagram as CVs and RS CVns. 
As discussed in \citet{cheng04}, identifying candidate systems among the 
point sources would be difficult, but millisecond pulsars could account
for extended features seen in the field \citep{mor03}.

\subsection{Accreting Sources}

\subsubsection{Low-Mass X-ray Binaries}

Neutron stars and black holes accreting from low-mass companions that 
over-fill their Roche lobes are typically identified in outburst with 
$L_{\rm X} > 10^{36}$~\ergsec, although the majority of their
time is spent in quiescence with $L_{\rm X} < 10^{34}$~\ergsec.
LMXBs have been observed extensively in 
quiescence. The spectra of quiescent neutron star systems have been 
described with a $kT \approx 0.3$~keV black body producing 
$L_{\rm X} \sim 10^{32}$~\ergsec, plus a $\Gamma \approx 1-2$ 
power-law tail that contributes $L_{\rm X} \sim 10^{31}$~\ergsec\
\citep{asa98,kon02}. 
The black hole systems have $L_{\rm X} \lesssim 10^{31}$~\ergsec\ 
and exhibit $\Gamma \approx 1-2$ power-law spectra
\citep{rut01,wij02,cam02}. The thermal emission 
from a neutron star would be unobservable behind $6\times10^{22}$ cm$^{-2}$ of
absorption, so is of little relevance to the current observations. The 
power-law components of both the neutron star and black hole systems 
would produce $-0.1 < HR < 0.2$. 

However, LMXBs are rare --- theoretical models predict that $\sim 10^4$ 
should currently be in quiescence in the entire Galaxy, whereas only 
only $\sim 100$  LMXBs, or 1\%, have been identified 
\citep[compare][]{itf97,bt04}. 
Thus, if LMXBs form at the Galactic center in a similar manner as in the disk,
our observation should encompass $\sim 20$ of them \citep{bt04}. 
Transient outbursts from three LMXBs already 
have been identified within 10\arcmin\ of the Galactic center 
\citep{eyl75,pgs94,mae96}. If these truly represent $\sim 20$\% of the total
number there, then it would appear that LMXBs near the Galactic center are 
considerably more active than those in the Galactic disk. Alternatively, 
LMXBs could be concentrated near the 
Galactic center through dynamical settling \citep[e.g.,][]{mor93,poz03}.
In order to better constrain the numbers of LMXBs within the nuclear bulge, 
it is important to 
continue to monitor this region in order to search for transient 
outbursts from additional systems.

Prior to the Roche-lobe overflow phase, accretion onto the compact objects 
should also proceed at low rates from the winds of the low-mass companions
\citep{ble02,wk03,bt04}. These
pre-LMXBs should have $L_{\rm X} = 10^{28} - 10^{32}$~\ergsec, and would
probably resemble Roche-lobe overflow systems in quiescence. 
Up to $10^5$ systems could be present in the Galaxy, and $\sim 20-100$ in our
image of \sgrastar \citep{wk03,bt04}. 

\subsubsection{High-Mass X-ray Binaries}

Neutron stars and black holes accreting from the winds of massive 
companions should be about as common as LMXBs, because although the massive
companions have much shorter lifetimes, accretion can occur when the 
separations between the binary components are much larger 
($\sim 1$ AU compared to $\sim R_\odot$; see Pfahl, Podsiadlowski, 
\& Rappaport 2002). They could be particularly abundant near the Galactic
center, because it appears that 10\% of Galactic star formation is currently
occurring within the nuclear bulge \citep{lzm02}. Our observations 
encompass $\sim 5$\% of the nuclear bulge, so it would not be 
unreasonable to assume that, of the $\sim 10^{4}$ HMXBs in the Galaxy, 
 $0.1\cdot0.05\cdot10^{4} \sim 50$ could be present in the field 
around \sgrastar\ \citep[see][]{pfa02}.

In both 
outburst and quiescence, black hole HMXBs generally resemble LMXBs, 
because their X-ray emission is produced entirely in
the accretion flow.
Neutron star HMXBs, on the other hand, usually look much different from their
LMXB counterparts, because the neutron stars in the young, high-mass systems
tend to be more highly magnetized ($B \ga 10^{12}$ Gauss). Neutron
star HMXBs in outburst produce X-rays from shocks that form 
in the magnetically-channeled column of accreted material. 
At the location of the shocks, the accretion flow is 
optically-thick, so the resulting spectra are flat, and can be described with 
a $\Gamma < 1$ power law between 2--8~keV \citep[e.g.][]{cam01}. 
Therefore, neutron star HMXBs should have $HR > 0.1$ in Figure~\ref{fig:indiv}.
HMXBs also sometimes exhibit line emission at 6.4 keV from fluorescent
neutral material in the companion's wind, as well as weaker emission from 
He-like Fe at 6.7 keV that is produced by photo-ionized plasma in the 
wind. Although large equivalent
widths have been reported from low-resolution 
measurements with gas proportional counters \citep[e.g.,][]{app94}, 
the few measurements of these lines 
with CCD resolution spectra indicate that they have equivalent widths 
$\lesssim 100$~eV \citep[e.g.,][]{nag94,shr99}. 

Since the strong magnetic fields around the neutron stars in HMXBs channel
the accreted material onto the star's polar caps, the surest way 
to identify neutron star HMXBs is through periodic 
modulations in their X-ray emission. We have found that seven hard Galactic 
center sources in our field exhibit periodic variability 
\citep{mun03c}. 
However, the periods are all 
$> 300$~s, which makes it impossible to rule out that they are accreting white
dwarfs. On the other hand, modulations with shorter periods would
be rendered undetectable by Doppler shifts from orbital motion, so the lengths
of the periods observed are not necessarily a strong constraint on the entire
population of sources at the Galactic center.
 
Other variability is also seen from HMXBs.
Short-term (several ks) flares are 
seen infrequently and are ascribed to instabilities in the accretion 
flow \citep[e.g.,][]{aug03, mew03}. Long-term variations are more
common and are often caused either by changes in the density of the wind 
at the 
location of compact objects that have eccentric orbits around the donor 
star, or by instabilities in the excretion disks around the Be stars that
are the mass donors in half of the known HMXBs \citep[e.g.][]{app94}. 

The above considerations suggest that faint, neutron star HMXBs  
can account for some fraction of the hard Galactic center point sources. 
The main problem with this hypothesis is that few HMXBs have been observed at 
$L_{\rm X} < 10^{34}$~\ergsec, and the ones that have can be described with 
much softer $\Gamma \sim 2$ spectra \citep[e.g.,][]{cam02}. 
Nonetheless, the physics of X-ray production at low accretion rates is
uncertain, so we cannot be certain of what X-ray properties to expect
from faint HMXBs. 

\subsubsection{Cataclysmic Variables}

Cataclysmic variables (CVs) are the most numerous accretion-powered X-ray 
sources. Their local space density is $\sim 3 \times 10^{-5}$ pc$^{-3}$ 
\citep{sch02}, so that if we scale their number to the stellar density 
at the Galactic center, within 20 pc of \sgrastar\ we would expect 
$\sim 9\times10^{3}$ CVs, and within a cylinder
centered on the Galactic center that is 20 pc in radius and 440 pc deep
we would expect $\sim 4\times10^{4}$. About 50\% of CVs are luminous enough
to be observed from the Galactic center \citep{ver97}, so they could 
account for the majority of the X-ray sources detected there.

Systems with non-magnetized white dwarfs, which comprise 80\% of CVs, have 
luminosities between $10^{29.5} - 10^{32}$~\ergsec, and spectra that
can be described with $kT = 1-25$~keV plasma from an accretion shock 
\citep[e.g.,][]{ehp91,ms93,ver97}. Thus, these systems should have hard colors 
$HR < 0$ in Figure~\ref{fig:indiv}, and would be located in a similar
portion of the color-intensity diagram as RS CVns and YSOs.

CVs containing magnetized white dwarfs, which are referred to as polars and 
intermediate polars depending on whether or not the rotational period of the 
white dwarf is synchronized to its orbital period,
comprise about 20\% of all CVs \citep[e.g.,][see also the 
CVcat database\footnote{\html{minerva.uni-sw.gwdg.de/cvcat/tpp3.pl}; 
Kube et al. (2003)}]{war95}.
Polars have similar spectra and luminosities as un-magnetized CVs, with 
the addition of a $kT \sim 50$~eV ``soft excess'' that is attributed to 
``blobs'' of accreted material that penetrate deeply into the 
photosphere \citep[e.g.,][]{ram94,ver97,ei99}. 
The soft component would be unobservable above 2~keV, so polars should 
also have $HR < 0.1$ in Figure~\ref{fig:indiv}. 
Polars also commonly exhibit variations in their
average luminosity on time scales of years: $\approx 50$\% of the polars
surveyed by \citet{ram04} changed in intensity by factors of $\ga 4$
between observations taken 
with \rosat\ (1990--1999) and \xmm\ (2000--present). Such variations would
be detectable from most of the Galactic center sources 
(Figure~\ref{fig:shortterm}). Therefore, in the two 
years spanned by the \chandra\ observations of the Galactic center, 
we would expect $\sim 15\%$ of the polars to exhibit long-term variations. 
Since only 2\% of the sources located at or beyond the Galactic center 
are variable, at most 20\% could be polars.

The intermediate polars are typically more luminous than 
other CVs, with $L_{\rm X} = 10^{31} - 10^{32.6}$~\ergsec, and represent 
about 5\% of the total population \citep[see CVcat;][]{kub03}. This is 
thought to be related to the fact that they tend to have longer orbital periods
($> 2$ h), which could result in a higher mass transfer rate; however, the 
high $\dot{M}$ could also be a selection effect, because if a CV is 
bright, it is easier to detect modulations in the X-ray and optical emission 
at the rotational and orbital periods \citep{war95}.
Intermediate polars also typically have much harder spectra than other CVs:
when approximated as a power law, the optically thin thermal plasma usually 
seen from CVs should have $\Gamma \approx 1.5$, whereas the spectra of
intermediate polars usually have $\Gamma \approx 0$. This is probably a 
result of the geometry of the accretion flow, because, as in other CVs, 
prominent line emission from 
He-like and H-like Fe indicates that the X-rays are produced either by
plasma with $kT \approx 1-20$~keV or by a plasma photo-ionized by continuum
X-rays that are not observed directly \citep[e.g.,][]{ei99,muk03}. 
In either case, the X-ray emitting regions would have to be
partially absorbed by material in the accretion flows, which removes 
low-energy photons from the spectra, thus making them flatter. 
Intermediate polars should have
$HR > 0.1$ in Figure~\ref{fig:indiv}, which makes them the best candidates 
among CVs for the hard Galactic center sources.

The detailed spectral properties of intermediate polars are broadly 
consistent with the 
average spectra of the point sources in Figure~\ref{fig:psmod}. 
Weak emission at 6.4~keV is observed from these systems, and is 
attributed to X-rays 
that reflect off of the white dwarf's surface \citep[e.g.,][]{ms93,ei99}. 
Moreover, when the spectra of intermediate polars are modeled as emission 
from thermal plasma, the derived Fe abundances are often near or 
below the solar values \citep[e.g.,][]{do97,fi97,ish97}. This is similar to 
what we infer for the point sources in Table~\ref{tab:twokt}. 

Finally, the general lack of variability in the X-ray emission from the 
Galactic center sources (aside from periodic modulations) is also 
consistent with the stable emission usually 
seen from intermediate polars. On long time scales, the optical luminosity
of intermediate polars usually remains constant for many decades 
\citep[e.g.,][]{gs88}; because the optical and X-ray flux are correlated in 
polars, we would expect the X-ray emission from intermediate polars also 
remains constant. Flares lasting several hours, presumably from accretion
events, are sometimes observed from 
magnetic CVs, but appear to be rare and most prominent in
the soft X-ray band \citep[$< 2$~keV; e.g.,][]{ps93,cda99, sm01}. 
The predominant short time scale variability in intermediate polars
is due to modulations of the emitting regions as the white dwarfs rotate
\citep[e.g.,][]{nw89,sch02,rc03}.
We have detected periodic modulations from seven of the brightest 285
Galactic center sources \citep{mun03c}. 
Since we were only sensitive to high-amplitude modulations, it is likely 
that many sources with low-amplitude modulations went undetected.
Therefore, although the faintness of the 
Galactic center X-ray sources is probably the main cause of the lack of 
observed short-term variability, it is also 
plausible that the sources are intrinsically steady X-ray emitters like 
intermediate polars. 

Since the properties of the Galactic center sources change little as 
a function of their luminosity 	between $10^{31}$ and $10^{33}$~\ergsec\
(Figures~\ref{fig:rat} and \ref{fig:indiv}), we believe that the majority 
of the Galactic center sources are intermediate polars. Intermediate polars
comprise 5\% of all known CVs \citep{kub03}, so given that 
there could be $4\times10^4$ CVs within a pencil-beam 
centered on the Galactic center that is 20 pc in radius and 440 pc deep,
they could reasonably account for the 1000 X-ray sources with $HR > 0$.

\subsection{Supernova Ejecta}

Bykov (2002, 2003)\nocite{byk02,byk03} has suggested that the 
point sources in the Galactic center may not be stellar, but could be
iron-rich fragments of supernova explosions that are interacting with
molecular clouds. On order $10^{3}$ X-ray emitting knots could plausibly
be produced by just 3 supernova occurring within the last 1000 y within 
20~pc of the Galactic center; already, Sgr A East \citep{mae02} and the 
radio wisp 'E' \citep{ho85} are thought to be remnants of recent supernova. 
The observational properties of the point sources
can be reproduced by choosing several parameters in the 
ejecta model \citep{byk03}: the slope $\log N - \log S$ distribution 
of the knots ($\alpha \approx 1.7$) is determined by their sizes and 
velocities, the slopes of their continuum X-ray emission ($0 < \Gamma < 1.5$) 
is set by the amplitudes of magneto-hydrodynamic turbulence in the 
shocks they produce, and the equivalent widths of the Fe emission 
(up to 1~keV) by their iron abundances. Future observations of known 
supernova remnants will better constrain the properties of the 
X-ray emitting knots, which in turn could make it possible to distinguish 
such knots from the stellar sources in the field.

\subsection{Unusual Sources}

A handful of the Galactic center sources resemble unusual objects that have 
been found through shallower \asca, \bepposax, \xmm, and 
\integral\ surveys of the Galactic plane. These sources are important,
because they could represent stellar remnants that are in short-lived 
states of accretion. 
We list the properties of 14 unusual sources from other
surveys in Table~\ref{tab:odd}. 
The first three are polars that were identified with \asca\ as having 
unusually strong emission lines from He-like Fe 
(equivalent widths $> 1$ keV); the fourth \xmm\ source has similarly
strong Fe emission at 6.7 keV, but its nature is uncertain. 
We find that 6 out of 183 
Galactic center sources searched for Fe emission have 6.7 keV 
lines with equivalent 
widths greater than 1 keV, which is similar to the fraction of 
such sources identified in the \asca\ Galactic plane survey.
The next four are highly-absorbed 
($N_{\rm H} > 10^{23}$~cm$^{-2}$) sources identified with \integral\ and
\xmm, one 
of which has strong low-ionization Fe emission with an equivalent width 
$> 1$~keV. We find that 30\% of the Galactic center sources have similarly
high absorption, and two systems exhibit 6.4~keV Fe 
lines with equivalent 
widths $> 1$ keV (CXOGC J174613.7--290662 and GXOCG J174617.2--285449 in 
Table~\ref{tab:iron}). 
The final five are hard X-ray sources 
with slow ($> 100$~s), high-amplitude periodic
modulations in their X-ray emission. We find seven hard sources near 
the Galactic center (and one
foreground source) with similar periodic X-ray modulations
\citep{mun03c}. 

These sources would have been difficult
to identify with the soft X-ray detectors on \rosat\ (0.1--2.4 keV), 
which was the last observatory that systematically
surveyed the sky for faint X-ray sources. Our study of the Galactic center
suggests that they account for a few percent of all faint X-ray sources. 

\section{Conclusions}

We have established that, on average, the X-ray sources detected in 626~ks 
of \chandra\ ACIS-I observations of the field around \sgrastar\ have hard, 
$\Gamma < 1$ spectra with prominent emission from He-like Fe at 6.7~keV
(Figure~\ref{fig:psmod} and Table~\ref{tab:psint}).
They also generally do not vary by more than factors of a few on time scales
of hours or months. 
The best candidates for these hard X-ray sources are intermediate polars, 
which represent the most luminous and spectrally hardest 5\% of all CVs. 
Therefore, the Galactic center X-ray sources are likely to be only a 
sub-sample of a population of $\sim 10^{4}$ CVs located near the Galactic 
center. 

Although a single population of sources may dominate the image, there are 
certainly many classes of objects present in smaller numbers in the field.
Determining the numbers of rare objects is particularly important. For 
instance, the numbers of massive Wolf-Rayet and O stars
and faint neutron star high-mass X-ray binaries can constrain the recent rate 
of massive star formation near the Galactic center, 
while the numbers of LMXBS provide direct tests of the validity of unusual 
pathways for binary stellar evolution. 
For this reason, we are carrying out 
deep infrared observations of the Galactic center to identify counterparts
to the X-ray sources. These observations will be useful for distinguishing
CVs from, for example, HMXBs and WR/O stars. At at a distance of 8~kpc and 
with an extinction of $A_K \approx 5$ \citep{td03}, CVs should have $K$ 
magnitudes of 22--25, and 
therefore would be among the faintest detectable sources at the Galactic 
center \citep{war95,hoa02}. In contrast, HMXBs and WR/O stars should have $K$ 
magnitudes brighter than 15 \citep{zom90,weg94} and will be 
very easy to detect. Therefore, the prospects for identifying the natures
of the Galactic center X-ray sources are promising.

\acknowledgments
We thank C. Belczyski, 
A. Bykov, M. Eracleous, C. Heinke, K. Mukai, F. Paerels, J. Sokoloski, 
and R. Taam for 
helpful discussions about the natures of the Galactic center X-ray sources,
and the referee for comments that helped to clarify the text.
We are also grateful to M. Nowak for providing us his implementation of the 
Bayesian Blocks algorithm.
MPM was supported by a Hubble 
Fellowship from the Space Telescope Science Institute, which is operated
by the Association of Universities for Research in Astronomy, Inc.,
under NASA contract NAS 5-26555. WNB was acknowledges an NSF CAREER award
AST-9983783.

\begin{deluxetable}{lccccc}
\tablecolumns{6}
\tablewidth{0pc}
\tablecaption{Observations of the Inner 20 pc of the Galaxy\label{tab:obs}}
\tablehead{
\colhead{} & \colhead{} & \colhead{} & 
\multicolumn{2}{c}{Aim Point} & \colhead{} \\
\colhead{Start Time} & \colhead{Sequence} & \colhead{Exposure} & 
\colhead{RA} & \colhead{DEC} & \colhead{Roll} \\
\colhead{(UT)} & \colhead{} & \colhead{(s)} 
& \multicolumn{2}{c}{(degrees J2000)} & \colhead{(degrees)}
} 
\startdata
1999 Sep 21 02:43:00 & 0242  & 40,872 & 266.41382 & $-$29.0130 & 268 \\
2000 Oct 26 18:15:11 & 1561 & 35,705 & 266.41344 & $-$29.0128 & 265 \\
2001 Jul 14 01:51:10 & 1561 & 13,504 & 266.41344 & $-$29.0128 & 265 \\
2002 Feb 19 14:27:32 & 2951  & 12,370 & 266.41867 & $-$29.0033 & 91 \\
2002 Mar 23 12:25:04 & 2952  & 11,859 & 266.41897 & $-$29.0034 & 88 \\
2002 Apr 19 10:39:01 & 2953  & 11,632 & 266.41923 & $-$29.0034 & 85 \\
2002 May 07 09:25:07 & 2954  & 12,455 & 266.41938 & $-$29.0037 & 82 \\
2002 May 22 22:59:15 & 2943  & 34,651 & 266.41991 & $-$29.0041 & 76 \\
2002 May 24 11:50:13 & 3663  & 37,959 & 266.41993 & $-$29.0041 & 76 \\
2002 May 25 15:16:03 & 3392  & 166,690 & 266.41992 & $-$29.0041 & 76 \\
2002 May 28 05:34:44 & 3393  & 158,026 & 266.41992 & $-$29.0041 & 76 \\
2002 Jun 03 01:24:37 & 3665  & 89,928 & 266.41992 & $-$29.0041 & 76 
\enddata
\end{deluxetable}

\begin{deluxetable}{lcccccccccccccc}
\tabletypesize{\scriptsize}
\rotate
\tablecolumns{15}
\tablewidth{0pc}
\tablecaption{Spectra of Individual Point Sources\label{tab:indiv}} 
\tablehead{ 
\colhead{} & \colhead{} & \multicolumn{5}{c}{Power Law} & 
\multicolumn{5}{c}{Plasma (mekal)} & \multicolumn{2}{c}{Fe Lines} &
\colhead{} \\
\colhead{Name} & \colhead{Net} & \colhead{$N_{\rm H}$} & \colhead{$\Gamma$} & 
\colhead{$F_{\rm X}$} & \colhead{$uF_{\rm X}$} &
\colhead{$\chi^2/\nu$} & 
\colhead{$N_{\rm H}$} & \colhead{$kT$} & 
\colhead{$F_{\rm X}$} & \colhead{$uF_{\rm X}$} &
\colhead{$\chi^2/\nu$} & 
\colhead{6.4 keV} & \colhead{6.7 keV} & \colhead{Flags} \\
\colhead{CXOGC~J} & \colhead{Counts} &
\colhead{$10^{22}$ cm$^{-2}$} & \colhead{} & 
\multicolumn{2}{c}{($10^{-14}$ erg cm$^{-2}$ s$^{-1}$)} & \colhead{} &
\colhead{$10^{22}$ cm$^{-2}$} & \colhead{(keV)} & 
\multicolumn{2}{c}{($10^{-14}$ erg cm$^{-2}$ s$^{-1}$)} & \colhead{} &
\colhead{} & \colhead{} & \colhead{}
}
\startdata
174521.9--290519 &  499 & $  8_{-  3}^{+  5}$ & $ 0.1_{- 0.5}^{+ 0.9}$ &   4.0 &     6.5 &   24/27 & $ 14_{-  2}^{+  1}$ & $63.2(e)$ &   3.6 &     9.7 &   27/27 & n & n & - \\
174521.9--290616 &  366 & $ 10_{-  6}^{+ 10}$ & $ 0.3_{- 0.9}^{+ 1.6}$ &   3.3 &     5.9 &   14/20 & $ 17_{-  4}^{+  4}$ & $15.6(e)$ &   3.1 &    10.6 &   15/20 & n & n & l \\
174522.3--290322 &   97 & $<  8$ & $-1.0_{- 1.4}^{+ 2.2}$ &   0.8 &     0.9 &    6/4 & $ 15_{-  7}^{+ 17}$ & $>  1.6$ &   0.6 &     2.0 &    7/4 & \nodata & \nodata & - \\
174522.9--285718 &  139 & $  3$ & -0.3(e) &   0.9 &     1.1 &   15/7 & $  8$ & $79.9(e)$ &   0.8 &     1.5 &   19/7 & \nodata & \nodata & - \\
174522.9--290706 &  158 & $ 16_{- 13}^{+ 17}$ & $ 3.3_{- 2.7}^{+ 3.6}$ &   0.8 &     5.2 &   14/16 & $ 18_{-  9}^{+ 11}$ &  1.8$_{- 0.8}^{+ 4.4}$ &   0.9 &     6.5 &   10/16 & \nodata & \nodata & - \\ [5pt]
174523.1--290205 &  123 & $ 12_{-  8}^{+ 12}$ & $ 1.4_{- 2.1}^{+ 3.0}$ &   0.8 &     2.0 &    9/5 & $ 16_{-  6}^{+ 10}$ &  3.3$_{- 1.9}^{+16.1}$ &   0.8 &     3.4 &    5/5 & \nodata & \nodata & l \\
174523.2--290116 &   93 & $  6$ &  1.3(e) &   0.5 &     0.9 &    7/3 & $  8_{-  3}^{+  3}$ & $>  1.9$ &   0.5 &     1.1 &    6/3 & \nodata & \nodata & - \\
174523.3--290637 &  127 & $< 12$ & $-1.6_{- 1.2}^{+ 1.9}$ &   1.1 &     1.2 &    9/13 & $ 21_{- 11}^{+ 16}$ & $>  1.7$ &   0.9 &     3.9 &   12/13 & \nodata & \nodata & - \\
174523.4--290248 &   80 & $< 10$ & $-1.9_{- 1.1}^{+ 1.2}$ &   0.7 &     0.7 &    1/3 & $ 22_{-  9}^{+ 23}$ & $>  1.3$ &   0.6 &     2.8 &    3/3 & \nodata & \nodata & - \\
174523.8--290514 &   92 & $  8_{-  6}^{+ 46}$ & $>-0.1$ &   0.6 &     1.1 &    9/7 & $  8_{-  3}^{+  7}$ & $79.9(e)$ &   0.6 &     1.1 &    9/7 & \nodata & \nodata & - \\ [5pt]
174523.8--290652 &   94 & $< 26$ & $-2.1_{- 0.9}^{+ 5.8}$ &   0.9 &     0.9 &   15/13 & $ 25_{- 12}^{+ 19}$ & $>  0.9$ &   0.7 &     3.7 &   14/13 & \nodata & \nodata & - \\
174524.0--285947 &   82 & $<  4$ & $-0.9_{- 1.0}^{+ 1.8}$ &   0.6 &     0.7 &    2/2 & $  6$ & $79.7(e)$ &   0.5 &     0.8 &    6/2 & \nodata & \nodata & - \\
174524.1--285845 &  224 & $  0.8_{-  0.7}^{+  1.0}$ & $ 2.9_{- 0.4}^{+ 0.6}$ &   0.2 &     0.2 &    2/9 & $  0.6_{-  0.5}^{+  0.7}$ &  2.5$_{- 0.6}^{+ 1.0}$ &   0.2 &     0.2 &    5/9 & n & n & - \\
174524.7--290038 &  121 & $  3$ & -0.6(e) &   0.9 &     1.0 &   11/4 & $ 10$ & $ 6.0(e)$ &   0.7 &     1.7 &   13/4 & \nodata & \nodata & - \\
174525.1--285703 &  152 & $<  0.5$ & $ 1.7_{- 0.3}^{+ 0.4}$ &   0.2 &     0.2 &    4/6 & $<  0$ &  5.2$_{- 2.4}^{+ 9.3}$ &   0.2 &     0.2 &    5/6 & \nodata & \nodata & se 
\enddata
\tablecomments{Spectral parameters are marked with $(e)$ when the uncertainty
calculation failed to converge. 
$F_{\rm X}$ refers to the observed 2--8 keV flux, and $uF_{\rm X}$ 
refers to the de-absorbed 2--8 keV flux. The ``Flags'' column indicates 
those sources
for which the spectra may be suspect, because they exhibit variability on 
the long-term (l) or short-term (s), lie near the edge of a CCD (e), or 
may be blended with a nearby source (c). The table in the printed version 
is only a sample of the full table, which 
is available through the electronic edition. The 
spectra, response functions, effective area functions, background estimates, 
and event lists for each source are available
from 
\html{www.astro.psu.edu/users/niel/galcen-xray-data/galcen-xray-data.html}.
}
\end{deluxetable}

\begin{deluxetable}{lcccccccccccc}
\tabletypesize{\scriptsize}
\rotate
\tablecolumns{13}
\tablewidth{0pc}
\tablecaption{Iron Emission from Sources with $> 160$ Net Counts \label{tab:iron}} 
\tablehead{ 
\colhead{} & \multicolumn{6}{c}{Low-ionization Fe (6.4 keV)} &
\multicolumn{6}{c}{He-like Fe (6.7 keV)} \\
\colhead{Source} & \colhead{$\Gamma$} & \colhead{$I_{\rm Fe}$} & 
\colhead{$EW_{\rm Fe}$} & \colhead{$\Delta\chi^2$} & \colhead{$P_{\rm Fe}$} & 
\colhead{$\chi^2/\nu$} & 
\colhead{$\Gamma$} & \colhead{$I_{\rm Fe}$} & 
\colhead{$EW_{\rm Fe}$} & \colhead{$\Delta\chi^2$} & \colhead{$P_{\rm Fe}$} & 
\colhead{$\chi^2/\nu$} \\
\colhead{} & \colhead{} & \colhead{($10^{-7}$ \phcms)} & \colhead{(eV)} &
\colhead{} & \colhead{} & \colhead{} &
\colhead{} & \colhead{($10^{-7}$ \phcms)} & \colhead{(eV)} &
\colhead{} & \colhead{} & \colhead{} 
}
\startdata
174508.7--290324 & $ 2.1_{- 1.2}^{+ 3.6}$ & $<   0.2$ & $<   830$  &  0.9 & 0.064  &   11.8/ 11  & $ 3.4_{- 1.7}^{+ 4.4}$ & $ 0.3_{- 0.3}^{+ 0.5}$ &  1581  &  3.3 & 0.000  &    9.3/ 11  \\
174510.3--285435 & $ 3.0_{- 2.0}^{+ 1.2}$ & $<   0.5$ & $<  1425$  &  1.2 & 0.100  &   21.6/ 17  & $ 3.0_{- 1.8}^{+ 1.0}$ & $ 0.2_{- 0.2}^{+ 0.3}$ &   732  &  2.0 & 0.007  &   20.7/ 17  \\
174510.5--290645 & $ 1.5_{- 1.1}^{+ 1.3}$ & $ 0.9_{- 0.6}^{+ 0.3}$ &   411  &  7.7 & 0.000  &   42.4/ 39  & $ 1.6_{- 1.0}^{+ 0.7}$ & $ 1.0_{- 0.5}^{+ 0.4}$ &   474  &  8.6 & 0.000  &   41.2/ 39  \\
174512.4--290604 & $ 1.3_{- 1.3}^{+ 3.4}$ & $<   0.5$ & $<  1447$  &  4.2 & 0.024  &   11.0/ 20  & $ 2.2_{- 2.0}^{+ 3.1}$ & $ 0.3_{- 0.3}^{+ 0.3}$ &  1085  &  5.4 & 0.000  &   10.2/ 20  \\
174517.3--290440 & $-0.3_{- 0.4}^{+ 0.6}$ & $<   0.4$ & $<   558$  &  5.3 & 0.011  &   21.6/ 23  & $-0.2_{- 0.4}^{+ 0.6}$ & $ 0.4_{- 0.2}^{+ 0.3}$ &   467  &  6.7 & 0.002  &   20.0/ 23  \\ [5pt]
174519.8--290114 & $ 0.9_{- 0.9}^{+-0.9}$ & \nodata & \nodata & \nodata & \nodata  &   36.5/ 15  & $ 1.4_{- 1.4}^{+-1.4}$ & $ 0.5_{- 0.5}^{+-0.5}$ &   948  &  4.9 & 0.006  &   28.3/ 15  \\
174520.9--285818 & $ 4.2_{- 0.7}^{+ 1.5}$ & $ 0.3_{- 0.1}^{+ 0.1}$ & $>10^4$  &  6.1 & 0.003  &   10.3/ 11  & $ 4.8_{- 1.3}^{+ 1.0}$ & $ 0.3_{- 0.2}^{+ 0.1}$ & $>10^4$  &  5.9 & 0.003  &   10.6/ 11  \\
174525.5--290028 & $ 0.3_{- 0.5}^{+ 0.5}$ & $<   0.6$ & $<   449$  &  4.8 & 0.012  &   49.9/ 28  & $ 0.5_{- 0.4}^{+ 0.2}$ & $ 1.0_{- 0.4}^{+ 0.2}$ &   818  & 13.0 & 0.000  &   33.0/ 28  \\
174527.6--285258 & $ 0.8_{- 0.6}^{+ 0.5}$ & $ 0.7_{- 0.4}^{+ 0.5}$ &   223  &  6.2 & 0.009  &   48.2/ 40  & $ 1.2_{- 0.8}^{+ 0.6}$ & $ 0.9_{- 0.3}^{+ 0.7}$ &   310  & 10.0 & 0.000  &   42.9/ 40  \\
174527.8--290542 & $ 1.0_{- 1.5}^{+ 1.0}$ & $<   0.2$ & $<   394$  &  1.1 & 0.152  &    6.7/ 13  & $ 2.0_{- 2.1}^{+ 0.6}$ & $ 0.4_{- 0.4}^{+ 0.4}$ &   761  &  4.2 & 0.003  &    5.1/ 13  \\ [5pt]
174529.0--290406 & $ 2.5_{- 1.4}^{+ 0.6}$ & $<   0.5$ & $<   699$  &  2.9 & 0.028  &   13.7/ 11  & $ 2.9_{- 1.3}^{+ 2.1}$ & $ 0.4_{- 0.3}^{+ 0.7}$ &   716  &  4.7 & 0.006  &   11.1/ 11  \\
174529.6--285432 & $ 1.5_{- 1.1}^{+ 0.5}$ & $ 0.2_{- 0.2}^{+ 0.2}$ &   238  &  3.3 & 0.007  &   11.4/ 18  & $ 1.3_{- 0.8}^{+ 1.2}$ & $<   0.4$ & $<   642$  &  1.8 & 0.050  &   12.5/ 18  \\
174531.1--290219 & $-1.5_{--1.5}^{+ 1.5}$ & \nodata & \nodata & \nodata & \nodata  &   14.0/  5  & $ 1.6_{- 1.9}^{+ 2.7}$ & $ 0.7_{- 0.3}^{+ 0.7}$ &  2450  &  4.8 & 0.003  &    3.4/  5  \\
174532.3--290251 & $ 1.6_{- 1.5}^{+ 0.7}$ & $<   0.5$ & $<   345$  &  2.1 & 0.060  &   24.9/ 14  & $ 1.7_{- 1.3}^{+ 1.9}$ & $ 0.8_{- 0.3}^{+ 0.4}$ &   589  &  7.5 & 0.000  &   14.5/ 14  \\
174532.4--290259 & $-0.4_{- 1.0}^{+ 0.6}$ & $ 0.4_{- 0.3}^{+ 0.2}$ &   434  &  4.9 & 0.001  &   17.5/ 12  & $-0.4_{- 0.4}^{+ 1.2}$ & $ 0.5_{- 0.2}^{+ 0.3}$ &   542  &  6.5 & 0.002  &   14.1/ 12  \\ [5pt]
174534.5--285523 & $-0.0_{- 0.5}^{+ 1.6}$ & $<   0.2$ & $<   280$  &  0.9 & 0.228  &   22.2/ 16  & $ 1.3_{- 1.4}^{+ 0.7}$ & $ 0.7_{- 0.3}^{+ 0.3}$ &   943  & 12.3 & 0.000  &    6.5/ 16  \\
174534.5--290201 & $ 0.3_{- 0.4}^{+ 0.4}$ & $ 0.6_{- 0.3}^{+ 0.3}$ &   230  &  8.4 & 0.001  &   43.7/ 39  & $ 0.4_{- 0.6}^{+ 0.4}$ & $ 0.9_{- 0.3}^{+ 0.4}$ &   344  & 14.3 & 0.000  &   35.7/ 39  \\
174534.9--290118 & $ 0.7_{- 0.9}^{+ 0.8}$ & $<   0.5$ & $<   321$  &  2.6 & 0.053  &   20.0/ 17  & $ 1.3_{- 1.4}^{+ 0.7}$ & $ 0.6_{- 0.3}^{+ 0.3}$ &   439  &  7.5 & 0.001  &   13.7/ 17  \\
174535.6--290034 & $ 0.8_{- 0.8}^{+ 0.9}$ & $<   0.2$ & $<   122$  &  1.0 & 0.148  &   26.2/ 24  & $ 1.5_{- 1.1}^{+ 0.9}$ & $ 0.6_{- 0.4}^{+ 0.2}$ &   513  &  7.5 & 0.000  &   19.1/ 24  \\
174536.1--285638 & $ 3.4_{- 3.4}^{+-3.4}$ & \nodata & \nodata & \nodata & \nodata  &  259.9/111  & $ 3.6_{- 0.2}^{+ 0.1}$ & $ 2.7_{- 0.4}^{+ 0.4}$ &  2173  & 48.5 & 0.000  &  156.4/111  \\ [5pt]
174537.6--290144 & $ 0.4_{- 0.5}^{+ 0.6}$ & $<   0.4$ & $<   410$  &  3.5 & 0.018  &   23.5/ 21  & $ 0.4_{- 0.5}^{+ 0.6}$ & $ 0.2_{- 0.2}^{+ 0.2}$ &   249  &  3.7 & 0.007  &   23.3/ 21  \\
174537.7--290002 & $-0.0_{- 1.0}^{+ 0.4}$ & $<   0.4$ & $<   815$  &  4.8 & 0.020  &   22.7/ 15  & $ 0.2_{- 0.5}^{+ 0.6}$ & $ 0.5_{- 0.2}^{+ 0.3}$ &   890  &  6.7 & 0.000  &   18.9/ 15  \\
174537.9--290134 & $ 4.4_{- 0.6}^{+ 1.0}$ & $<   0.3$ & $< 15803$  &  3.9 & 0.010  &   12.2/  9  & $ 4.6_{- 0.9}^{+ 0.9}$ & $ 0.2_{- 0.2}^{+ 0.1}$ & 16521  &  3.9 & 0.007  &   12.3/  9  \\
174540.1--290055 & $ 2.2_{- 0.5}^{+ 0.8}$ & $ 0.2_{- 0.2}^{+ 0.3}$ &   130  &  3.4 & 0.000  &   41.1/ 51  & $ 2.1_{- 0.6}^{+ 0.7}$ & $<   0.2$ & $<   129$  &  1.0 & 0.050  &   43.1/ 51  \\
174540.5--285550 & $ 0.5_{- 0.3}^{+ 0.9}$ & $<   0.4$ & $<   249$  &  1.7 & 0.092  &   42.2/ 26  & $ 1.0_{- 0.9}^{+ 0.4}$ & $ 0.7_{- 0.3}^{+ 0.3}$ &   448  &  8.0 & 0.002  &   31.7/ 26  \\ [5pt]
174541.2--290210 & $-1.0_{- 0.7}^{+ 1.0}$ & $<   0.7$ & $<   279$  &  2.1 & 0.055  &   26.1/ 17  & $-0.9_{- 0.6}^{+ 0.5}$ & $ 0.8_{- 0.4}^{+ 0.5}$ &   361  &  6.4 & 0.001  &   19.1/ 17  \\
174541.5--285814 & $ 1.0_{- 0.2}^{+ 0.1}$ & $ 0.3_{- 0.2}^{+ 0.2}$ &   174  &  6.7 & 0.003  &   78.1/ 78  & $ 0.9_{- 0.1}^{+ 0.3}$ & $<   0.5$ & $<   271$  &  3.9 & 0.020  &   81.1/ 78  \\
174541.6--285952 & $-0.1_{- 0.6}^{+ 0.5}$ & $ 0.4_{- 0.3}^{+ 0.4}$ &   365  &  5.1 & 0.003  &   12.2/ 14  & $ 0.3_{- 0.8}^{+ 0.8}$ & $<   1.1$ & $<  1060$  &  2.3 & 0.021  &   15.7/ 14  \\
174541.8--290037 & $ 0.4_{- 1.5}^{+ 1.1}$ & $<   0.7$ & $<   432$  &  4.6 & 0.019  &   41.5/ 27  & $ 0.3_{- 1.1}^{+ 1.6}$ & $ 1.0_{- 0.4}^{+ 0.6}$ &   678  & 11.1 & 0.001  &   30.0/ 27  \\
174542.0--285824 & $ 0.3_{- 2.4}^{+ 1.4}$ & $<   0.4$ & $<   451$  &  1.6 & 0.112  &    8.8/  6  & $ 0.3_{- 1.9}^{+ 1.1}$ & $ 0.4_{- 0.3}^{+ 0.3}$ &   584  &  4.4 & 0.002  &    4.3/  6  \\
174542.2--285732 & $-0.2_{- 0.4}^{+ 0.7}$ & $ 0.2_{- 0.1}^{+ 0.3}$ &   229  &  4.1 & 0.007  &   23.0/ 18  & $ 0.1_{- 0.3}^{+ 0.4}$ & $<   0.5$ & $<   601$  &  4.9 & 0.020  &   21.8/ 18  \\ [5pt]
174543.3--285605 & $-0.2_{- 0.6}^{+ 1.4}$ & $<   0.2$ & $<   342$  &  1.7 & 0.083  &    9.6/ 12  & $ 1.0_{- 1.1}^{+ 1.6}$ & $ 0.5_{- 0.3}^{+ 0.4}$ &   782  &  8.3 & 0.000  &    4.0/ 12  \\
174543.4--285841 & $ 1.2_{- 1.1}^{+ 0.4}$ & $<   0.5$ & $<   312$  &  3.0 & 0.029  &   26.8/ 21  & $ 1.4_{- 1.0}^{+ 0.6}$ & $ 0.7_{- 0.4}^{+ 0.3}$ &   409  &  7.7 & 0.000  &   20.2/ 21  \\
174543.7--285946 & $ 0.7_{- 1.0}^{+ 0.6}$ & $ 1.3_{- 0.5}^{+ 0.5}$ &   654  & 14.1 & 0.000  &   23.9/ 28  & $ 0.7_{- 0.9}^{+ 0.6}$ & $ 1.0_{- 0.5}^{+ 0.5}$ &   493  &  7.2 & 0.001  &   34.9/ 28  \\
174544.3--290156 & $ 0.7_{- 1.7}^{+ 1.5}$ & $ 0.2_{- 0.2}^{+ 0.3}$ &   355  &  6.0 & 0.001  &    1.2/  7  & $ 0.9_{- 1.9}^{+ 3.9}$ & $<   1.3$ & $<  1890$  &  1.8 & 0.060  &    3.8/  7  \\
174544.8--285953 & $ 0.3_{- 1.5}^{+ 5.0}$ & $<   0.3$ & $<  1151$  &  0.9 & 0.200  &   32.1/ 17  & $ 4.3_{- 2.7}^{+ 3.8}$ & $ 0.9_{- 0.3}^{+ 1.8}$ &  5241  &  8.7 & 0.002  &   17.5/ 17  \\ [5pt]
\newtablebreak
174544.9--290027 & $ 0.5_{- 0.3}^{+ 0.2}$ & $ 0.4_{- 0.3}^{+ 0.2}$ &   237  &  6.7 & 0.000  &   33.0/ 38  & $ 0.4_{- 0.3}^{+ 0.2}$ & $ 0.4_{- 0.4}^{+ 0.2}$ &   248  &  3.9 & 0.001  &   35.8/ 38  \\
174546.1--290057 & $-1.7_{- 0.5}^{+ 0.7}$ & $<   0.3$ & $<   485$  &  2.6 & 0.050  &   16.4/ 11  & $-1.2_{- 0.5}^{+ 0.4}$ & $ 0.4_{- 0.2}^{+ 0.2}$ &   674  &  6.5 & 0.001  &    9.6/ 11  \\
174546.2--285906 & $ 0.6_{- 0.4}^{+ 0.3}$ & $<   0.3$ & $<   292$  &  3.1 & 0.034  &   21.6/ 31  & $ 0.7_{- 0.4}^{+ 0.5}$ & $ 0.2_{- 0.2}^{+ 0.2}$ &   217  &  5.8 & 0.000  &   19.6/ 31  \\
174546.9--285903 & $ 0.5_{- 1.5}^{+ 1.9}$ & $<   0.4$ & $<   558$  &  3.2 & 0.014  &   11.0/ 11  & $ 2.8_{- 2.9}^{+ 2.5}$ & $ 0.7_{- 0.5}^{+ 1.1}$ &   823  &  4.4 & 0.004  &    9.6/ 11  \\
174547.0--285333 & $ 0.3_{- 0.6}^{+ 0.8}$ & $<   0.4$ & $<   188$  &  1.4 & 0.124  &   60.5/ 43  & $ 0.7_{- 0.4}^{+ 0.4}$ & $ 0.8_{- 0.4}^{+ 0.4}$ &   423  & 10.8 & 0.000  &   47.3/ 43  \\ [5pt]
174547.2--290000 & $ 0.1_{- 0.8}^{+ 2.3}$ & $ 0.5_{- 0.2}^{+ 0.3}$ &   561  &  6.8 & 0.001  &   25.4/ 18  & $ 1.0_{- 1.4}^{+ 1.4}$ & $ 0.4_{- 0.3}^{+ 0.3}$ &   327  &  3.0 & 0.008  &   33.2/ 18  \\
174548.9--285751 & $ 0.9_{- 0.4}^{+ 0.4}$ & $ 0.5_{- 0.3}^{+ 0.2}$ &   261  &  7.7 & 0.000  &   66.4/ 49  & $ 1.0_{- 0.3}^{+ 0.3}$ & $ 0.8_{- 0.3}^{+ 0.3}$ &   446  & 13.8 & 0.000  &   56.8/ 49  \\
174549.3--285557 & $-0.4_{- 0.3}^{+ 0.4}$ & $ 0.7_{- 0.3}^{+ 0.3}$ &   364  & 10.4 & 0.000  &   44.9/ 37  & $-0.2_{- 0.4}^{+ 0.3}$ & $ 0.8_{- 0.3}^{+ 0.3}$ &   413  &  9.7 & 0.000  &   46.1/ 37  \\
174549.6--290457 & $ 0.7_{- 1.4}^{+ 0.7}$ & $ 0.4_{- 0.2}^{+ 0.2}$ &   414  &  6.5 & 0.000  &   12.2/ 14  & $ 0.4_{- 0.5}^{+ 1.6}$ & $<   0.8$ & $<   862$  &  2.8 & 0.013  &   17.5/ 14  \\
174550.9--285430 & $ 0.9_{- 0.4}^{+ 0.5}$ & $ 0.5_{- 0.4}^{+ 0.3}$ &   426  &  6.5 & 0.000  &   18.0/ 23  & $ 0.7_{- 0.4}^{+ 1.0}$ & $<   0.6$ & $<   509$  &  1.9 & 0.030  &   22.7/ 23  \\ [5pt]
174552.0--285312 & $ 0.6_{- 0.8}^{+ 0.6}$ & $ 1.1_{- 0.5}^{+ 0.8}$ &   238  &  6.9 & 0.001  &   59.1/ 43  & $ 0.7_{- 0.8}^{+ 0.7}$ & $ 0.8_{- 0.6}^{+ 0.8}$ &   183  &  3.9 & 0.002  &   63.8/ 43  \\
174554.4--285816 & $ 0.1_{- 1.0}^{+ 0.3}$ & $ 0.9_{- 0.4}^{+ 0.3}$ &   420  &  9.2 & 0.000  &   39.0/ 26  & $-0.2_{- 0.7}^{+ 0.4}$ & $ 0.9_{- 0.3}^{+ 0.3}$ &   417  &  8.4 & 0.000  &   40.8/ 26  \\
174555.6--285600 & $ 2.8_{- 2.5}^{+ 3.6}$ & $<   0.5$ & $<  1595$  &  3.4 & 0.020  &    6.9/ 12  & $ 5.5_{- 3.6}^{+ 0.5}$ & $ 0.5_{- 0.4}^{+ 0.4}$ &  2640  &  5.5 & 0.003  &    5.5/ 12  \\
174558.9--290724 & $ 0.3_{- 0.3}^{+ 0.3}$ & $ 1.1_{- 0.4}^{+ 0.4}$ &   338  & 13.5 & 0.000  &   96.4/ 66  & $ 0.4_{- 0.3}^{+ 0.3}$ & $ 1.4_{- 0.5}^{+ 0.4}$ &   416  & 13.7 & 0.000  &   96.0/ 66  \\
174559.5--290601 & $ 1.7_{- 1.9}^{+ 0.6}$ & $<   0.5$ & $<   552$  &  3.0 & 0.050  &   29.6/ 20  & $ 1.4_{- 1.0}^{+ 1.0}$ & $ 0.4_{- 0.3}^{+ 0.4}$ &   508  &  5.0 & 0.002  &   26.3/ 20  \\ [5pt]
174601.0--285854 & $ 1.5_{- 0.9}^{+ 0.8}$ & $<   1.1$ & $<   284$  &  3.3 & 0.030  &   50.6/ 31  & $ 1.7_{- 1.1}^{+ 0.5}$ & $ 1.8_{- 0.8}^{+ 0.4}$ &   526  & 12.2 & 0.000  &   34.8/ 31  \\
174601.1--285953 & $ 1.0_{- 1.3}^{+ 0.6}$ & $ 0.4_{- 0.3}^{+ 0.3}$ &   749  &  4.5 & 0.006  &    4.2/  7  & $-0.4_{- 0.7}^{+ 2.3}$ & $<   0.4$ & $<   601$  &  0.9 & 0.170  &    8.5/  7  \\
174606.3--285810 & $ 1.1_{- 1.7}^{+ 0.6}$ & $ 1.1_{- 0.4}^{+ 0.4}$ &   841  & 14.6 & 0.000  &   25.9/ 29  & $ 1.0_{- 1.4}^{+ 0.7}$ & $ 0.6_{- 0.3}^{+ 0.3}$ &   458  &  6.1 & 0.004  &   40.2/ 29  \\
174608.4--290623 & $ 2.2_{- 1.3}^{+ 0.5}$ & $ 1.3_{- 0.5}^{+ 0.9}$ &   455  &  9.9 & 0.000  &   48.3/ 36  & $ 2.3_{- 1.4}^{+ 0.8}$ & $ 1.4_{- 0.6}^{+ 1.1}$ &   510  &  8.9 & 0.002  &   50.0/ 36  \\
174609.8--290321 & $ 1.6_{- 2.6}^{+ 3.5}$ & $<  11.6$ & $<  1560$  &  4.1 & 0.037  &   26.4/ 15  & $ 1.6_{- 2.4}^{+ 4.1}$ & $ 2.9_{- 0.7}^{+ 7.0}$ &   623  &  5.6 & 0.004  &   23.1/ 15  \\ [5pt]
174610.9--285345 & $ 1.3_{- 1.0}^{+ 1.1}$ & $ 0.7_{- 0.6}^{+ 0.6}$ &   336  &  4.8 & 0.001  &   48.4/ 51  & $ 1.2_{- 0.9}^{+ 1.1}$ & $<   0.7$ & $<   346$  &  1.8 & 0.100  &   51.5/ 51  \\
174612.3--285706 & $-0.6_{- 1.1}^{+ 1.9}$ & $<   0.2$ & $<   450$  &  1.2 & 0.180  &   27.0/ 23  & $ 0.1_{- 1.4}^{+ 1.4}$ & $ 0.3_{- 0.2}^{+ 0.2}$ &   817  &  5.8 & 0.000  &   21.5/ 23  \\
174613.7--290622 & $ 1.3_{- 1.1}^{+ 1.2}$ & $ 0.3_{- 0.2}^{+ 0.2}$ &  1594  &  6.8 & 0.006  &   15.3/ 22  & $ 1.1_{- 1.3}^{+ 1.2}$ & $<   0.5$ & $<  2798$  &  1.7 & 0.060  &   20.1/ 22  \\
174614.5--285428 & $ 0.1_{- 2.4}^{+ 7.3}$ & $ 0.5_{- 0.5}^{+ 2.1}$ &   923  &  3.7 & 0.004  &   62.7/ 51  & $ 0.1_{- 2.5}^{+ 5.0}$ & $<  18.6$ & $< 26553$  &  1.0 & 0.360  &   66.2/ 51  \\
174616.5--285846 & $-1.3_{- 0.5}^{+ 0.5}$ & $ 0.3_{- 0.3}^{+ 0.3}$ &   344  &  5.5 & 0.006  &   23.2/ 27  & $-1.4_{- 0.5}^{+ 1.9}$ & $<   0.4$ & $<   385$  &  1.3 & 0.160  &   27.6/ 27  \\
174617.2--285449 & $ 1.0_{- 0.5}^{+ 1.0}$ & $ 0.2_{- 0.2}^{+ 0.4}$ &  3021  &  1.7 & 0.000  &   35.2/ 26  & $ 0.9_{- 0.5}^{+ 0.9}$ & $<   0.2$ & $<  2647$  &  1.0 & 0.130  &   36.1/ 26  \\ [5pt]
174619.4--290213 & $-0.3_{- 0.6}^{+ 0.5}$ & $ 0.7_{- 0.5}^{+ 0.6}$ &   369  &  7.6 & 0.001  &   13.1/ 22  & $-0.4_{- 0.4}^{+ 1.5}$ & $<   0.9$ & $<   432$  &  1.6 & 0.090  &   18.2/ 22  \\
174623.6--285629 & $ 1.4_{- 1.0}^{+ 1.9}$ & $<   1.7$ & $<   582$  &  4.4 & 0.020  &   76.8/ 51  & $ 1.6_{- 1.2}^{+ 2.2}$ & $ 1.2_{- 0.7}^{+ 1.0}$ &   440  &  5.8 & 0.002  &   74.6/ 51
\enddata
\tablecomments{$P_{\rm Fe}$ refers to the chance probability that the line 
emission could be produce by random variations in a featureless continuum. 
Only sources with $P_{\rm Fe} < 0.01$ for a line at either 6.4 or 6.7 keV 
are listed in the table. Note that the line emission for those sources with 
a best-fit power laws with $\Gamma > 4$ should be viewed extremely skeptically,
because these steep spectra are background-dominated above 6~keV.}
\end{deluxetable}

\begin{deluxetable}{lcccccc}
\tabletypesize{\scriptsize}
\tablecolumns{7}
\tablewidth{0pc}
\tablecaption{Line Emission from Galactic Center Point Sources\label{tab:psint}} 
\tablehead{ 
\colhead{} & \colhead{$C < 500$} & \colhead{$C < 80$} & 
\colhead{$80 < C < 500$} & \colhead{$HR > 0.1$} & \colhead{$-0.1 < HR < 0.1$} &
\colhead{$HR < -0.1$} 
}
\startdata
$N_{\rm H}$ ($10^{22}$ cm$^{-2}$) & 1.6$_{-0.1}^{+0.1}$ & 1.6$_{-0.2}^{+0.1}$ & 2.2$_{-0.1}^{+0.2}$ & 1.4$_{-0.1}^{+0.2}$ & 2.3$_{-0.1}^{+0.3}$ & 4.1$_{-0.1}^{+0.4}$  \\
$N_{\rm pc,H}$ ($10^{22}$ cm$^{-2}$) &   7.1$_{-  0.5}^{+  0.2}$ &   7.1$_{-  0.6}^{+  0.1}$ &   7.5$_{-  0.2}^{+  0.5}$ &   8.3$_{-  0.2}^{+  0.7}$ &   8.2$_{-  0.2}^{+  0.8}$ &   8.7$_{-  0.4}^{+  0.9}$  \\
$f_{\rm pc}$ & 0.95  & 0.95  & 0.95  & 0.95  & 0.95  & 0.95   \\
$\Gamma$ & 0.86$_{-0.08}^{+0.03}$ & 0.85$_{-0.14}^{+0.01}$ & 1.04$_{-0.00}^{+0.10}$ & 0.34$_{-0.01}^{+0.12}$ & 1.51$_{-0.04}^{+0.15}$ & 3.31$_{-0.12}^{+0.25}$  \\
$N_\Gamma$ ($10^{-6}$ ph cm$^{-2}$ s$^{-1}$ arcmin$^{-2})$ &  5.0$_{- 0.7}^{+ 0.3}$ &  2.3$_{- 0.5}^{+ 0.1}$ &  3.2$_{- 0.1}^{+ 0.8}$ &  1.2$_{- 0.1}^{+ 0.3}$ &  5.7$_{- 0.4}^{+ 1.9}$ & 29.2$_{- 3.2}^{+18.1}$  \\
Si XIII He-$\alpha$ (eV) & 181 $\pm$ 29& 138 $\pm$ 38& 248 $\pm$ 57& 242 $\pm$ 65&  97 $\pm$ 41& 253 $\pm$ 123 \\
Si XIV Ly-$\alpha$ (eV) &  53 $\pm$ 14&  32 $\pm$ 20&  81 $\pm$ 26&  88 $\pm$ 34& $< 37$ &  85 $\pm$ 41 \\
Si XIII He-$\beta$ (eV) &  57 $\pm$ 15&  64 $\pm$ 24&  50 $\pm$ 24&  78 $\pm$ 35&  56 $\pm$ 27&  67 $\pm$ 42 \\
S XV He-$\alpha$ (eV) & 117 $\pm$ 19&  45 $\pm$ 20& 184 $\pm$ 33& 131 $\pm$ 35&  69 $\pm$ 26& 259 $\pm$ 128 \\
S XVI Ly-$\alpha$ (eV) &  21 $\pm$ 8& $< 18$ &  33 $\pm$ 13&  67 $\pm$ 20&  10 $\pm$ 13& $< 23$  \\
Ar XVII He-$\alpha$ (eV) &  17 $\pm$ 5& $< 14$ &  23 $\pm$ 8&  35 $\pm$ 11& $<  4$ &  34 $\pm$ 18 \\
Ca XIX He-$\alpha$ (eV) &   7 $\pm$ 4&   5 $\pm$ 5&   7 $\pm$ 5& $<  2$ &  17 $\pm$ 7&  24 $\pm$ 14 \\
Fe K-$\alpha$ (eV) & 137 $\pm$ 21&  96 $\pm$ 21& 180 $\pm$ 32& 134 $\pm$ 27& 128 $\pm$ 37& 226 $\pm$ 120 \\
Fe XXV He-$\alpha$ (eV) & 404 $\pm$ 59& 465 $\pm$ 90& 350 $\pm$ 61& 396 $\pm$ 76& 388 $\pm$ 108& 411 $\pm$ 213 \\
Fe XXVI Ly-$\alpha$ (eV) & 225 $\pm$ 34& 266 $\pm$ 53& 195 $\pm$ 36& 209 $\pm$ 41& 217 $\pm$ 62& 335 $\pm$ 180 \\
$F$ ($10^{-13}$ erg cm$^{-2}$ s$^{-1}$ arcmin$^{-2}$) &  0.45 &  0.22 &  0.21 &  0.26 &  0.15 &  0.04  \\
$uF$ ($10^{-13}$ erg cm$^{-2}$ s$^{-1}$ arcmin$^{-2}$) &  0.45 &  0.22 &  0.21 &  0.26 &  0.15 &  0.04  \\
$\chi^2/\nu$ &   490/464 &   448/464 &   523/464 &   593/464 &   526/464 &   518/436 
\enddata
\tablecomments{The line strengths are given in units of equivalent width. The 
fluxes are listed with units of arcmin$^{-2}$ to indicate the amount of 
flux that is produced by point sources per unit area on the sky.}
\end{deluxetable}

\begin{deluxetable}{lcccccc}
\tabletypesize{\scriptsize}
\tablecolumns{7}
\tablewidth{0pc}
\tablecaption{Two-$kT$ Plasma Models of Galactic Center Point Sources\label{tab:twokt}} 
\tablehead{ 
\colhead{} & \colhead{$C < 500$} & \colhead{$C < 80$} & 
\colhead{$80 < C < 500$} & \colhead{$HR > 0.1$} & \colhead{$-0.1 < HR < 0.1$} &
\colhead{$HR < -0.1$} 
}
\startdata
$N_{\rm H1}$ ($10^{22}$ cm$^{-2}$) & 4.5$_{-0.2}^{+0.4}$ & 1.4$_{-2.8}^{+3.9}$ & 4.8$_{-0.3}^{+0.5}$ & 1.2$_{-0.4}^{+2.1}$ & 1.1$_{-0.3}^{+0.5}$ & 7.6$_{-1.6}^{+0.9}$  \\
$N_{\rm pc,H1}$ ($10^{22}$ cm$^{-2}$) &   4.6$_{-  0.4}^{+  0.8}$ &   3.2$_{-  0.8}^{+  3.0}$ &   4.3$_{-  0.8}^{+  1.2}$ &   3.1$_{-  1.0}^{+  0.8}$ &   4.5$_{-  1.2}^{+  2.0}$ &  22.8$_{-  3.6}^{+  0.8}$  \\
$f_{{\rm pc},1}$ & 0.95 & 0.95 & 0.95 & 0.95 & 0.95 & 0.95  \\
$kT_1$ (keV) &  0.56$_{-0.08}^{+0.04}$ &  0.67$_{-0.08}^{+0.08}$ &  0.63$_{-0.09}^{+0.10}$ &  0.69$_{-0.11}^{+0.13}$ &  0.69$_{-0.09}^{+0.19}$ &  0.58$_{-0.09}^{+0.23}$  \\
$EM_1$ ($10^{-4}$ cm$^{-6}$ pc) &   2.0$_{-  0.6}^{+  2.0}$ &   0.0$_{-  0.8}^{+  4.6}$ &   0.7$_{-  0.4}^{+  1.2}$ &   0.0$_{-  0.0}^{+  0.2}$ &   0.0$_{-  0.0}^{+  0.0}$ &   6.7$_{-  5.1}^{+  3.8}$  \\
$F_1$ ($10^{-13}$ erg cm$^{-2}$ s$^{-1}$ arcmin$^{-2}$) &  0.03 &  0.01 &  0.02 &  0.01 &  0.01 &  0.01  \\
$uF_1$ ($10^{-13}$ erg cm$^{-2}$ s$^{-1}$ arcmin$^{-2}$) &  0.2 &  0.1 &  0.1 &  0.0 &  0.0 &  1.7  \\
$N_{\rm H2}$ ($10^{22}$ cm$^{-2}$) & 15.3$_{- 1.2}^{+ 0.6}$ &  7.5$_{- 6.4}^{+ 7.4}$ & 16.2$_{- 0.7}^{+ 1.4}$ &  8.5$_{- 0.4}^{+ 3.2}$ &  4.9$_{- 0.2}^{+ 0.8}$ &  9.7$_{- 8.4}^{+ 7.8}$  \\
$N_{\rm pc,H2}$ ($10^{22}$ cm$^{-2}$) &   117$_{-   18f}^{+   13}$ &    24$_{-   84f}^{+  101}$ &   119$_{-    5f}^{+   44}$ &    39$_{-    1f}^{+    4}$ &    10$_{-    1f}^{+    1}$ &     2$_{-    9f}^{+   16}$  \\
$f_{{\rm pc},2}$ & 0.77 & 0.66 & 0.74 & 0.81 & 0.81 & 0.95  \\
$kT_2$ (keV) &  7.8$_{-0.1}^{+0.4}$ &  9.0$_{-0.3}^{+0.3}$ &  7.8$_{-0.5}^{+0.3}$ &  8.4$_{-0.2}^{+0.2}$ &  8.6$_{-0.3}^{+0.4}$ &  8.7$_{-0.4}^{+1.3}$  \\
$EM_2$ ($10^{-4}$ cm$^{-6}$ pc) &   1.71$_{-  0.24}^{+  0.47}$ &   0.24$_{-  0.43}^{+  0.88}$ &   0.76$_{-  0.22}^{+  0.55}$ &   0.44$_{-  0.02}^{+  0.04}$ &   0.14$_{-  0.01}^{+  0.01}$ &   0.02$_{-  0.01}^{+  0.01}$  \\
$F_2$ ($10^{-13}$ erg cm$^{-2}$ s$^{-1}$ arcmin$^{-2}$) &   0.41 &   0.20 &   0.18 &   0.24 &   0.15 &   0.03  \\
$uF_2$ ($10^{-13}$ erg cm$^{-2}$ s$^{-1}$ arcmin$^{-2}$) &   3.60 &   1.71 &   2.10 &   1.01 &   0.32 &   0.06  \\
$Z_{\rm Si}/Z_{{\rm Si},\odot}$ & 0.19$_{-0.05}^{+0.06}$ & 0.82$_{-0.15}^{+0.45}$ & 0.26$_{-0.04}^{+0.09}$ & 1.64$_{-1.21}^{+0.78}$ & 0.57$_{-0.57}^{+0.58}$ & 0.46$_{-0.00}^{+0.27}$  \\
$Z_{\rm S}/Z_{{\rm S},\odot}$ & 0.30$_{-0.06}^{+0.07}$ & 0.00$_{-0.00}^{+0.38}$ & 0.47$_{-0.09}^{+0.10}$ & 1.51$_{-0.90}^{+0.87}$ & 1.14$_{-1.00}^{+0.79}$ & 0.82$_{-0.27}^{+0.52}$  \\
$Z_{\rm Fe}/Z_{{\rm Fe},\odot}$ & 0.45$_{-0.05}^{+0.05}$ & 0.97$_{-0.06}^{+0.03}$ & 0.39$_{-0.10}^{+0.05}$ & 0.67$_{-0.05}^{+0.04}$ & 0.81$_{-0.05}^{+0.05}$ & 0.72$_{-0.02}^{+0.29}$  \\
Fe K-$\alpha$ ($10^{-7}$ ph cm$^{-2}$ s$^{-1}$ arcmin$^{-1}$) &    3.8$_{-   0.8}^{+   1.1}$ &    0.8$_{-   0.3}^{+   1.4}$ &    2.4$_{-   0.4}^{+   1.3}$ &    1.4$_{-   0.1}^{+   0.1}$ &    0.6$_{-   0.1}^{+   0.1}$ &    0.1$_{-   0.0}^{+   0.0}$  \\
$\chi^2/\nu$ &  598/465 &  564/465 &  495/465 &  622/465 &  546/465 &  520/437  \\
\enddata
\tablecomments{The 
fluxes are listed with units of arcmin$^{-2}$ to indicate the amount of 
flux that is produced by point sources per unit area on the sky.
The abundances of Ar and Ca were held to their solar values.
Note that the soft component of the spectrum does not improve the fit for 
the hard sources, with $HR > 0.1$ and $-0.1 < HR < 0.1$; we have included it 
only for the sake of comparison with the other groups of sources.}
\end{deluxetable}

\begin{deluxetable}{lccccccc}
\tabletypesize{\scriptsize}
\tablecolumns{8}
\tablewidth{0pc}
\tablecaption{Sources with Short-term Variability\label{tab:shortvar}} 
\tablehead{ 
\colhead{Source} & \colhead{Loc.} & \colhead{ObsID} & \colhead{Var.} & 
\colhead{$\Delta t$} & \colhead{$F_{\rm min}$} & \colhead{$F_{\rm max}$} & 
\colhead{$F_{\rm max}/F_{\rm min}$} \\
\colhead{} & \colhead{} & \colhead{} & \colhead{Type} &
\colhead{(ks)} & \colhead{($10^{-7}$ \phcms)} & \colhead{($10^{-7}$ \phcms)} &
\colhead{}
}
\startdata
174517.4--290650 & gc & 3392 & step & \nodata & $ 28^{+  9}_{-  7}$ & $ 105^{+ 14}_{- 12}$ & $   3$ \\
174520.3--290143 & gc & 2943 & step & \nodata & $  3^{+  5}_{-  3}$ & $  85^{+ 53}_{- 39}$ & $  25$ \\
174520.6--290152 & f & 3392 & flare &  91.0 & $ 28^{+  9}_{-  7}$ & $1160^{+329}_{-256}$ & $  40$ \\
  &   & 3393 & flare &  34.1 & $ 16^{+  5}_{-  4}$ & $ 819^{+ 84}_{- 78}$ & $  51$ \\
174521.8--285912 & f & 3392 & flare &  24.2 & $  4^{+  3}_{-  3}$ & $ 189^{+ 63}_{- 56}$ & $  44$ \\ [5pt]
174525.1--285703 & f & 3665 & step & \nodata & $ 10^{+  5}_{-  4}$ & $ 288^{+137}_{-110}$ & $  28$ \\
174530.3--290341 & gc & 3392 & flare &  22.0 & $  6^{+  9}_{-  5}$ & $  86^{+ 22}_{- 18}$ & $  14$ \\
174531.0--285605 & gc & 3393 & step & \nodata & $  2^{+  2}_{-  2}$ & $  40^{+ 11}_{-  9}$ & $  18$ \\
174533.4--285328 & f & 0242 & flare &   8.2 & $ 26^{+ 11}_{-  9}$ & $ 217^{+ 67}_{- 51}$ & $   8$ \\
174534.5--290236 & gc & 3392 & step & \nodata & $<  1$ & $  10^{+  4}_{-  4}$ & $>   5$ \\ [5pt]
174535.6--290133 & gc & 3665 & flare &  61.2 & $ 14^{+  8}_{-  6}$ & $ 162^{+ 21}_{- 20}$ & $  11$ \\
174535.8--290159 & gc & 3393 & step & \nodata & $<  3$ & $  12^{+  4}_{-  3}$ & $>   2$ \\
174535.9--290806 & gc & 3393 & step & \nodata & $  4^{+  5}_{-  4}$ & $  30^{+  9}_{-  8}$ & $   6$ \\
174536.3--285545 & f & 3392 & flare &   3.8 & $  3^{+  2}_{-  2}$ & $ 182^{+ 77}_{- 64}$ & $  63$ \\
174538.2--285602 & f & 3665 & flare &  28.4 & $ 13^{+ 15}_{-  9}$ & $ 116^{+ 21}_{- 19}$ & $   9 $ \\ [5pt]
174538.3--290048 & gc & 3392 & step & \nodata & $ 26^{+  6}_{-  5}$ & $  56^{+  9}_{-  8}$ & $   2$ \\
174540.1--290804 & f & 3392 & flare &  87.2 & $<  3$ & $ 276^{+ 36}_{- 32}$ & $>  76$ \\
  &   & 3393 & step & \nodata & $ 24^{+ 10}_{-  8}$ & $  62^{+  8}_{-  8}$ & $   2$ \\
174540.4--285831 & gc & 3392 & step & \nodata & $  5^{+  3}_{-  2}$ & $  23^{+  7}_{-  6}$ & $   4$ \\
174541.4--290348 & gc & 3392 & step & \nodata & $<  2$ & $  20^{+  5}_{-  4}$ & $>   8$ \\ [5pt]
174541.5--290752 & f & 2943 & flare &   7.6 & $< 11$ & $ 124^{+ 45}_{- 40}$ & $>   7$ \\
174541.8--290319\tablenotemark{a} & gc & 1561a & \nodata & \nodata & \nodata & \nodata & \nodata \\
174542.8--285352 & gc & 0242 & step & \nodata & $<  3$ & $  36^{+ 18}_{- 15}$ & $>   6$ \\
174542.9--285522 & f & 3665 & flare &   3.3 & $<  4$ & $ 190^{+ 95}_{- 75}$ & $>  30$ \\
174543.4--290347\tablenotemark{a} & gc & 1561a & \nodata & \nodata & \nodata & \nodata & \nodata \\ [5pt]
174543.9--290456 & f & 3392 & flare &  11.1 & $ 81^{+  8}_{-  7}$ & $1397^{+466}_{-370}$ & $  17$ \\
  &   & 3663 & flare &   6.1 & $113^{+ 21}_{- 19}$ & $ 433^{+ 87}_{- 77}$ & $   3$ \\
174545.0--290336 & gc & 3663 & step & \nodata & $<  8$ & $  37^{+ 13}_{- 12}$ & $>   3$ \\
174546.8--290252\tablenotemark{a} & gc & 1561b & \nodata & \nodata & \nodata & \nodata & \nodata \\
174547.4--290817 & f & 3393 & step & \nodata & $  7^{+  4}_{-  4}$ & $  54^{+ 18}_{- 16}$ & $   8$ \\ [5pt]
174548.0--290352 & f & 3665 & flare &  11.0 & $  4^{+  5}_{-  3}$ & $  84^{+ 29}_{- 25}$ & $  22 $ \\
174548.4--290234 & gc & 2943 & flare &   9.9 & $< 22$ & $  84^{+ 30}_{- 26}$ & $>   2$ \\
174548.4--290832 & gc & 3393 & step & \nodata & $<  3$ & $  21^{+ 10}_{-  9}$ & $>   4$ \\
174548.6--290522 & f & 3392 & flare &  35.3 & $ 12^{+  5}_{-  4}$ & $ 226^{+ 53}_{- 44}$ & $  18 $ \\
174550.7--290434 & f & 3393 & flare &  15.0 & $<  4$ & $  46^{+ 21}_{- 17}$ & $>   7$ \\ [5pt]
174552.1--290422 & gc & 3393 & step & \nodata & $  5^{+  3}_{-  3}$ & $  26^{+  9}_{-  8}$ & $   5$ \\
174552.2--290744 & gc & 3392 & flare &   0.1 & $ 99^{+ 14}_{- 13}$ & $2123$\tablenotemark{b} & $  21$ \\
174552.9--290358 & f & 0242 & flare &   0.7 & $ 34^{+ 15}_{- 13}$ & $ 652^{+330}_{-264}$ & $  19$ \\
  &   & 3393 & flare &   7.3 & $ 20^{+ 13}_{- 10}$ & $ 404^{+116}_{-105}$ & $  20$ \\
174556.9--285819\tablenotemark{a} & f & 2952 & \nodata & \nodata & \nodata & \nodata & \nodata \\ [5pt]
174558.5--290451 & f & 3392 & flare &  12.6 & $ 14^{+  5}_{-  4}$ & $ 253^{+ 50}_{- 46}$ & $  18$ \\
174559.0--290418 & f & 3392 & step & \nodata & $<  3$ & $  24^{+  6}_{-  6}$ & $>   5$ \\
174605.2--290700 & gc & 3393 & flare &  24.8 & $<  4$ & $ 269^{+ 98}_{- 86}$ & $>  47$ \\
174612.4--290234 & f & 3393 & step & \nodata & $<  2$ & $  17^{+  8}_{-  7}$ & $>   3$ \\
\enddata
\tablenotetext{a}{The variability from these sources was identified via the 
KS-test, but not by the Bayesian Blocks algorithm.}
\tablenotetext{b}{This flare consisted of 5 photons received within 100 s.}
\end{deluxetable}

\begin{deluxetable}{lccccccc}
\tabletypesize{\scriptsize}
\tablecolumns{8}
\tablewidth{0pc}
\tablecaption{Sources with Long-term Variability\label{tab:longvar}} 
\tablehead{ 
\colhead{Source} & \colhead{Loc.} & \colhead{ObsID}
& \colhead{$F_{\rm min}$} & \colhead{ObsID} & \colhead{$F_{\rm max}$} & 
\colhead{$F_{\rm max}/F_{\rm min}$} \\
\colhead{} & \colhead{} & \colhead{of Min.} & \colhead{($10^{-7}$ \phcms)} &
\colhead{of Max.} & \colhead{($10^{-7}$ \phcms)} & \colhead{} &
\colhead{}
}
\startdata
174503.9--290051 & gc & 3392 & $ 44^{+  6}_{-  6}$ & 0242 & $  88^{+ 24}_{- 20}$ & $   2$ \\
174507.0--290356 & gc & 1561b & $<  9$ & 2952 & $  41^{+ 25}_{- 19}$ & $>   2$ \\
174514.1--285426 & gc & 2953 & $< 13$ & 3392 & $  30^{+  6}_{-  5}$ & $>   1$ \\
174517.5--285646 & gc & 3663 & $<  3$ & 3665 & $  14^{+  5}_{-  4}$ & $>   2$ \\
174519.8--290114 & gc & 3663 & $ 12^{+  7}_{-  6}$ & 0242 & $  35^{+ 10}_{-  9}$ & $   3$ \\ [5pt]
174520.5--285927 & gc & 3665 & $<  2$ & 2953 & $  22^{+ 29}_{- 16}$ & $>   2$ \\
174520.6--285712 & gc & 0242 & $<  8$ & 2954 & $ 150^{+ 37}_{- 32}$ & $>  15$ \\
174520.8--285304 & f & 3665 & $<  3$ & 3393 & $  15^{+  4}_{-  4}$ & $>   3$ \\
174521.7--285812 & gc & 3665 & $<  2$ & 2952 & $  13^{+ 19}_{- 11}$ & $>   1$ \\
174521.9--290616 & gc & 1561b & $< 19$ & 2953 & $  68^{+ 30}_{- 24}$ & $>   2$ \\ [5pt]
174522.4--285707 & gc & 3665 & $<  2$ & 2953 & $  21^{+ 22}_{- 13}$ & $>   3$ \\
174523.1--290205 & gc & 1561a & $<  3$ & 2954 & $  11^{+ 13}_{-  8}$ & $>   1$ \\
174526.4--290148 & gc & 3392 & $<  1$ & 2951 & $   8^{+ 12}_{-  7}$ & $>   1$ \\
174526.7--290220 & gc & 3665 & $<  2$ & 2952 & $  18^{+ 15}_{- 10}$ & $>   4$ \\
174527.4--285938 & gc & 0242 & $<  4$ & 2943 & $  28^{+ 10}_{-  8}$ & $>   4$ \\ [5pt]
174529.0--290406 & gc & 1561b & $<  8$ & 3663 & $  25^{+  9}_{-  7}$ & $>   2$ \\
174529.6--285432 & gc & 2952 & $<  9$ & 3393 & $  40^{+  6}_{-  5}$ & $>   3$ \\
174530.5--290323 & gc & 3392 & $<  1$ & 3663 & $   8^{+  6}_{-  4}$ & $>   5$ \\
174531.3--285949 & gc & 3393 & $<  1$ & 2953 & $  25^{+ 26}_{- 16}$ & $>   6$ \\
174531.8--290000 & gc & 3392 & $<  1$ & 2951 & $  11^{+ 28}_{- 11}$ & $>   0$ \\ [5pt]
174532.7--290552 & f & 0242 & $<  3$ & 2954 & $ 108^{+ 33}_{- 27}$ & $>  28$ \\
174532.9--285823 & gc & 3393 & $  4^{+  2}_{-  2}$ & 2952 & $  20^{+ 16}_{- 11}$ & $   5$ \\
174533.0--285355 & gc & 0242 & $<  5$ & 3393 & $  38^{+  6}_{-  5}$ & $>   6$ \\
174534.2--290119 & gc & 3393 & $<  3$ & 2951 & $  74^{+ 26}_{- 21}$ & $>  20$ \\
174535.5--290124 & gc & 0242 & $<  3$ & 2951 & $ 255^{+ 45}_{- 40}$ & $>  80$ \\ [5pt]
174535.9--285825 & gc & 3392 & $<  1$ & 1561b & $   9^{+ 16}_{-  7}$ & $>   1$ \\
174536.1--285638 & f & 2952 & $ 65^{+ 26}_{- 21}$ & 1561b & $ 199^{+ 39}_{- 35}$ & $   3$ \\
174536.6--290109 & gc & 3393 & $<  2$ & 1561a & $  13^{+ 13}_{-  8}$ & $>   2$ \\
174537.2--285459 & f & 1561a & $<  3$ & 2951 & $  59^{+ 27}_{- 21}$ & $>  11$ \\
174537.5--290125 & gc & 3665 & $<  2$ & 2951 & $  24^{+ 16}_{- 11}$ & $>   5$ \\ [5pt]
174538.0--290022 & gc & 3665 & $ 39^{+  7}_{-  6}$ & 0242 & $ 290^{+ 31}_{- 29}$ & $   7 $ \\
174538.4--290044 & gc & 1561a & $<  2$ & 2954 & $ 108^{+ 30}_{- 26}$ & $>  33$ \\
174538.7--290134 & gc & 1561a & $<  5$ & 3392 & $  19^{+  4}_{-  3}$ & $>   2$ \\
174539.1--290112 & gc & 1561a & $<  4$ & 3392 & $  10^{+  3}_{-  3}$ & $>   2$ \\
174539.5--285454 & f & 3392 & $  3^{+  2}_{-  2}$ & 2951 & $  25^{+ 18}_{- 13}$ & $   8 $ \\ [5pt]
174540.1--290055 & gc & 2954 & $ 16^{+ 17}_{- 12}$ & 3663 & $  86^{+ 16}_{- 14}$ & $   5$ \\
174540.6--290001 & gc & 0242 & $<  9$ & 1561b & $ 134^{+ 33}_{- 28}$ & $>  11$ \\
174540.8--290040 & gc & 0242 & $<  3$ & 2953 & $  10^{+ 15}_{- 10}$ & $>   0$ \\
174541.0--290014 & gc & 1561a & $<  5$ & 3392 & $ 204^{+  0}_{-  0}$ & $>  37$ \\
174541.5--285148 & f & 3393 & $<  5$ & 2953 & $  40^{+ 32}_{- 24}$ & $>   3$ \\ [5pt]
174541.5--285814 & f & 2951 & $< 14$ & 3663 & $ 170^{+ 21}_{- 19}$ & $>  11$ \\
174541.7--285555 & gc & 3663 & $<  4$ & 3392 & $  24^{+  4}_{-  4}$ & $>   4$ \\
174542.2--285732 & gc & 3665 & $  2^{+  2}_{-  2}$ & 3392 & $  57^{+  6}_{-  6}$ & $  25 $ \\
174542.2--290132 & gc & 3393 & $<  1$ & 2951 & $  14^{+ 15}_{-  9}$ & $>   4$ \\
174542.5--285722 & gc & 3665 & $<  1$ & 1561b & $  10^{+ 11}_{-  7}$ & $>   2$ \\
174543.4--285742 & gc & 0242 & $<  2$ & 2953 & $  37^{+ 30}_{- 20}$ & $>   7$ \\ [5pt]
174543.4--285900 & gc & 0242 & $<  3$ & 2951 & $  80^{+ 43}_{- 32}$ & $>  15$ \\
174543.6--285629 & gc & 3665 & $  2^{+  3}_{-  2}$ & 2952 & $  23^{+ 45}_{- 20}$ & $   9 $ \\
174543.9--290245 & gc & 0242 & $<  3$ & 2952 & $  16^{+ 15}_{- 10}$ & $>   2$ \\
174544.2--290644 & gc & 3665 & $<  4$ & 3392 & $  13^{+  4}_{-  3}$ & $>   2$ \\
174545.2--285828 & f & 1561b & $ 68^{+ 26}_{- 21}$ & 2951 & $ 208^{+ 43}_{- 38}$ & $   3$ \\ [5pt]
174546.1--285831 & gc & 3665 & $<  5$ & 1561b & $  39^{+ 21}_{- 16}$ & $>   4$ \\
174546.6--290356 & gc & 3393 & $<  1$ & 2953 & $  11^{+ 13}_{-  8}$ & $>   3$ \\
174547.8--290145 & gc & 1561a & $<  5$ & 3663 & $  21^{+  8}_{-  7}$ & $>   2$ \\
174549.5--285815 & gc & 3392 & $<  2$ & 1561a & $  11^{+  7}_{-  5}$ & $>   2$ \\
174550.5--285239 & f & 1561a & $ 23^{+ 26}_{- 19}$ & 1561b & $ 112^{+ 40}_{- 34}$ & $   4 $ \\ [5pt]
174550.9--285430 & gc & 3393 & $ 29^{+  7}_{-  6}$ & 2951 & $  88^{+ 33}_{- 27}$ & $   3$ \\
174552.0--290324 & gc & 0242 & $<  6$ & 2953 & $  22^{+ 17}_{- 12}$ & $>   1$ \\
174552.5--285759 & gc & 3393 & $<  1$ & 2943 & $  10^{+  7}_{-  5}$ & $>   4$ \\
174553.3--290444 & f & 2943 & $  5^{+  6}_{-  4}$ & 3393 & $  46^{+  6}_{-  5}$ & $   8 $ \\
174553.3--290632 & gc & 3665 & $  4^{+  4}_{-  3}$ & 2954 & $  38^{+ 21}_{- 16}$ & $  10$ \\
174554.2--285729 & gc & 3665 & $  2^{+  3}_{-  2}$ & 3392 & $  10^{+  3}_{-  3}$ & $   4$ \\ [5pt]
\newtablebreak
174558.4--290120 & f & 3665 & $<  3$ & 1561a & $  36^{+ 18}_{- 14}$ & $>   7$ \\
174558.9--290724 & gc & 3665 & $ 65^{+ 10}_{-  9}$ & 0242 & $ 138^{+ 22}_{- 20}$ & $   2$ \\
174601.0--285854 & gc & 2952 & $  9^{+ 13}_{-  8}$ & 1561b & $  46^{+ 21}_{- 17}$ & $   4$ \\
174601.4--285416 & f & 3392 & $<  5$ & 2952 & $  77^{+ 34}_{- 28}$ & $>   9$ \\
174603.7--290247 & f & 1561a & $<  7$ & 3393 & $  30^{+  5}_{-  5}$ & $>   3$ \\ [5pt]
174606.2--290941 & gc & 3665 & $< 11$ & 3392 & $  44^{+  7}_{-  7}$ & $>   3$ \\
174607.5--285951 & f & 2952 & $ 34^{+ 21}_{- 16}$ & 1561a & $ 209^{+ 26}_{- 24}$ & $   6 $ \\
174610.8--290019 & gc & 3665 & $<  4$ & 1561a & $  60^{+ 15}_{- 13}$ & $>  10$ \\
174612.3--285706 & gc & 3665 & $<  5$ & 0242 & $  78^{+ 17}_{- 15}$ & $>  13$ \\
174613.9--285924 & gc & 3393 & $<  4$ & 2951 & $  47^{+ 25}_{- 20}$ & $>   7$ \\ [5pt]
174614.0--290220 & f & 3665 & $<  5$ & 0242 & $  74^{+ 16}_{- 14}$ & $>  11$ \\
174615.9--290257 & f & 2951 & $< 11$ & 2943 & $  35^{+ 13}_{- 11}$ & $>   2$ \\
174624.4--285712 & f & 3393 & $ 25^{+  8}_{-  7}$ & 2954 & $  88^{+ 72}_{- 52}$ & $   3$ \\
\enddata
\end{deluxetable}

\begin{deluxetable}{lccccccccc}
\tabletypesize{\scriptsize}
\tablecolumns{10}
\tablewidth{0pc}
\tablecaption{Hard X-ray Sources with Unusual Properties \label{tab:odd}}
\tablehead{
\colhead{} & \colhead{$F_{\rm X}$\tablenotemark{a}} & 
\multicolumn{3}{c}{Continuum} & 
\multicolumn{2}{c}{Iron Emission} &
\multicolumn{2}{c}{Periodicity}  & \colhead{} \\
\colhead{Object} & \colhead{($10^{-12}$ } &
\colhead{$N_{\rm H}$} & \colhead{$\Gamma$} & \colhead{$kT$} & 
\colhead{$E_{\rm Fe}$} & \colhead{$EW_{\rm Fe}$} &
\colhead{$P$} & \colhead{$A$}  & \colhead{Ref.} \\
\colhead{} & \colhead{erg cm$^{-2}$ s$^{-1}$)} & 
\colhead{($10^{22}$ cm$^{-2}$)} & \colhead{} & \colhead{(keV)} & 
\colhead{(keV)} & \colhead{(eV)} &
\colhead{(s)} & \colhead{(\%)} & \colhead{} }
\startdata
AX J2315--0592 & 50 & 0.07 & \nodata & 17 & 6.8 & 900 & 5360 & 90\tablenotemark{b} & [1] \\
RX J1802.1$+$1804 & 0.5 & 13\tablenotemark{c} & \nodata & $>7$ & 6.7 & 4000 & 6840 & 100\tablenotemark{b} & [2] \\
AX J1842.8--0423 & 4 & 5 & 2.9 & 5.1 & 6.7 & 4000 & \nodata & \nodata & [3] \\ 
XMM J174457--2850.3 & 6.5 & 6 & 0.98 & \nodata & 6.7 & 180 & \nodata & \nodata & [4] \\ [5pt]
IGR J16358--4726 & 70 & 33 & 0.5 & \nodata & 6.4 & 130 & 5580 & 37 & [5,6] \\
IGR J16318--4848 & 8 & 196 & 1.6 & \nodata & 6.4 & 2000 & \nodata & \nodata & [7,8] \\
IGR J16320--4751 & 400 & 21 & 2.5 & \nodata & \nodata & \nodata & \nodata & \nodata & [9,10] \\ 
XMM J174544--2913.0 & 6.5 & 12 & \nodata & \nodata & 6.7 & 2000 & \nodata & \nodata & [4] \\ [5pt]
AX J1820.5--1434 & 23 & 9.8 & 0.9 & \nodata & 6.4 & 90 & 152 & 57 & [11] \\
AX J170006--4157 & 5 & 6 & 0.2 & \nodata & \nodata & $<$1200 & 715 & 50 & [12] \\
AX J1740.1--2847 & 4 & 2.5 & 0.7 & \nodata & \nodata & $< 500$ & 729 & 100 & [13] \\
1SAX J1452.8--5949 & 0.6 & 1.9 & 1.4 & \nodata & 6.4 & $> 1300$ & 437 & 74 & [14] \\
AX J183220--0840 & 11 & 1.3 & 0.8 & \nodata & 6.7 & 450 & 1549 & 63 & [15] \\
\enddata
\tablenotetext{a}{Observed flux, 2--10 keV.}
\tablenotetext{b}{Modulation is only present below 2~keV.}
\tablenotetext{c}{Partial-covering, intrinsic absorption.}
\tablerefs{[1] \citet{mis96} ; [2] \citet{ish98}; [3] \citet{ter99} ; 
[4] \citet{sak04} ; [5] \citet{pat03} ; [6] \citet{rev03} ; [7] \citet{mg03} ; 
[8] \citet{wal03} ; [9] \citet{rod03} ; [10] \citet{zan03} ; 
[11] \citet{kin98} ;
[12] \citet{tor99} ; [13] \citet{sak00} ; [14] \citet{oos99} ; 
[15] \citet{sug00}}
\end{deluxetable}


\begin{thebibliography}{0}
\bibitem[Apparao \etal(1994)]{app94} Apparao, K. M. V. 1994, {\it SSRev}, 
  69, 255
\bibitem[Arabadjis \etal(2003)]{ara03} Arabadjis, J. S., Bautz, M. W., \&
  Arabadjis, G. 2003, submitted to \apj, astro-ph/0305547
\bibitem[Asai \etal(1998)]{asa98} Asai, K., Dotani, T., Hoshi, R., 
  Tanaka, Y., Robinson, C. R., \& Terada, K. 1998, \pasj, 50, 611
\bibitem[Augello \etal(2003)]{aug03} Augello, G., Iaria, R., Robba, N. R., 
  Di Salvo, T., Burderi, L., Lavagetto, G., \& Stella 2003, \apj, 596, L63
\bibitem[Baganoff \etal(2003)]{bag03} Baganoff, F. K. \etal\ 2003, 
  \apj, 591, 891
\bibitem[Becker \etal(2003)]{bec03} Becker, Werner, Swartz, D. A., Pavlov, 
  G. G., Elsner, R. F., Grindlay, J., Mignani, R., Tennant, A. F., 
  Backer, D., Pulone, L., Testa, V., \& Weisskopf, M. C. 2003, \apj, 594, 798
\bibitem[Belczynski \& Taam(2004)]{bt04} Belczynski, K. \& Taam, R. E. 2004,
  submitted to \apj, astro-ph/0307492
\bibitem[Binney \& Tremaine(1994)]{bt94} Binney, J. \& Tremaine, S. 1994, 
  {\em Galactic Dynamics}, Princeton University Press
\bibitem[Bleach(2002)]{ble02} Bleach, J. N. 2002, \mnras, 332, 689
\bibitem[Bykov(2002)]{byk02} Bykov, A. M. 2002, \aap, 390, 327
\bibitem[Bykov(2003)]{byk03} Bykov, A. M. 2003, \aap, 410, L5
\bibitem[Campana \etal(2001)]{cam01} Campana, S., Gastaldello, F., Stella, L.,
  Israel, G. L., Colpi, M., Pizzolato, F., Orlandini, M., \& Dal Fiume, D.
  2001, \apj, 561, 924
\bibitem[Campana \etal(2002a)]{cam02} Campana, S., Stella, L., Gastaldello, F.,
  Mereghetti, S., Colpi, M., Israel, G. L., Burderi, L., Di Salvo, T., \&
  Robba, R. N. 2002, \apj, 575, L15
\bibitem[Campana \etal(2002b)]{cam02b} {Campana}, S., {Stella}, L., {Israel}, 
  G.~L., {Moretti}, A., {Parmar}, A.~N., \& {Orlandini}, M. 2002b, 
  \apj, 580, 389
\bibitem[Cheng \etal(2004)]{cheng04} Cheng, K. S., Taam, R. E., Wang, W., \&
  Belczynski, K. 2004, submitted to \apj.
\bibitem[Chlebowski \& Garmany(1991)]{cg91} 
  Chlebowski, T. \& Garmany, C. D. 1991, \apj, 368, 241
\bibitem[Choi, Dotani, \& Argawal(1999)]{cda99} 
  Choi, C.-S., Dotani, T., \& Agrawal, P. C. 1999, \apj, 525, 399
\bibitem[Cotera \etal(1999)]{cot99} Cotera, A. S., Simpson, J. P., 
  Erickson. E. F., Colgan, S. W. J., Burton, M. G., \& Allen, D. A. 1999, 
  \apj, 510, 747
\bibitem[Dempsey \etal(1993a)]{dem93a} Dempsey, R. C., Linsky, J. L., Fleming,
  T. A., \& Schmitt, J. H. M. M. 1993a, \apjs, 86, 599
\bibitem[Dempsey \etal(1993b)]{dem93b} Dempsey, R. C., Linsky, J. L., Schmitt,
  J. H. M. M., \& Fleming, T. A. 1993b, \apj, 413, 333
\bibitem[Done \& Osborne(1997)]{do97} Done, C. \&  Osbrone, J. P. 1997, 
  \mnras, 288, 649
\bibitem[Eckart \etal(2004)]{eck04} Eckardt, A. \etal\ 2004, astro-ph/0403577
\bibitem[Eracleous, Halpern, \& Patterson(1991)]{ehp91} Eracleous, M., 
  Halpern, J., \& Patterson, J. 1991, \apj, 382, 290
\bibitem[Eyles \etal(1975)]{eyl75} Eyles, C. J., Skinner, G. K., Willmore, A.
  P., \& Rosenberg, F. D. 1975, \nat, 257, 291
\bibitem[{Ezuka} \& {Ishida}(1999)]{ei99} {Ezuka}, H. \& {Ishida}, M. 1999,
  \apjs, 120, 277
\bibitem[Favata \etal(1995)]{fms95} Favata, F., Micela, G., \& Sciortino, S.
  1995, \aap, 298, 482
\bibitem[Feigelson(2004)]{fei04} Feigelson, E. D. 2004, in Stars As Suns:
  Activity, Evolution, and Planets, A. Benz \& A. Dupree (eds.) IAU Symposium
  219, in press
\bibitem[Feigelson \etal(2002)]{fei02} Feigelson, E. D., Broos, P., Gaffney, 
  J. A. III, Garmire, G., Hillenbrand, L. A., Pravdo, S. H., Townsley, 
  L., \& Tsuboi, Y. 2002, \apj, 574, 258
\bibitem[Figer(1995)]{fig95} Figer, D. F. 1995, Ph.D. thesis, University of 
  California, Los Angeles
\bibitem[Figer \etal(1999)]{fig99} Figer, D. F., Kim, S. S., Morris, M., 
  Serabyn, E., Rich, R. M., \& McLean, I. S. 1999, \apj, 525, 750
\bibitem[Figer \etal(2004)]{fig03} Figer, D. F., Rich, R. M., Kim, S. S., 
  Morris, M., \& Serabyn, E. 2004, \apj, 601, 319
\bibitem[Franciosini, Pallavicini, \& Tagliaferri(2001)]{fpt01} 
  Franciosini, E., Pallavicini, R., \& Tagliaferri, G. 2001, \aap, 375, 196
\bibitem[Franciosini, Pallavicini, \& Tagliaferri(2003)]{fpt03} 
  Franciosini, E., Pallavicini, R., \& Tagliaferri, G. 2003, \aap, 399, 279
\bibitem[Freeman \etal(2002)]{fre02} Freeman, P. E., Kashyap, V., Rosner, R., 
  \& Lamb, D. Q. 2002, \apjs, 138, 185
\bibitem[Fujimoto \& Ishida(1997)]{fi97} Fujimoto, R., \& Ishida, M. 1997, 
  \apj, 474, 774
\bibitem[Garnavich \& Szkody(1988)]{gs88} Garnavich, P. \& Szkody, P. 1988, 
  \pasp, 100, 1522
\bibitem[Grindlay \etal(2001)]{gri01} Grindlay, J. E., Heinke, C., Edmonds, 
  P. D., \& Murray, S. S. 2001, {\it Science}, 292, 2290
\bibitem[Grosso \etal(2004)]{gro04} Grosso, N., Montmerle, T., Feigelson, 
 E. D., \& Forbes, T. G. 2004, submitted to \aap, astro-ph/0402672
\bibitem[Heinke \etal(2003a)]{hei03} Heinke, C. O., Edmonds, P. D., Grindlay, 
  J. E., Lloyd, D. A., Cohn, H. N., \& Lugger, P. M. 2003a, \apj, 590, 809
\bibitem[Heinke \etal(2003b)]{hei03b} Heinke, C. O., Edmonds, P. D., Grindlay, 
  J. E., Lloyd, D. A., Murray, S. S., Cohn, H. N., \& Lugger, P. M. 
  2003b, \apj, 598, 516
\bibitem[Ho \etal(1985)]{ho85} Ho, P. T. P., Jackson, J. M., Barret, A. H., \&
  Armstrong, J. T. 1985, \apj, 288, 575
\bibitem[Hoard \etal(2002)]{hoa02} Hoard, D. W., Wachter, S., Clark, L. L, 
  \& Bowers, T. P. 2002, \apj, 565, 511
\bibitem[in 't Zand \etal(2003)]{zan03} in 't Zand, J. J. M., Ubertini, P., 
  Capitanio,  F., \& Del Santo, M. 2003, IAUC 8077
\bibitem[Ishida \etal(1997)]{ish97} Ishida, M., Matsuzaki, K., Fujimoto, R., Mukai, K., \& Osborne, J. P. 1997, \mnras, 287, 651
\bibitem[Ishida \etal(1998)]{ish98} Ishida, M., Greiner, J., Remillard, R. A.,
  \& Motch, C. 1998, \aap, 336, 200
\bibitem[Iben, Tutukov, \& Fedoroval(1997)]{itf97} Iben, Icko, Jr., 
  Tutukov, A. V., \& Fedoroval, A. V. 1997, \apj, 486, 955
\bibitem[{Kinugasa} et al.(1998)]{kin98} {Kinugasa}, K. \etal\ 1998, \apj, 
  495, 435
\bibitem[Kohno, Koyama, \& Hamaguchi(2002)]{koh02} 
  Kohno, M., Koyama, K., \& Hamaguchi, K. 2002, \apj, 567, 423
\bibitem[Kong \etal(2002a)]{kon02} Kong, A. K. H., McClintock, J. E.,
  Garcia, M. R., Murray, S. S., \& Barret, D. 2002a, \apj, 570, 277
\bibitem[Krabbe et al.(1995)]{kra95} Krabbe, A. \etal\ 1995, \apj, 447, L95
\bibitem[Krishnamurthi \etal(2001)]{kri01} Krishnamurthi, A., Reynolds, C. S., 
  Linsky, J. L., Mart\'{\i}n, E., \& Gagn\'{e}, M. 2001, \aj, 121, 337
\bibitem[Kube \etal(2003)]{kub03} Kube, J., G\"{a}nsicke, B. T., Euchner, 
  F., \& Hoffmann, B. 2003, \aap, 404, 1159
\bibitem[Launhardt \etal(2002)]{lzm02} Launhardt, R., Zylka, R., \&
  Mezger, P. G. 2002, \aap, 384, 112
\bibitem[Maeda \etal(2002)]{mae02} Maeda, Y. \etal\ 2002, \apj, 570, 671
\bibitem[Maeda \etal(1996)]{mae96} Maeda, Y., Koyama, K., Sakano, M., 
  Takeshima, T., \& Yamauchi, S. 1996, \pasj, 48, 417
\bibitem[Matt \& Guainazzi(2003)]{mg03} Matt, G. \& Guainazzi, M. 2003, 
  \mnras, 341, L13
\bibitem[McNamara \etal(2000)]{mcn00} McNamara, D. H., Madsen, J. B., 
  Barnes, J., \& Ericksen, B. F. 2000, \pasp, 112, 202
\bibitem[Mewe \etal(1985)]{mgo85} Mewe, R., Gronenschild, E. H. B. M., \& 
  van den Oord, G. H. J. 1985, \aaps, 62, 197
\bibitem[Mewe, Lemen, \& van den Oord(1986)]{mew86} Mewe, R., Lemen, J. R., \&
  van den Oord, G. H. J. 1986, \aaps, 65, 511
\bibitem[Misaki \etal(1996)]{mis96} Misaki, K., Terashima, Y., Kamata, Y., 
  Ishida, M., Kunieda, H., \& Tawara, Y. 1996, \apj, 470, L53
\bibitem[Moon, Eikenberry, \& Wasserman(2003)]{mew03} Moon, D.-S., 
  Eikenberry, S. S., \& Wasserman, I. M. 2003, \apj, 582, L91
\bibitem[Morris(1993)]{mor93} Morris, M. 1993, \apj, 408, 496
\bibitem[Morris \etal(2003)]{mor03} Morris, M. Baganoff, F., Muno, M., Howard, 
  C., Maeda, Y., Feigelson, E., Bautz, M., Brandt, W. N., Chartas, G., 
  Garmire, G., \& Townsley, L. 2003, Astronomische Nachrichten, 324, S1, 167 
\bibitem[Mukai \etal(2003)]{muk03} Mukai, K., Kinkhabwala, A., Peterson, J. R.,
  Kahn, S. M., \& Pearels, F. 2003, \apj, 586, L77
\bibitem[Mukai \& Shiokawa(1993)]{ms93} Mukai, K. \& Shiokawa, K. 1993, 
  \apj, 418, 863
\bibitem[Muno \etal(2003a)]{mun03} Muno, M. P., Baganoff, F. K., Bautz, M. W., 
  Brandt, W. N., Broos, P. S., Feigelson, E. D., Garmire, G. P., Morris, M.,
  Ricker, G. R., \& Townsely, L. K.  2003a, \apj, 589, 225
\bibitem[Muno \etal(2003b)]{mun03b} Muno, M. P., Baganoff, F. K., \& 
  Arabadjis, J. A. 2003b, \apj, 598, 474
\bibitem[Muno \etal(2003c)]{mun03c} Muno, M. P., Baganoff, F. K., 
  Bautz, M. W., 
  Brandt, W. N., Garmire, G. P., \& Ricker, G. R.  2003c, \apj, 599, 465
\bibitem[Muno \etal(2004)]{mun04} Muno, M. P., 
  Baganoff, F. K., Bautz, M. W., Feigelson, E. D., Garmire, G. P., 
  Morris, M. R., Park, S. , Ricker, G. R., \& Townsley, L. K. 2004, submitted
  to \apj, astro-ph/0402087
\bibitem[Nagase \etal(1994)]{nag94} Nagase, F., Zylstra, G., Sonobe, T., 
  Kotani, T., Inoue, H., \& Woo, J. 1994, \apj, 436, L1
\bibitem[Negueruela \etal(2000)]{neg00} Negueruela, I., Reig, P., Finger, 
  M. H., \& Roche, P. 2000, \aap, 356, 1003
\bibitem[Norton \& Watson(1989)]{nw89} Norton, A. J. \& Watson, M. G. 1989,
  \mnras, 237, 853
\bibitem[{Oosterbroek} et al.(1999)]{oos99} {Oosterbroek}, T., {Orlandini}, 
  M., {Parmar}, A.~N., {Angelini}, L., {Israel},
  G.~L., {Dal Fiume}, D., {Mereghetti}, S., {Santangelo}, A., \& {Cusumano}, G.
  1999, \aap, 351, L33
\bibitem[Orlandini \etal(2003)]{orl03} Orlandini, M. \etal\ 2003, 
  astro-ph/0309819
\bibitem[Park \etal(2004)]{par04} Park, S., Baganoff, F. K., Morris, M., 
  Maeda, Y., Muno, M. P., Howard, C., Bautz, M. W., \& Garmire, G. P. 2003,
  \apj\ in press, astro-ph/0311460
\bibitem[Patel \etal(2004)]{pat03} Patel, S. K. \etal\ 2004, \apj, 602, L45
\bibitem[Patterson \& Szkody(1993)]{ps93} 
  Patterson, J. \& Szkody, P. 1993, PASP, 105, 1116
\bibitem[Paumard \etal(2001)]{pau01} Paumard, T., Maillard, J. P., Morris, 
  M., \& Rigaut, F. 2001, \aap, 366, 466
\bibitem[Pavlinsky, Grebenev, \& Sunyaev(1994)]{pgs94} Pavlinsky, M. N., 
  Grebenev, S. A., \& Sunyaev, R. A. 1994, \apj, 425, 110
\bibitem[{Pfahl} et al.(2002)]{pfa02} {Pfahl}, E., {Rappaport}, S., \& 
  {Podsiadlowski}, P. 2002, \apjl, 571, L37
\bibitem[Pollock(1987)]{pol87} Pollock, A. M. T. 1987, \apj, 320, 283
\bibitem[Pooley \etal(2002)]{pool02} Pooley, D. \etal 2002, \apj, 569, 405
\bibitem[Portegies-Zwart \etal(2002)]{poz02} Portegies-Zwart, S. F.,
   Pooley, D., \& Lewin, W. H. G. 2002, \apj, 574, 762
\bibitem[Portegies-Zwart, McMillan, \& Gerhard\etal(2003)]{poz03} 
  Portegies-Zwart, S. F. McMillan,
  S. L. W., \& Gerhard, O. 2003, \apj, 593, 352
\bibitem[Possenti \etal(2002)]{pos02} Possenti, A., Cerutti, R., Colpi, M., 
  \& Mereghetti, S. 2002, \aap, 387, 993
\bibitem[Priebisch \& Zinnecker(2002)]{pz02} Priebisch, T. \& Zinnecker, H.
  2002, \aj, 123, 1613
\bibitem[Protassov \etal(2002)]{pro02} Protassov, R., van Dyk, D. A., Connors,
  A. Kashyap, V. L. \& Siemiginowska, A. 2002, \apj, 571, 545
\bibitem[Ramsay \etal(2004a)]{ram03} Ramsay, G., Cropper, M., Mason, K. O., 
  C\'{o}rdova, F. A., \& Priedhorsky, W. 2004a, \mnras, 347, 95
\bibitem[Ramsay \& Cropper(2003)]{rc03} Ramsay, G. \& Cropper, M. 2003, 
  \mnras, 338, 219
\bibitem[Ramsay \etal(2004b)]{ram04} Ramsay, G., Cropper, M., Wu, K., 
  Mason, K. O., C\'{o}rdova, F. A., \& Priedhorsky, W. 2004, astro-ph/0402526
\bibitem[Ramsay \etal(1994)]{ram94} Ramsay, G., Mason, K. O., Cropper, M., 
  Watson, M. G., \& Clayton, K. L. 1994, \mnras, 270, 692
\bibitem[Revnivtsev \etal(2003a)]{rev03} Revnivtsev, M., Tuerler, M., Del 
  Santo, M., Westergaard,  N. J., Gehrels, N., \& Winkler, C. 2003a, IAUC, 8097
\bibitem[Revnivtsev \etal(2003b)]{rev03b} Revnivtsev, M., Sazonov, S.,  
  Gilfanov, M., \& Sunyaev, R. 2003, astro-ph/0303274
\bibitem[Rodriguez \etal(2003)]{rod03} Rodriguez, J., Tomsick, J. A., 
  Foschini, L., Walter, R. \& Goldwurm, A. 2003 IAUC 8096
\bibitem[Rutledge \etal(2001)]{rut01} Rutledge, R. E., Bildsten, L., 
 Brown, E. F., Pavlov, G. G., \& Zavlin, V. E. 2001, \apj, 551, 921
\bibitem[{Sakano} et al.(2000)]{sak00} {Sakano}, M., {Torii}, K., {Koyama}, K.,
  {Maeda}, Y., \& {Yamauchi}, S. 2000, \pasj, 52, 1141
\bibitem[Sakano \etal(2004)]{sak04} Sakano, M., Warwick, R. S., 
  Decourchelle, A., \& Wang, W. D. 2004, to appear in 
  Young Neutron Stars and Their Environments, eds. 
  Camilo, F. \& Gaensler, B. M., IAUS, 218, 183
\bibitem[Sako \etal(1999)]{sak99} Sako, M., Liedahl, D. A., Kahn, S. M., \& 
  Paerels, F. 1999, \apj, 525, 921
\bibitem[Scargle(1998)]{sca98} Scargle, J. D. 1998, \apj, 504, 405
\bibitem[Schlickeiser(2002)]{schl02} Schlickeiser, R. 2002, ``Cosmic Ray 
  Astrophysics'', Springer-Verlag, Berlin
\bibitem[Schwope \etal(2002)]{sch02} Schwope, A. D., Brunner, H., Buckley, D., 
  Greiner, J., Heyden, K. V. D., 
  Neizvestny, S., Potter, S., \& Schwarz, R. 2002, \aap, 396, 895
\bibitem[Shrader \etal(1999)]{shr99} Shrader, C. R., Sutaria, F. K., 
  Singh, K. P., \& Macomb, D. J. 1999, \apj, 512, 920
\bibitem[Singh, Drake, \& White(1996)]{sdw96} 
  Singh, K. P., Drake, S. A., \& White, N. E. 1996, \aj, 111, 2415
\bibitem[Still \& Mukai(2001)]{sm01} Still, M. \& Mukai, K. 2001, ApJ, 562, L71
\bibitem[{Sugizaki} et al.(2000)]{sug00} {Sugizaki}, M., {Kinugasa}, K., 
  {Matsuzaki}, K., {Terada}, Y., {Yamauchi}, S., \& {Yokogawa}, J. 2000, \apj,
  534, L181
\bibitem[Sugizaki \etal(2001)]{sug01} Sugizaki, M., Mitsuda, K., Kaneda, H., 
  Matsuzaki, K., Yamauchi, S., \& Koyama, K. 2001, \apjs, 134, 77
\bibitem[Swank \& Markwardt(2003)]{sm03} Swank, J. H. \&  Markwardt, C. B.
  2003, ATEL 128
\bibitem[Tan \& Draine(2003)]{td03} Tan, J. D. \& Draine, B. T. 2003, 
  astro-ph/0310442
\bibitem[Terada \etal(1999)]{ter99} Terada, Y., Kaneda, H., Makishima, K., 
  Ishida, M., Matsuzaki, K., Nagase, F., \& Kotani, T. 1999, \pasj, 51, 39
\bibitem[{Torii} et al.(1999)]{tor99} {Torii}, K., {Sugizaki}, M., 
  {Kohmura}, T., {Endo}, T., \& {Nagase}, F. 1999, \apj, 523, L65
\bibitem[Townsley \etal(2002a)]{tow02a} Townsley, L. K. \etal, 2002a, 
  NIM-A, 486, 716
\bibitem[Townsley \etal(2002b)]{tow02b} Townsley, L. K. \etal, 2002b, 
  NIM-A, 486, 751
\bibitem[Townsley \etal(2003)]{tow03} Townsley, L. K., Feigelson, E. D., 
  Montmerle, T., Broos, P. S., Chu, Y.-H., \& Garmire, G. P. 2003, \apj,
  593, 874
\bibitem[Tsuru \etal(1989)]{tsu89} Tsuru, T. \etal\ 1989, \pasj, 41, 679
\bibitem[Verbunt \etal(1997)]{ver97} Verbunt, F., Bunk, W. H., Ritter, H., \&
  Pfeffermann, E. 1997, \aap, 327, 602
\bibitem[Waldron \& Cassinelli(2001)]{wc01} Waldron, W. L. \& 
  Cassinelli, J. P. 2001, \apj, 548, L45
\bibitem[Walter \etal(2003)]{wal03} Walter, R., Rodriguez, J., Foschini, L., 
  de Plaa, J., Corbel, S., Courvoisier, T. J.-L., den Hartog, P. R., Lebrun, 
  F., Parmar, A. N., Tomsick, J. A. \& Ubertini, P. 2003, \aap, 411, L427
\bibitem[Wang \etal(2002)]{wgl02} Wang, Q. D., Gotthelf, E. V., \& 
   Lang, C. C. 2002, \nat, 415, 148
\bibitem[{Warner}(1995)]{war95} {Warner}, B. 1995, 
  {\em {Cataclysmic Variable Stars}}, Cambridge University Press
\bibitem[Wegner(1994)]{weg94} Wegner, W. 1994, \mnras, 270, 229 
\bibitem[Weisskopf \etal(2002)]{wei02} Weisskopf, M. C., Brinkman, B., 
  Canizares, C., Garmire, G., Murray, S., van Speybroeck, L. P. 2002, 
  \pasp, 114, 1
\bibitem[Wijnands \etal(2002)]{wij02} Wijnands, R., Guainazzi, M.,
  van der Klis, M., \& M\'{e}ndez, M. 2002, \apj, 573, L45
\bibitem[Willems \& Kolb(2003)]{wk03} Willems, B. \& Kolb, U. 2003, 
  \mnras, 343, 949
\bibitem[Wojdowski \etal(2003)]{woj03} Wojdowski, P. S., Liedahl, D. A., 
  Sako, M., Kahn, S. M., \& Paerels, F. 2003, \apj, 582, 959
\bibitem[Zombeck(1990)]{zom90} {Zombeck}, M.~V. 1990, 
  {\em {Handbook of Space Astronomy and Astrophysics}},
  Cambridge University Press
\end{thebibliography}
\end{document}